\documentclass[fleqn,usenatbib]{mnras}

\usepackage{newtxtext,newtxmath}

\usepackage[T1]{fontenc}
\usepackage{ae,aecompl}

\usepackage{graphicx}	
\usepackage{amsmath}	
\usepackage{amssymb}	
\usepackage{times}
\usepackage{caption}
\usepackage{subcaption}
\usepackage{fixltx2e}
\usepackage{mathrsfs}
\usepackage[section,above,below]{placeins}

\usepackage{epstopdf}

\newcommand{\FF}{\textsc{Firefly}}%

\title[\FF: a full spectral fitting code]{\FF~(Fitting IteRativEly For Likelihood analYsis): a full spectral fitting code}
\author[David M. Wilkinson et al.]{
David M. Wilkinson$^{1}$,
Claudia Maraston$^{1}$,\thanks{E-mail: claudia.maraston@port.ac.uk}
Daniel Goddard$^{1}$,
Daniel Thomas$^{1}$,
Taniya Parikh$^{1}$
\\
$^{1}$Institute of Cosmology and Gravitation, University of Portsmouth, Portsmouth, PO1 3FX, United Kingdom.}

\date{Accepted 2017 August 25. Received 2017 August 24; in original form 2015 December 18}

\pubyear{2017}

\begin{document}
\label{firstpage}
\pagerange{\pageref{firstpage}--\pageref{lastpage}}
\maketitle

\begin{abstract}
\noindent We present a new spectral fitting code, \FF, for deriving the stellar population properties of stellar systems. \FF~is a chi-squared minimisation fitting code that fits combinations of single-burst stellar population models to spectroscopic data, following an iterative best-fitting process controlled by the Bayesian Information Criterion. No priors are applied, rather all solutions within a statistical cut are retained with their weight. Moreover, no additive or multiplicative polynomia are employed to adjust the spectral shape. This fitting freedom is envisaged in order to map out the effect of intrinsic spectral energy distribution (SED) degeneracies, such as age, metallicity, dust reddening on galaxy properties, and to quantify the effect of varying input model components on such properties. Dust attenuation is included using a new procedure, which was tested on Integral Field Spectroscopic (IFS) data in a previous paper. The fitting method is extensively tested with a comprehensive suite of mock galaxies, real galaxies from the Sloan Digital Sky Survey and Milky Way globular clusters. We also assess the robustness of the derived properties as a function of signal-to-noise ratio and adopted wavelength range. We show that \FF\ is able to recover age, metallicity, stellar mass and even the star formation history remarkably well down to a $S/N\sim5$, for moderately dusty systems. Code and results are publicly available at \textsc{www.icg.port.ac.uk/firefly}.
\vspace{1.5cm}
\end{abstract}
\footnotetext[1]{E-mail: claudia.maraston@port.ac.uk, daniel.thomas@port.ac.uk}

\epstopdfsetup{outdir=./}
\section{Introduction}\label{introduction}
The comparison of model spectra to observed galaxy and star clusters spectra is essential to the interpretation of galaxy evolution problems and for determining the gravitational potential of galaxies via velocity dispersion estimates. Models at an increasingly high spectral resolution are now available (e.g. \citet{2003MNRAS.344.1000B}, \cite{2010MNRAS.404.1639V}, \citet{2011MNRAS.412.2183T}, \citet{2011MNRAS.418.2785M}), which also contain sophisticated prescriptions of stellar evolution. As the models grow in complexity and precision we are increasingly able to derive more robust stellar evolution histories of galaxies. In the recent past, several algorithms for matching models to observed spectra have been published (e.g. PPXF by \citet{2004PASP..116..138C} and \citet{2006MNRAS.366.1151S}; STARLIGHT by \citet{2005MNRAS.358..363C}; STEC(K)MAP by \citet{2006MNRAS.365...46O}VESPA by \citet{2009ApJS..185....1T}; NBURSTS by \cite{2007IAUS..241..175C}; CIGALE by \citet{2009A&A...507.1793N}.; \citet{Conroy14};  BEAGLE by \citet{2016MNRAS.462.1415C}.

In this paper, we introduce a new full spectral fitting code called \FF\ ({\bf F}itting {\bf I}te{\bf R}ativ{\bf E}ly {\bf F}or {\bf L}ikelihood anal{\bf Y}sis), designed to tackle many specific challenges in the interpretation of model spectra into galaxy properties. We shall now summarise these challenges before we introduce the specifics of our code.
\\
\\
The problem of degeneracies is one that pervades the study of spectra, especially at optical wavelengths (e.g. \citet{1994ApJS...95..107W}). Three main parameters are particularly degenerate with respect to each other, namely age, metallicity and dust. These degeneracies are inherent to theoretical spectra and can be difficult to disentangle even at high spectral resolution and accuracy. Due to this, many local minima (in a $\chi$-squared sense) may exist across the multi-dimensional parameter space, which arise from the inherent spectral degeneracies in physical properties. It is therefore important to ``map out our ignorance'' when providing an estimate of physical properties and have algorithms in place to avoid becoming `trapped' into local maxima in likelihood. In a galaxy, multiple stellar populations co-exist and so the degeneracies become even more difficult to break. Nearby globular clusters, whose stars we can individually resolve, can provide sanity checks on the modelling of galaxies by allowing us to check the results of spectral fitting with the properties determined independently through techniques such as colour-magnitude-diagram (CMD) fitting.
\\
\\
More technically, model spectra do not vary smoothly as a function of a given stellar population property such as age or metallicity, due to short-lived stellar phases contributing in different ways to the total spectral energy distribution (SED) at various wavelengths. What this means is that we use as a fitting tool, a discrete set of model spectra which correspond to a grid which may or may not be a regular. In principle, given a fine enough grid of models in parameter space, one should in theory be able to use a spectral fitting code to approximately recover any possible star formation history in a galaxy, no matter how complex the observed SED. Modern stellar population models allow this to be possible to high precision, typically having very fine grids of SEDs, especially in age, corresponding to sets of hundreds of SEDs, each corresponding to a unique stellar population age, metallicity (e.g. [Z/H]), and initial mass function (IMF), with properties like dust content and emission lines typically added at the end of the computation. 
\\
\\
In practice, a perfect decomposition is not possible in astrophysics as stellar light is emitted inversely proportional to time and strongly non linear with stellar mass, which means that the most luminous components are those that might be the less relevant in a mass budget. For a recent discussion of this problem, see \citet{2010MNRAS.407..830M}, who define it as the `iceberg effect'. This effect means that even without setting priors and using an arbitrarily fine model grid, some stellar components may remain out-shined by the latest stellar generations. In addition to this, spectral fitting at high spectral resolution involves a significant amount of computational time. Ideally, one would wish to obtain the full posterior distribution of all possible star formation histories, for example via Monte-Carlo Markov Chain (MCMC) methods. However, the time taken to undertake MCMC for hundreds of possible model spectra (i.e. dimensions) in combination is completely impractical for even the most powerful modern computing clusters. Problems tackled by such approaches (e.g. \cite{2011ApJ...737...47A}, \cite{2003MNRAS.343.1145P}, \cite{2007A&A...471...71N}, \cite{2006MNRAS.369..939S}, \cite{2013MNRAS.436.2535J}, \cite{2011ApJ...740...22S}) typically have no more than 10 dimensions. CPU time can be reduced to manageable levels at the same time as covering adequate parameter space using a variety of methods when producing a spectral fitting code. We will briefly summarise these here and direct the reader to Section~\ref{fitting} for more details of how these are overcome in this work.
\\
\\
Firstly, one may opt to use a reduced (in number) set of base models, thus reducing the dimensionality of the problem. This can be done by summing models together, either by grouping base models that are spectrally similar (as in e.g. STEC(K)MAP \citep{2006MNRAS.365...46O}), or by constructing a set of base models with layers of complexity that are used as the data requires it (e.g. as in the fitting code VESPA, \citet{2007MNRAS.381.1252T}). Most of these methods have the advantage that the physical intuition of the solutions are retained, but reduces the accuracy of the solutions and requires an input prior on the possible star formation histories, which is a non-trivial choice. Secondly, one may use an automated procedure for dimensionality reduction, such as Principle Component Analysis (PCA) (as in e.g. \citet{2012MNRAS.421..314C}), before using a simple fitting method. This has the advantage of being extremely quick to fit once the base model components are chosen, but the solutions are difficult to interpret when converting back into measurable galaxy physical properties. Finally, one may choose to use the full set of base models available, but restrict the exploration of parameters, such that the posterior probability is calculated over a smaller area or to a lower precision. STARLIGHT \citep{2005MNRAS.358..363C} does this by compressing the stellar population parameters to a coarser grid after fitting using the full available grid, choosing the coarse grid through simulations. This approach is robust but may not exploit the full precision able to be achieved with modern advancements in wavelength resolution both in models and data. PPXF~\citep{2004PASP..116..138C} uses a similar approach, in which penalty terms are applied to the residuals found from least squares fitting, in order to bias the solutions to a Gaussian shape around the best-fit values. The output of this approach is therefore a best-fit combination of base models, with errors estimated from Gaussian exploration around this best fit. Optionally, PPXF can also preferentially choose solutions with smooth star formation histories. In either case however, this approach can be prone to falling into local minima in chi-squared space, which can be prolific when attempting to recover complex star formation histories. Thus, the full range of possible spectral degeneracies cannot be explored without applying Monte Carlo simulations on the input data. 
\\
\\
In this paper, we describe the development of a new analysis tool, \FF, that follows the latter approach described in the previous paragraph, i.e. a posterior distribution of galaxy physical properties is estimated by comparing to data a large set of model stellar populations in linear combinations. Our procedure employs a chi-squared minimisation approach, but to avoid the problem of falling into local minima we use liberal parameter searching with a convergence test. This choice becomes more important as the number of models used in linear combination increases. We are then able to derive a large set of plausible star formation histories that cover a large portion of parameter space. As we will demonstrate in this paper, trusting the best fit over all other solutions can be misleading when applied to galaxy data, especially since spectral degeneracies in the models are often prominent. We take the point of view that is preferable to explore model degeneracies rather than avoid them.
\\
\\
Model SEDs depend on their input stellar spectra. In this work, we fit observational data using models by Maraston \& Str\"omback (2011) which are calculated keeping the energetics fixed and varying the input stellar spectra, using three different empirical libraries, namely MILES \citep{2006MNRAS.371..703S}, STELIB \citep{2003A&A...402..433L}, and ELODIE \citep{2007astro.ph..3658P}. This way we can assess the effect of stellar spectra on the final result on galaxy evolution (Section \ref{models}). 

In order to validate our modelling technique, we perform extensive testing making use of observed spectra of simple Milky Way populations (e.g. globular clusters), mock galaxies with well-defined properties known in advance, involving both simple and complex star formation histories, and galaxy data which have been analysed previously in the literature.
\\
\\
The paper is structured as follows. In Section~\ref{models}, we describe the stellar population models used in this work. Section~\ref{fitting} provides the details of the \FF\ algorithm, the treatment of reddening and the visualisation of results and output format. Section~\ref{comparecodes} briefly compares the \FF  procedure to those of other widely used codes. The comprehensive testing with globular clusters data and mock galaxies is found in Section~\ref{calibration}. The testing with the SDSS DR7 galaxy survey \citep{2009ApJS..182..543A} and comparisons to results obtained with other codes are shown in Section~\ref{sdss}. In Section~\ref{conclusions} we provide a discussion and present our conclusions.

\section{Input Stellar Population Models}\label{models}
\begin{figure*}
\begin{center}
\includegraphics[width=15cm]{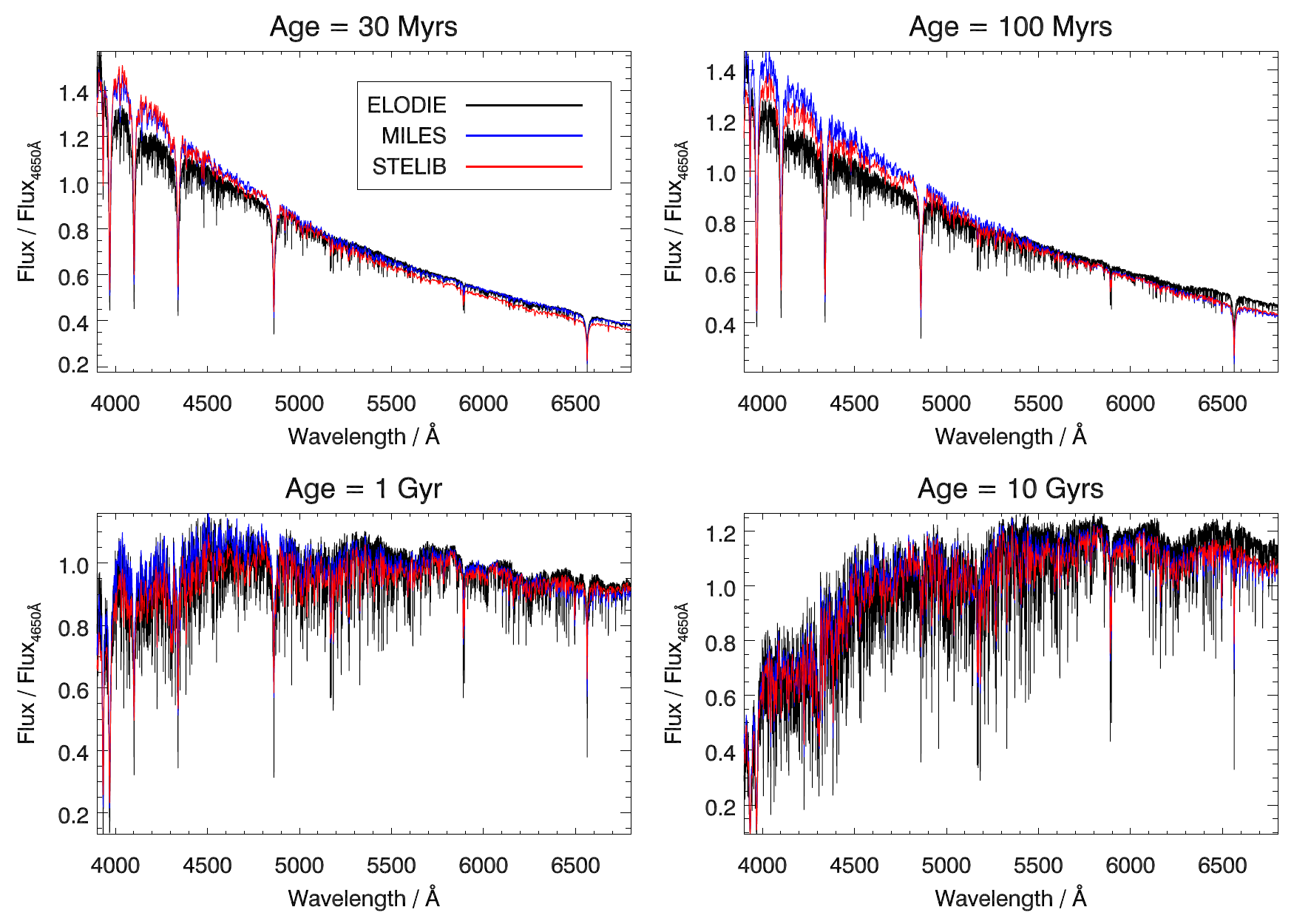}
\caption{Comparison of stellar population model spectra for different ages, 30 Myr, 100 Myr, 1 Gyr and 10 Gyr respectively, for a Kroupa initial mass function (\protect\citet{2001MNRAS.322..231K}) and solar metallicity, from the stellar population models of \protect\citet{2011MNRAS.418.2785M}.}
\label{modelspectra}
\end{center}
\end{figure*}
\FF\ can be used to fit any set of stellar population models to observational SEDs. Here we use the stellar population models of \cite{2011MNRAS.418.2785M}, hereafter M11, which are calculated for several input stellar spectra\footnote{Models are available at \url{www.maraston.eu/M11/}}. In this paper, we experiment with models based on three empirical stellar libraries\footnote{M11 also publish a set of models based on the theoretical stellar library MARCS, see detail in M11. The theoretical version of M11 models are also a standard input of \FF.}, namely MILES \citep{2006MNRAS.371..703S}, STELIB \citep{2003A&A...402..433L} and ELODIE \citep{2007astro.ph..3658P}. The M11 models make the same assumptions as \cite{2005MNRAS.362..799M} with regards to the fuel consumption in the post-Main sequence and stellar tracks and energetics, and the contribution from thermally-pulsating asymptotic giant branch (TP-AGB) stars. We summarise the differences between the models based on the stellar libraries here, but refer the reader to M11 for a more detailed description. 
\begin{table*}
\centering
{\footnotesize 
\begin{tabular}{ l | cccccc |}
\hline\hline
Model & Wavelength coverage & Age coverage & Age grid & Metallicity & HB morphology\\
 & (min -- max) / \AA & (min -- max) / Gyr & N ages & [Z/H] \\ \hline \hline
M11-MILES & 3500 -- 7429 & 5 -- 15 & 11 & [$-2.3$] & red\\ 
  &  & 5 -- 15 & 11 & [$-2.3$] & blue\\
            & & 2 -- 15 & 14 & [$-1.3$] & red \\ 
            &  & 2 -- 15 & 14 & [$-1.3$] & blue \\ 
            &  & 0.055 -- 15 & 34 & [$-0.3$] & red \\
            &  & 0.0065 -- 15 & 50 & [$+0.0$] & red \\ 
             &  & 0.1 -- 15 & 25 & [$+0.3$] & red \\ \hline 
M11-STELIB & 3201 -- 7900 & 0.2 -- 15 & 24 & [$-0.3$] & red \\ 
                      & 3201 - 9296.5 & 0.03 -- 15 & 39 & [$+0.0$] & red \\
                      
  & 3201 -- 7900 & 0.4 -- 15 & 22 & [$+0.3$] & red \\ \hline
M11-ELODIE & 3900 -- 6800 & 6 -- 15 & 10 & [$-2.3$] & red \\ 
 & & 6 -- 11 & 6 & [$-2.3$]  & blue \\
  & & 0.055 -- 15 & 34 & [$-0.3$]  & red \\
 & & 0.003 -- 15 & 57 & [$+0.0$]  & red \\ 
 & & 0.1 -- 15 & 25 & [$+0.3$] & red \\ \hline
\end{tabular}
\caption[Parameter space of the stellar population models used in this paper.]{Parameter space of the set of \cite{2011MNRAS.418.2785M} stellar population models used in this paper. For a given initial mass function M11-MILES, M11-STELIB and M11-ELODIE collect 159,  85 and  132 SSP models, respectively. These and other models are available at \protect\url{www.maraston.eu/M11}. \label{model_params}}
} \end{table*}
\\
\\
Models based on the MILES stellar library have the largest range and sampling of the stellar population parameter grid (in terms of age and metallicity), of all three models, and a high spectral resolution of 2.54 \AA~\citep{2011A&A...531A.109B}. In particular, M11-MILES models have coverage of both low and very-low metallicity populations ([Z/H] = $0.3$ and $-2.3$ respectively), albeit only for old ages (due to the Milky Way stellar age distribution at low-Z). The wavelength coverage encompasses the range 3500 -- 7429 \AA, and as discussed in M11 and \cite{2009MNRAS.394L.107M}, have a somewhat lower flux upwards of 6400 \AA~compared to STELIB and ELODIE-based models. This is due the differences in the assumed temperature scales for RGB-bump stars (see M11). M11 also published a version of MILES-based models where the temperature assigned was changed following other libraries and as a result it has a higher near-IR flux which matches the models based on other libraries (see M11, Appendix A). 
\\
\\
Models based on the STELIB stellar library cover three metallicities, half-solar, solar, and twice-solar, and lack some of the younger ages present in other models. However, they boast the largest wavelength range of 3200 -- 7900 \AA, and for solar metallicity 3200 -- 9300 \AA, with a resolution of $\sim$ 3 \AA. To keep the wavelength range consistent when comparing models at different metallicities, we limit the wavelength range of the spectral fit to 3200 -- 7900 \AA.
\\
\\
Lastly, models based on the ELODIE (v3.1) stellar library have a fair age and metallicity coverage, extending down to ultra-low metallicities ($[Z/H] = -2.3$) but lacking the low-metallicity ($[Z/H] = -1.3$) stellar populations compared to MILES-based ones. They contain the youngest stellar populations among the models used in this paper, down to just 3 Myr for solar metallicity, though older for non-solar metallicities. The observed stellar spectra used to calculate these very young populations can suffer from unknown dust reddening however, so fitting solutions containing these ages should be checked with care. ELODIE-based models have an exceptionally high resolution of 0.55 \AA, but in exchange have a fairly low wavelength coverage of 3900 -- 6800 \AA, and a somewhat too flat continuum shape at the edges (see M11, Figure~3).
\\
\\
The use of empirical libraries to calculate stellar population models comes with some drawbacks (see M11 for a detailed discussion). Specifically, there is a lack of M-type dwarf stars, limiting the coverage of temperature and surface gravity in the dwarf tail of the Main Sequence. To mitigate this, M11 used theoretical dwarf spectra from the MARCS library (Gustaffsson et al. 2008), smoothed to match the resolution of the specific empirical spectra. In addition, all these libraries lack Carbon and Oxygen-rich stars, necessary to model the TP-AGB stellar phase. To represent this phase, M11 use interpolated versions of low-resolution spectra from \cite{2000A&AS..146..217L}, as was used in \cite{2005MNRAS.362..799M}, which still provides enough resolution to identify the broad features from the component AGB stars in the resulting SEDs.
\\
\\
An additional caveat to the use of empirical libraries is that the element abundance ratios encoded in the observed stars are usually not known for all stars, making the element abundance ratio of the integrated model uncertain (see \citet{2011MNRAS.414.1227M} for an attempt to derive all chemical abundance ratios for the MILES stars). Elemental abundance ratios are probably solar-scaled at high-metallicity and $\alpha-$enhanced at low metallicity, following the pattern of chemical evolution known for the Milky Way (see extensive discussion in \citet{2003MNRAS.339..897T}). Furthermore, the flux calibration may carry additional uncertainties. The relative flux calibration of each empirical library has been qualitatively assessed in M11, finding, in summary, that the most significant difference is that cool stars are bluer, and hot stars are redder, in the MILES library compared with STELIB and ELODIE. We note already now that our method of dust and flux calibration treatment described in Section \ref{dust} should minimise flux calibration issues, as demonstrated in \cite{2015MNRAS.449..328W}.
\\
\\
The base stellar population model is referred to as simple stellar population (SSP), a model describing a coeval population of stars with a given age and chemical composition. In nature, globular clusters are generally well-represented by a single SSP and even those showing evidence of so-called multiple generations (e.g. Piotto et al. 2007)  remain simpler in terms of star formation history than galaxies.
Hence globular clusters provide a good `sanity check' on the combination of stellar population models and a fitting procedure. 
We shall fit globular cluster data in Section \ref{calibration}. 
In figure \ref{modelspectra} we have plotted the SSP spectra for the M11 empirical models at various ages, as a function of stellar library for a visual comparison. Each model clearly gives different SEDs, the differences of which change as a function of age and metallicity. In section \ref{sdss}, we describe how the differences in SEDs propagate to diverse galaxy properties.
\\
\\
Lastly we comment on the Initial Mass Function (IMF), which is a model input. The M11 models used in this paper are calculated for three IMFs for describing the distribution of stellar masses at birth, namely `Salpeter' \citep{1955ApJ...121..161S}, `Kroupa' \citep{2001MNRAS.322..231K}, and `Chabrier' \citep{2003PASP..115..763C}. Since Kroupa and Chabrier IMFs have lower amounts of low-mass stars, they have lower mass-to-light ratios, e.g. in the SDSS `i'-band Kroupa and Chabrier mass-to-light ratios are different by a factor $\sim$0.6 and $\sim$0.55 compared to Salpeter respectively. However, other than the scaling factor the impact on the SED fit is restricted to only some small changes in certain absorption features. In tests with models based on different IMFs, we did not measure any difference on the recovered SED fits greater than 0.1 \% in any of the properties.  As Kroupa IMFs are often used in the literature, in this paper we only show results from using models with a Kroupa IMF. 

\section{Fitting Method}\label{fitting}
\FF\ is a full spectroscopic fitting code of models to data, based on $\chi^2$-minimisation with treatment of the inherent spectral degeneracies and uncertainties in the data. \FF\ finds physical parameters, such as age ($t$), metallicity ($[Z/H]$) and dust ($E(B-V)$) by minimising the $\chi^2$ values of models with respect to data. Models are arbitrary combinations of individual bursts. The star formation history is reconstructed at the end of the procedure and the stellar mass is calculated. In addition to the `best fit', the code also outputs many other solutions and their associated values of $\chi^2$. Thus, it gives a likelihood surface with many maxima across a large amount of parameter space. Full details of the algorithm are given in Section \ref{algorithm}.
\\
\\
The main motivations driving the design of our code were the following:
\begin{itemize}
\item Be able to map out the inherent degeneracies in the spectra, and how they propagate into degeneracies in physical properties;
\item Be fast enough to analyse millions of galaxy spectra at high spectral resolution spanning a large wavelength range in a reasonable timeframe;
\item Allow a comparison of the different models to be conducted easily;
\item Have a method that visualises the parameters obtained conveniently so that improvements can be made incrementally;
\item Work relatively well also in low signal-to-noise (S/N) regimes;
\item Makes as few assumptions regarding star formation history and other parameters as possible, within the constraints of the available models.
\end{itemize}
In Section \ref{rationale} we explain the rationale for the first requirement in this list, using as an example case the Milky Way old open cluster M67. In Section \ref{algorithm} we detail the generalised method, with descriptions of how we treat additional complications in the spectra, including dust, complex star formation histories, emission lines, and variable wavelength ranges. In Section \ref{choosingsolution} we discuss how we obtain galaxy properties such as the average age, metallicity and total stellar mass. In Section \ref{visual} we describe the statistics of the fits obtained and how these lead to visualisation of physical properties. Lastly, in Section \ref{comparecodes} we compare our procedure with other fitting codes in the literature.

\subsection{The Age-Metallicity-Stellar Library Degeneracy}\label{rationale}

\begin{figure}
\begin{center}
\includegraphics[width=8.0cm]{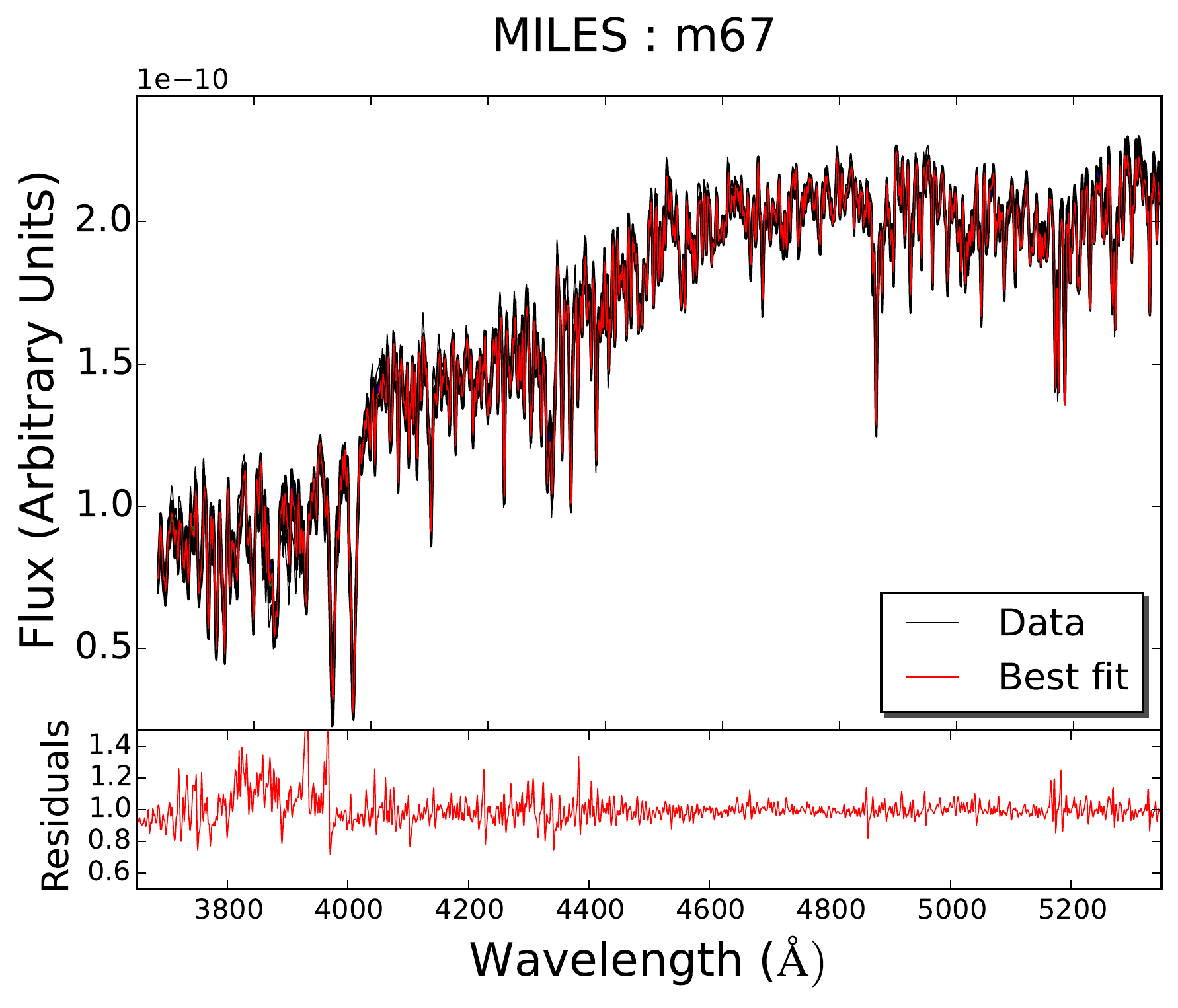}
\includegraphics[width=8.0cm]{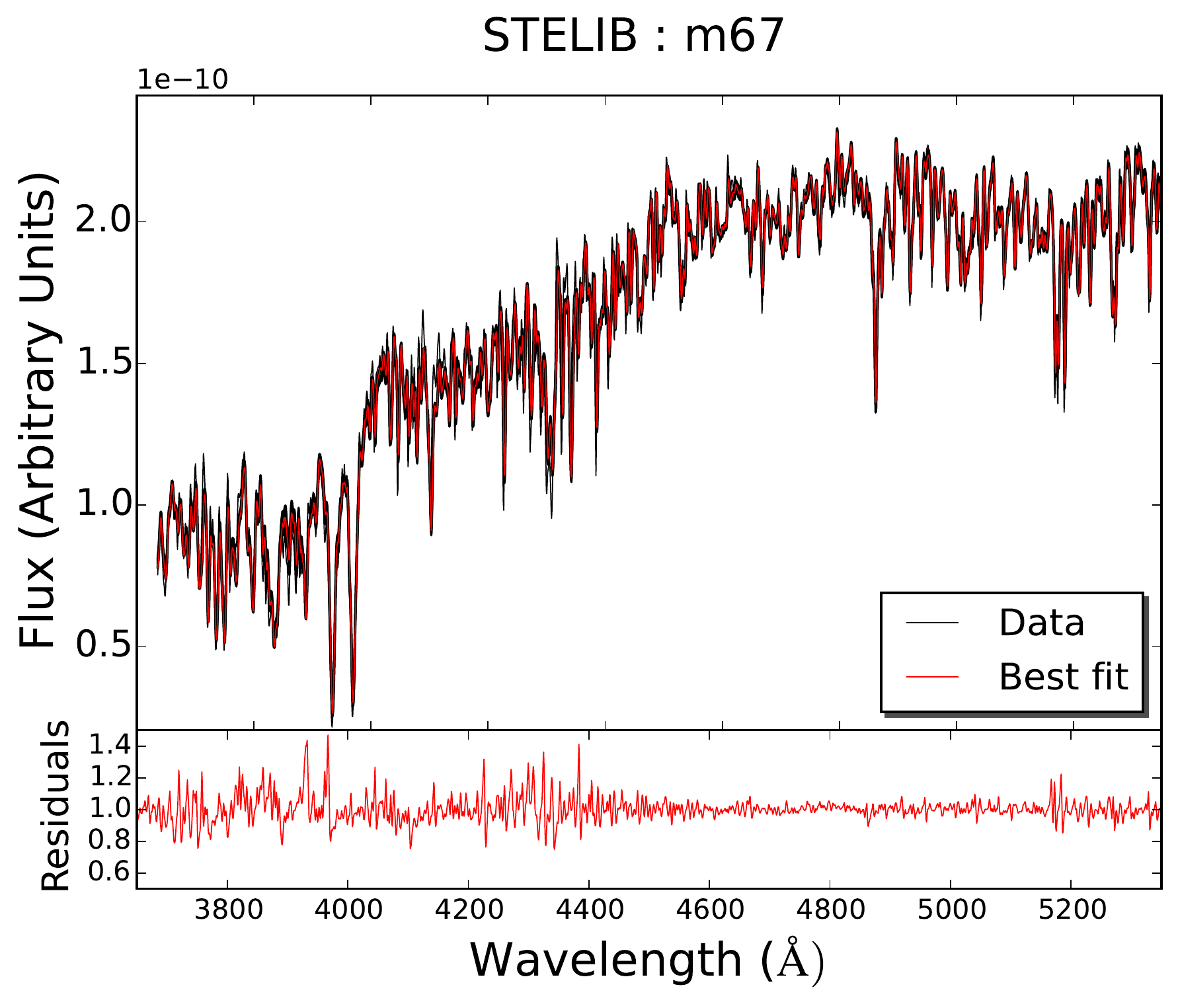}
\includegraphics[width=8.0cm]{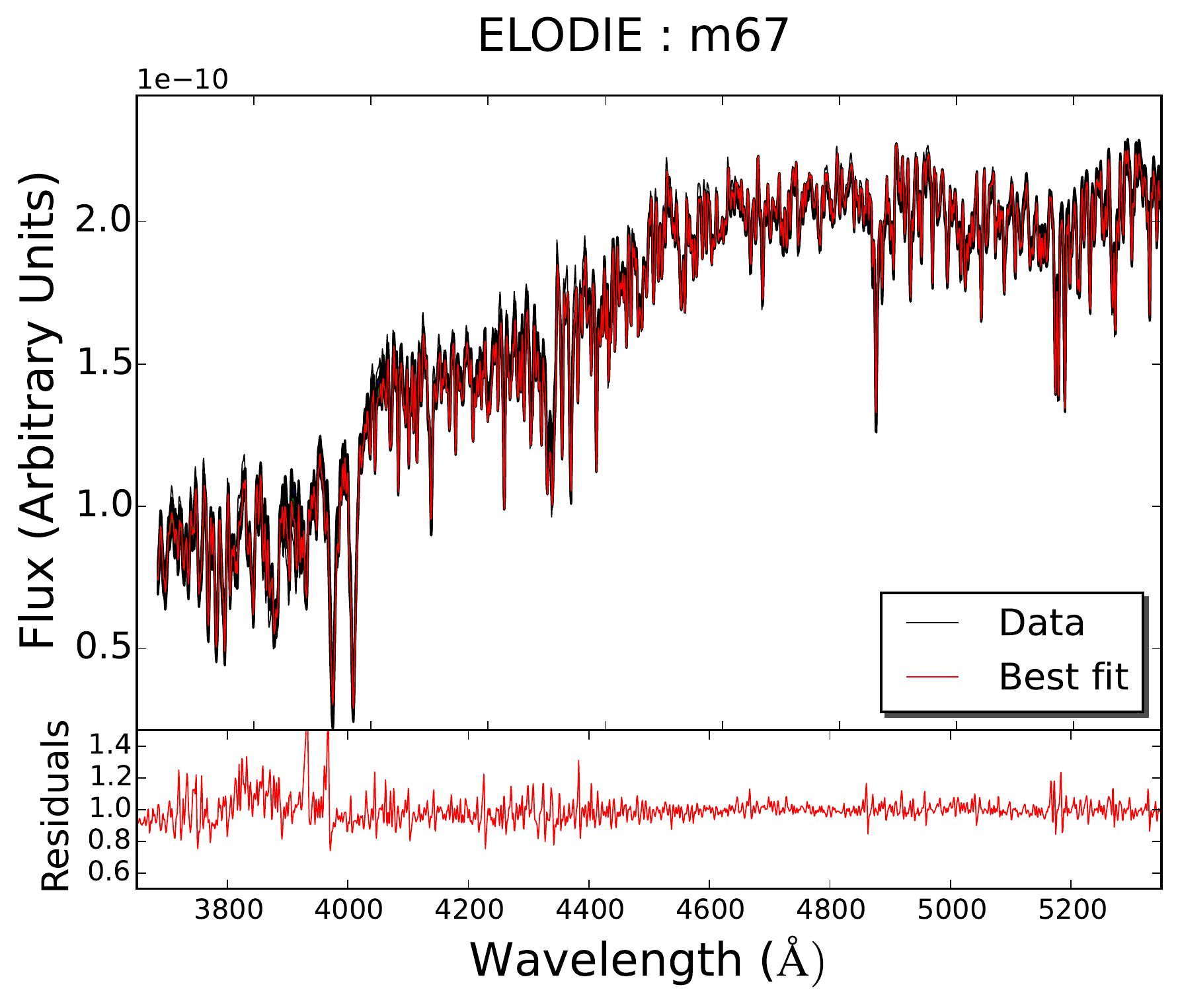}
\caption{Best fits for the star cluster M67, all of which are visually excellent. The top panel shows the fit using models based on the MILES library (M11-MILES), whereas the middle and the bottom plot show the fits using M11-STELIB and and M11-ELODIE, respectively. Each individual sub-panel shows the residual of the data and model. For the three models M11-MILES, M11-STELIB and M11-ELODIE we obtain 9, 3, and 10 Gyr and a half-solar, solar, and half-solar metallicity, respectively. These results match well with those obtained in M11.}
\label{m67_best_fits}
\end{center}
\end{figure}

\citet{2011MNRAS.418.2785M} performed a fit of their models to the optical spectrum of the Milky Way old open cluster M67 \citep{2005ApJS..160..163S}. M67 is an ideal object for the purpose of model calibration, because it is a simple stellar system, with age and metallicity determinations from fitting stellar models to the colour-magnitude diagram (CMD). A current measure of M67 age is 3.5 - 4 Gyr (see \cite{2009ApJ...698.1872S}). M67 is also a desirable object for these models because it is approximately solar in metallicity and has got solar scaled element abundance ratios, which matches with the models (see discussion in M11, and their Figure 21). Using MILES- and STELIB-based models, M11 find that different solutions corresponded to the minimum $\chi^2$, namely 9 Gyr, $0.5~Z_{\odot}$ for M11-MILES and 3 Gyr, $Z_{\odot}$, for M11-STELIB. The quality of fits in terms of absolute $\chi^2$~was the same.
As the energetics and the code are identical for the two sets of models, the difference was attributed to the effect of the input stellar library. The effect quantitatively mimics the well-known age-metallicity degeneracy (e.g. \citet{1994ApJS...95..107W}), hence we may worry about a complicated 'age-metallicity-stellar library' degeneracy. The M11 test was based on a simpler code and looked only at the minimum $\chi^2$. In this paper we shall re-address this problem comprehensively in Section \ref{m67_section}.
\\
\\
For a first test of our fitting code, we use \FF\ to repeat the M11 test and fit the M67 observed spectrum. We additionally use also the ELODIE-based M11 models. We restrict the fit to single-SSP solutions (see Section \ref{algorithm} for the general algorithm) since M67 is shown from colour-magnitude diagram fitting to be very well-represented by a simple population \citep{2009ApJ...698.1872S}. The wavelength range used for the best fits is the maximum used by the data ($3650 - 5350~\AA$) or the models. The spectra in this case are normalised to $4600-4700~\AA$~because this region is relatively free of absorption lines and matches the work of M11. Note that, in general, we normalise to as large a wavelength range as possible so that we do not need to worry about absorption features in particular regions, and so that this method still works when we are missing data from portions of the wavelength region (see Section \ref{mockwavelength}). 
\\
\\
Figure \ref{m67_best_fits} shows the results. All fits are visually very good, and their $\chi^2$'s lie within $0.1 \%$~of each other, but the recovered properties do vary. The age and metallicity of M67 are 9, 3, and 10 Gyr at half-solar, solar, and half-solar metallicity, for MILES-, STELIB-, and ELODIE-based models, respectively. Our results for MILES and STELIB are fully consistent with the M11 ones, with the STELIB-based models matching the CMD-derived results well, whereas MILES-based results suffer from the age-metallicity degeneracy, which places their lowest $\chi^2$~values at a higher age, but a lower metallicity. ELODIE-based models behave similarly to MILES-based ones. 
\\
\\
Figure \ref{m67_best_fits} demonstrates yet another problem in spectral model degeneracy, namely that the physical properties derived for stellar systems vary as a function of just one model ingredient, the input stellar library, and paint a different picture of the galaxy's current properties and star formation history.
\\
\\
In principle, the STELIB-M11 solution is the one matching the CMD solution, hence it should be retained as the only one physical. On the other hand, there is also a statistical problem, namely: is the solution for the minimum $\chi^2$~the only one we should consider? Indeed, it is easy to find combinations of age and metallicity that give very similar chi-squared values. In writing our code, we explore all solutions and retain those within a statistical significance from the best-fit.

\subsection{Algorithm}\label{algorithm}
When using a $\chi^2$-minimisation technique, as demonstrated here, in \cite{2010MNRAS.407..830M} and M11, depending on input model ingredients, the `best' solution  may not always correspond to the most realistic or precise galaxy properties. Clearly, this method of finding the physical parameters of a galaxy is inadequate, especially given the close proximity of the chi-squared values of many combinations of parameters. With this in mind we have developed our fitting code to give the additional information of not only the minimal chi-squared solution, but also a good sample of other solutions. \\

\begin{figure*}
\begin{center}
\includegraphics[width=18cm]{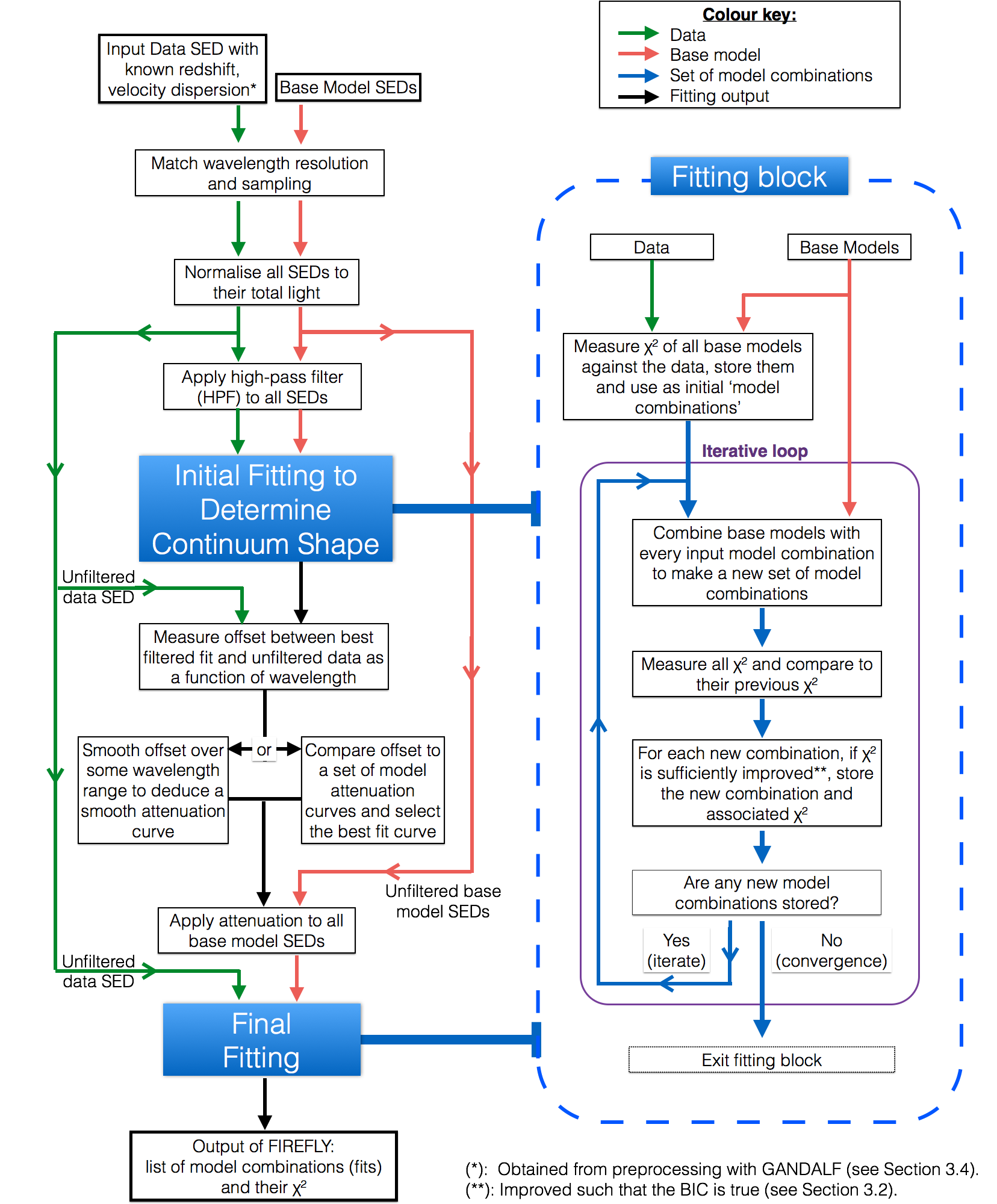}
\caption{Schematic description of \FF.}
\label{flowchart}
\end{center}
\end{figure*}

\begin{figure*}
\begin{center}
\includegraphics[width=18cm]{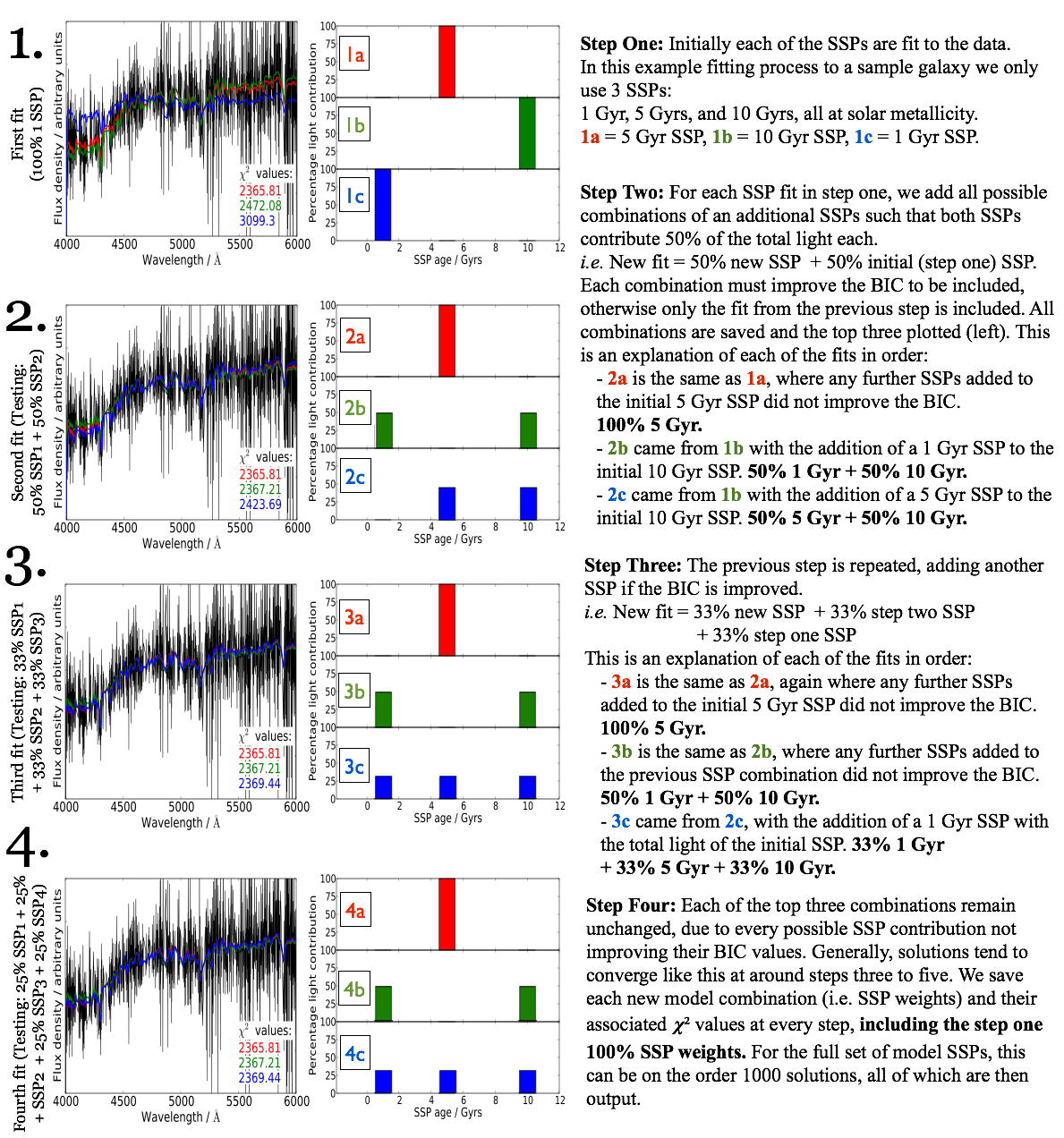}
\caption[Step-by-step example of fitting with \FF.]{Step-by-step example of running the \FF\ fitting block on an SDSS galaxy, using a reduced set of base SSPs for clarity.}
\label{examplefit}
\end{center}
\end{figure*}
Figure \ref{flowchart} describes our general procedure for fitting an object spectrum with a set of model spectra. A step-by-step visualisation of the fitting process of the full spectrum of an example SDSS galaxy is shown in Figure \ref{examplefit} for a subset of the model SEDs used. By using this parallel fitting approach we can obtain many combinations of simple stellar populations (SSPs), often giving on order 1000 solutions. The assumptions in this approach are:
\begin{itemize}
\item We are able to find good solutions by working in the basis of the SSPs (i.e. assuming the {\it individual} solutions can be modelled with a bursty star formation history). However, if we use a finely sampled in time grid of SSPs then we are able to obtain star formation histories that are effectively continuous.
\item We are able to first find a suitable SSP-based fit and then hone our overall composite star formation history by adding smaller proportions of other SSPs to this fit.
\item Solutions can be combined via $\chi^2$ likelihoods to give a physically realistic sum star formation history.
\end{itemize}
We first discuss the fitting block of the code in Figure \ref{flowchart}. The main component of this code is the iterative loop over equal-weights of base model SSPs. At each iteration of the loop, we progressively increase the size of the linear combinations used in the fits. To give a concrete example, suppose a 10 Gyr, solar metallicity SSP ($\mathrm{Fit}_{1} = M_{10 Gyr, Z_\odot}$) has a $\chi^2$ value that may be reduced by adding a 3 Gyr, half-solar metallicity component, thus creating a linear combination of SSPs as a fit to the data as $\mathrm{Fit}_{2} = \frac{1}{2}~M_{10 Gyr, Z_\odot} + \frac{1}{2}~M_{3 Gyr, 0.5 Z_\odot}$. Each of the possible combinations of two SSPs are checked for improvement on the one-SSP fits, and all fits from both one- and two-SSP fits are saved for the next iteration. This continues until convergence, see below. 

It is important to note that even if the initial combinations use equal weights to start with, not all combinations will pass the goodness threshold (see below) and the survivors will get recombined. Hence at the end solutions can have arbitrary proportions of the initial grid of SSPs.

With our approach we obtain a very large number of components, i.e. linear combinations of SSPs. This number is $\rm{N}_{\rm baseSSP}*(\rm{N}_{\rm baseSSP}-1)^{{\rm{N}_{\rm{iterations}-1}}}$. So, for example for M11-MILES containing 159 base SSPs and 4 iterations, the number of components is: $6.2\cdot10^{8}$. In practice, the majority of these components will not survive the iterations, because they will not lead to an improvement of the fit (see below). After a few iterations, the typical number of components is of the order of thousands.
\\
In order to avoid over-fitting and allow for convergence of solutions, we employ the Bayesian Information Criterion (BIC). The BIC quantifies the likelihood of the set of parameter values found given the goodness of the spectral fit (in our case, the $\chi^2$ value) and includes a penalty term that increases as the number of parameters used increases; see \cite{2007MNRAS.377L..74L} for a discussion on the BIC and other information criteria. In order to iteratively improve our fits, we require that the BIC must be improved (reduced) at each iteration of the fit in order for a new SSP contribution to be added. The BIC is defined as follows:
\begin{equation}\label{chi_test}
\	\textrm{BIC} = \chi^2 + k \cdot ln(n),
\end{equation}
where $k$ is number of fitting parameters used (in our case, the number of SSPs added in combination to make a fit), and $n$ is the number of observations (in our case, the number of flux points used in the fit). Therefore at each iteration step the $\Delta_i$BIC must be less than zero (where we take $\Delta_i x$ to mean the value of $x$ at step $i$ minus the value at step $i-1$, hence $\Delta_i k = 1$), which means that for a single iteration
\begin{equation}\label{chi_test}
\	\Delta_i\chi^2 < ln(n)
\end{equation}
must hold in order to contribute to the fit. This prevents extremely unimportant contributions to the star formation history from extending the fitting process, and hence prevents wasting CPU time for no tangible benefit to any of the physical properties obtained. 
\\
\\
We ensure that we cover adequate parameter space in order to avoid over-investigating local minima in $\chi^2$ at each iteration by allowing any combination that improves the fit beyond the median value of  $\chi^2$  computed at the previous stage. This means that as more solutions get more precise, this median value converges. 

We see that the solutions converge at around the fourth step of the iterative process, and for the vast majority of SDSS DR7 galaxies (as in Section \ref{sdss}) we find that solutions converge between the third and fifth iterations with the BIC employed. From this procedure we obtain a range of fits, each of which has a record of the linear combination of their SSP contribution, the luminosity of which is given by
\begin{equation}\label{basicfitformula}
 \  L (\lambda) = \sum\limits_{i}^{n_{SSP}} a_i L_{SSP_i}(\lambda),
\end{equation}
where $a_i$ are the weights to the base models $SSP_i$, which has luminosity $L_{SSP_i}$~and $n_{SSP}$ is the number of SSPs used in the fit. In the next subsections, we describe key steps of the fitting procedure more in-depth.

The sampling of a galaxy star formation history we can reach with this approach is just the number of entry SSPs (i.e. the base SSPs) as we do not interpolate between ages. So for example for the M11-MILES models the sampling is of 159 ages between 6 Myr and 15 Gyr.

Another interesting quantity is the smallest SSP fraction. Assuming the final solution contains 1,000 components, the minimum weight an SSP can have after 4 iterations is $1/4/1,000=0.004$, hence of the order $10^{-3}$. This is the smallest SSP fraction in this case. This smallest fraction varies according to the input grid, the number of iterations (which will also depend on the data), and how well the model fits the data. For SDSS-DR7 galaxies fitted with M11-MILES we find $~10^{-4}-10^{-3}$.

\subsection{Interstellar Reddening}\label{dust}
\FF\ takes into account both interstellar reddening of the object observed, and foreground reddening due to the Milky Way's interstellar material. We take account of the foreground reddening due to dust in the Milky Way by applying the attenuation law of \citet{1999PASP..111...63F} and using the $E(B-V)$~values from the Schlegel's maps \cite{1998ApJ...500..525S}, using the input right ascension and declination of each object. As shown in Figure \ref{flowchart} there are two methods used to fit for dust attenuation, both of which involve pre-processing the data and base models. This is described in detail in \cite{2015MNRAS.449..328W}, where we found that without a sophisticated method for treating dust attenuation, we were unable to retrieve physically realistic values for the dust properties in the case of high dust column density, and/or in data with flux calibration problems. However, this method is applicable in general to measure dust attenuation accurately, so we summarise it in the following section.
\\
\\
Note that in principle the stellar population properties of the stellar system under analysis could already be derived when performing the fit for dust. The reason for re-applying the fitting procedure to determine stellar population properties rather than using the results obtained from fitting the filtered data is that we are adding in the additional prior of assuming a smooth attenuation curve, as deduced in the fourth step of the process above. This helps us better constrain the stellar population properties, as we have the smoothed attenuation curve working as a prior. Hence the final properties we provide are those obtained by fitting again, now with the intrinsic knowledge of the attenuation curve (loop called `Fitting Block' in Figure \ref{flowchart}). 

\subsubsection{Producing Attenuation Curves from the Data}\label{default_dust}
We use an analytical function across all wavelengths to rectify the continuum before deriving the stellar population parameters. This function is called a high-pass filter (HPF). A high-pass filter applied to SEDs is able to remove the long-wavelength modes of the data, such as continuum shape and dust extinction, through the use of a window function applied to the Fourier transform of the spectra as follows:
\begin{equation}\label{hpf}
\	\mathrm{Flux}_{\lambda}^{\mathrm{output}} = \mathrm{Flux}_{\lambda}^{\mathrm{input}} \otimes \mathrm{W}_{\lambda},
\end{equation}
where this is the convolution in wavelength space, and the window function W$_\lambda = \mathscr{F}^{-1}$W$_{k}$ describes which modes, $k$, are removed. This window function is given by:
\begin{equation}\label{windowfunction}
\	\mathrm{W}_k = \begin{cases} 0 & k \le k_{crit} \\  1 & \text{otherwise,} \end{cases}
\end{equation}
where we have parameterised the form of the window function by a single value used in the filter, called $k_{crit}$. This parameter relates to masking features on wavelength scales greater than $\sim$ number of wavelength points in the spectrum divided by $k_{crit}$. Hence, an SED with 4000 wavelength points between 3000 and 7000 \AA~and a $k_{crit}$ of 40 will remove all wavelength modes greater than 100 \AA~in size. In \cite{2015MNRAS.449..328W}, we used a fixed value of $k_{crit}$ since the data always had the same wavelength range and sampling, but for a general application we set $k_{crit}$ to the number of flux points / 100, which masks any modes greater than 100 \AA~in size. This approximately corresponds to that used in \cite{2015MNRAS.449..328W} and the tests done on the consistency of solutions when varying $k_{crit}$ are the same.
\\
\\
We then use the parameters measured from this fit in the unfiltered models and data, dividing the best model fit by the data to give a residual attenuation curve. This is then smoothed over a wavelength range of at least 100 \AA~depending on the features one wishes to try to capture. Once one has fit the filtered model SEDs to the filtered data SED, the fitting block returns a measured attenuation array, which is smoothed over the wavelength scales one is trying to assess, i.e. 100 \AA~in our application. Then, in our default approach we measure the closest fit to a set of model attenuation curves to obtain an estimate of the extinction that is output by the code as a single number $(E(B-V))$, but apply the full derived attenuation array to all model SEDs and the data is refit assuming this attenuation. 
\begin{figure}
\centering
\begin{subfigure}[t]{0.99\linewidth}
	\includegraphics[width=\linewidth,clip=true,trim=0cm 0cm 0cm 1.5cm]{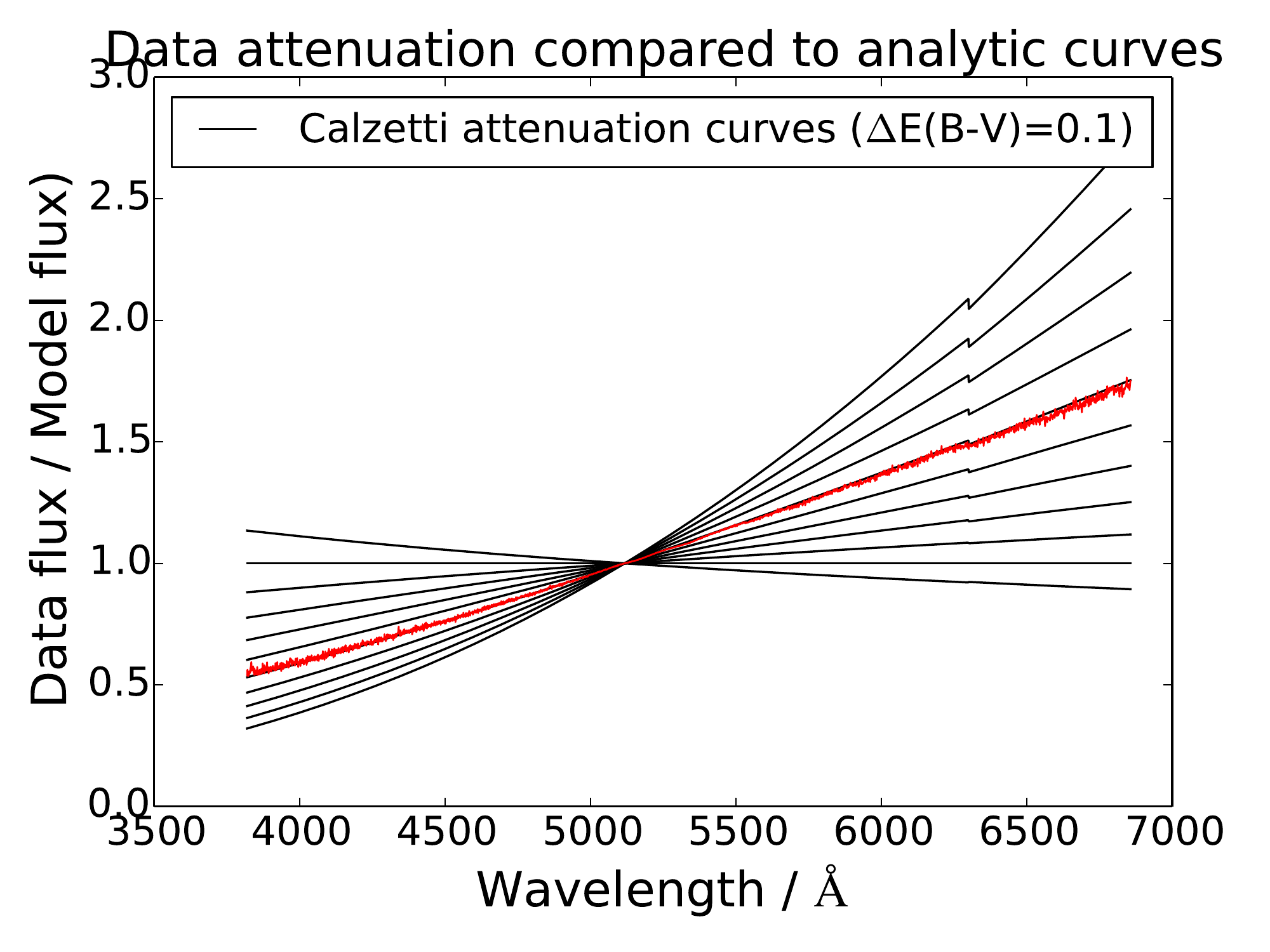}
\end{subfigure}
\begin{subfigure}[t]{0.99\linewidth}
	\includegraphics[width=\linewidth,clip=true,trim=0cm 0cm 0cm 1.5cm]{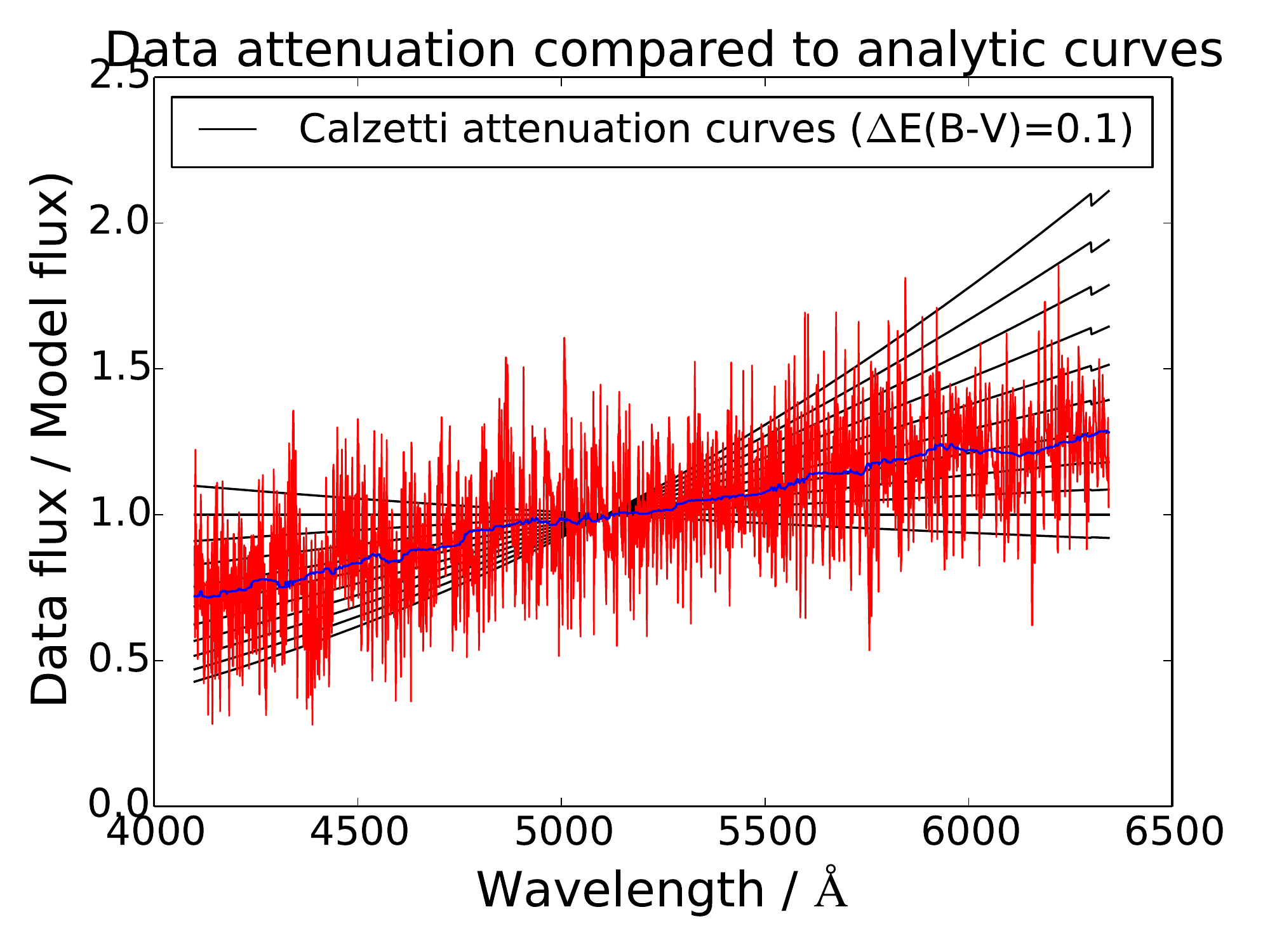}
\end{subfigure}
\caption[Measured attenuation arrays for 2 test SEDs.]{Attenuation arrays derived from \FF's fitting method of filtered data, as described in Section \protect\ref{dust}. The top panel shows the method applied to a mock galaxy spectrum consisting of a 100 Myr old SSP with an input extinction of $E(B-V)=0.5$ and $S/N=50$ . The bottom panel refers to an example SDSS DR7 galaxy (see Section \ref{sdss}) with S/N $\sim$ 5. The deduced attenuation is in red, the smoothed one, which is then applied to the models for refitting (see Section \protect\ref{default_dust}) is in blue. Calzetti et al.'s law curves for $E(B-V) = - 0.1~{\rm to}~1.0$, in steps of 0.1are shown in black. Note that in the upper panel there is no blue line as smoothing was unecessary.}
\label{hpf_examples}
\end{figure}
\\
\\
In Figure \ref{hpf_examples} we show two examples of fitting SEDs using this method. The returned attenuation arrays are shown in red, its smoothed version in blue. To guide the comparison, we also plot \cite{2000ApJ...533..682C} (see next subsection) attenuation curves for comparison from $-0.1~{\rm to}~1.0~{\rm in}~E(B-V)$, in steps of 0.1, where the zero value of extinction corresponds to a constant value of 1.0. 
\\
\\
In the top panel of Figure \ref{hpf_examples}, we have used a mock galaxy spectrum, consisting of a 100 Myr old SSP, to which we applied a Calzetti-type attenuation law with input $E(B-V) = 0.5$. The $S/N$ of this mock is 50. The recovery of the input Calzetti curve with $E(B-V) = 0.5$~is excellent as the recovered attenuation function lies almost exactly along a Calzetti's curve corresponding to $E(B-V) = 0.5$. In the lower panel we show the same exercise for an example galaxy from SDSS/DR7 galaxy (see Section \ref{sdss}) with a median S/N of $\sim$ 5. The noise is clearly seen in the raw attenuation array in red. Once smoothed however (blue curve), we can see that the Calzetti law represents the function derived from the data reasonably well, suggesting that it models the data well. In this case, the method described here and in the next section are equivalent. However, as shown in \cite{2015MNRAS.449..328W}, the method described in this section offers significant advantages when the flux calibration is uncertain.

\subsubsection{Fitting Attenuation Using General Curves}
The other method for including dust extinction in spectral fitting is to use general attenuation curves, which were derived independently of the data under analysis. Here we adopt the Calzetti law as in \citet{2000ApJ...533..682C}, using a range of values of $E(B-V)$ from 0.0 to 1.0. This law has been shown to be generally applicable to good accuracy to many different types of galaxies such as those from the SDSS surveys used in this work (\cite{2013MNRAS.432.2061C}, \cite{2012MNRAS.421.2002P}). We assume that a dust screen is uniform across the whole galaxy and thus is applied to the SSP combinations equally. This method may be used over the default HPF method of determining the attenuation curve when one wishes to assume a known law such as the Calzetti's one. Should any user of \FF\ wish to use a different extinction law for reddening, or use different parameters, the extinction law can be easily changed via swapping out the attenuation module. In both cases the attenuated SEDs are then refit to the data.
\\
\\
We tested both dust methods on $S/N>5$ mock spectra, finding good agreement between the two, with results within 0.02 dex for average age, metallicity and stellar mass. However, since the HPF method can be applied more generally (see Figure \ref{hpf_examples}) when the attenuation law is unknown and/or the flux calibration is uncertain, we set this as default in this paper

\subsection{Emission Lines}\label{elines}
\FF\ can in principle fit any spectrum either including emission-lines or not, depending on the input models. As the M11 models do not include emission-lines, to fit emission-line galaxies one should either mask out regions of emission, or find the strength of emission features and remove them before the fitting.\footnote{For emission-line removal of SDSS galaxy spectra we use the emission and absorption-line fitting code called Gas AND Absorption Line Fitting (GANDALF; \citet{2006MNRAS.366.1151S}), equipped with the same M11 stellar population models as described in \citet{2013MNRAS.431.1383T}. GANDALF accurately fits the emission lines of a galaxy, providing an emission line model spectrum. This can then be used to subtract the emission lines from the observed galaxy spectrum, to provide us with an emission-line free spectrum which we can then fit.}   Although our code can work under both circumstances, the latter method should be preferable because including as much of the SED as possible should allow an increased precision in the fits, except for cases of poor wavelength calibration (e.g. \citet{2008MNRAS.385.1998K}). This hypothesis is tested in Section \ref{calibration}.

In order to gain an idea on how critical the emission line removal is, in Appendix~\ref{App:AppendixA}, we performed the exercise of fitting a set of galaxy spectra in which emission-lines were removed or left. We find a surprisingly good agreement between the ages and metallicities derived in both cases. These results lead us to conclude that full spectral fitting of a wide enough portion of a galaxy spectrum is relatively robust to the presence of a few narrow emission lines.
\\
\\
\subsection{Broadening of the Spectrum}\label{vdisp}
The combined effect of the galaxy intrinsic velocity dispersion and of the instrumental resolution broaden absorption features in the spectra. Models have their specific spectral resolution, which varies between models and maybe different from data. \FF\ needs as input the velocity dispersion of the data, such that models can be adapted to this resolution. In this study, we use a combination of GANDALF and the pPXF code (Cappellari \& Emsellem~2004) as in Thomas et al. (2013) to measure stellar and gas kinematics. 
\\
\\
\cite{2007MNRAS.378.1550P} argue that with 3 \AA~spectral resolution models and data, there is only negligible effects of velocity dispersions on their model fitting. Hence their fitting code VESPA (see Section~4.4.3) assumes a typical galaxy velocity dispersion of 170 km/s (for the SDSS data) and downgrades models to this resolution accordingly \citep{2009ApJS..185....1T}. However, our models use somewhat higher spectral resolution and we may in the future apply our code to even higher spectral resolution data, so this assumption should be carefully tested. \\
\begin{figure}
\centering
\begin{subfigure}[t]{0.99\linewidth}
	\includegraphics[width=\linewidth]{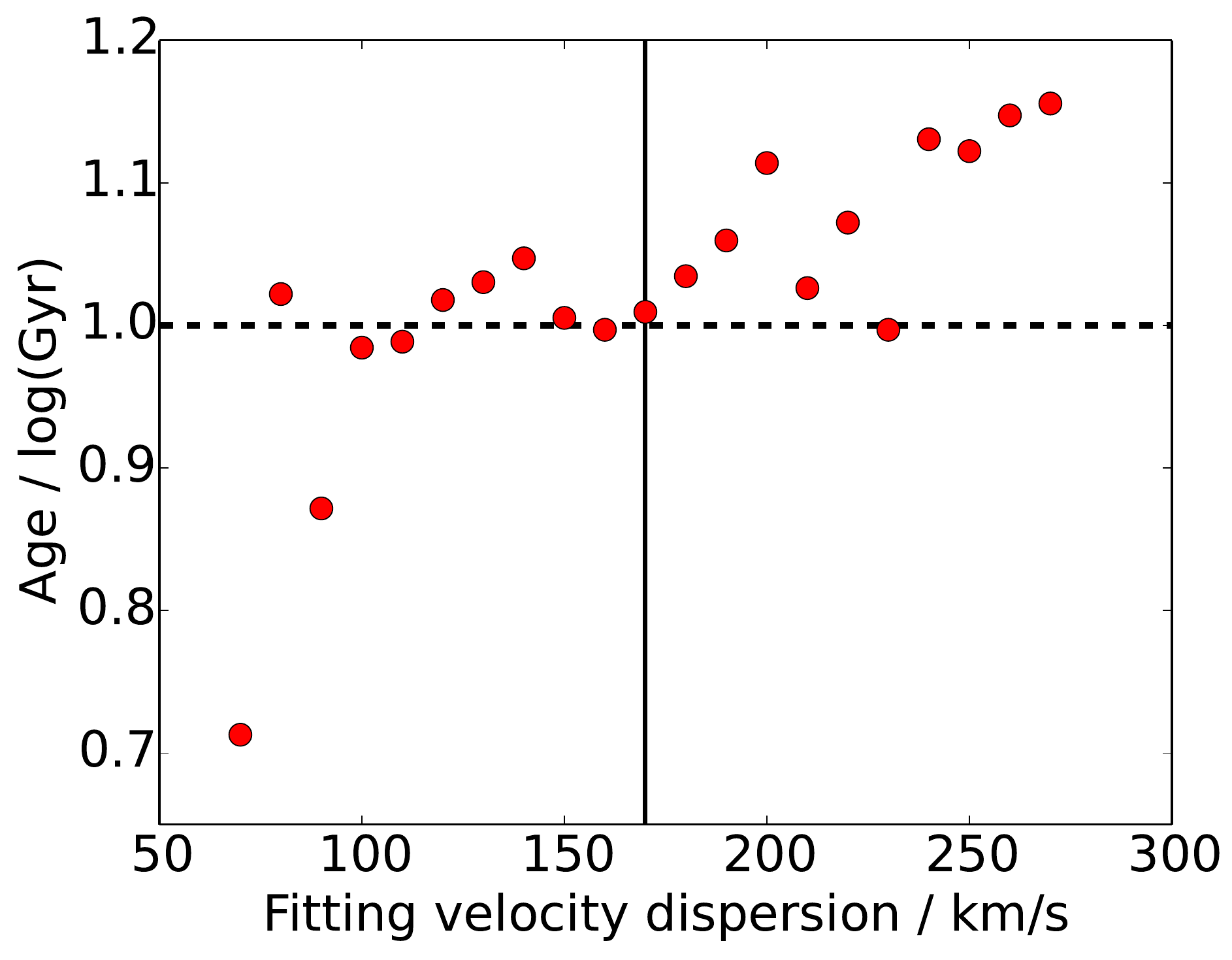}
\end{subfigure}
\begin{subfigure}[t]{0.99\linewidth}
	\includegraphics[width=\linewidth]{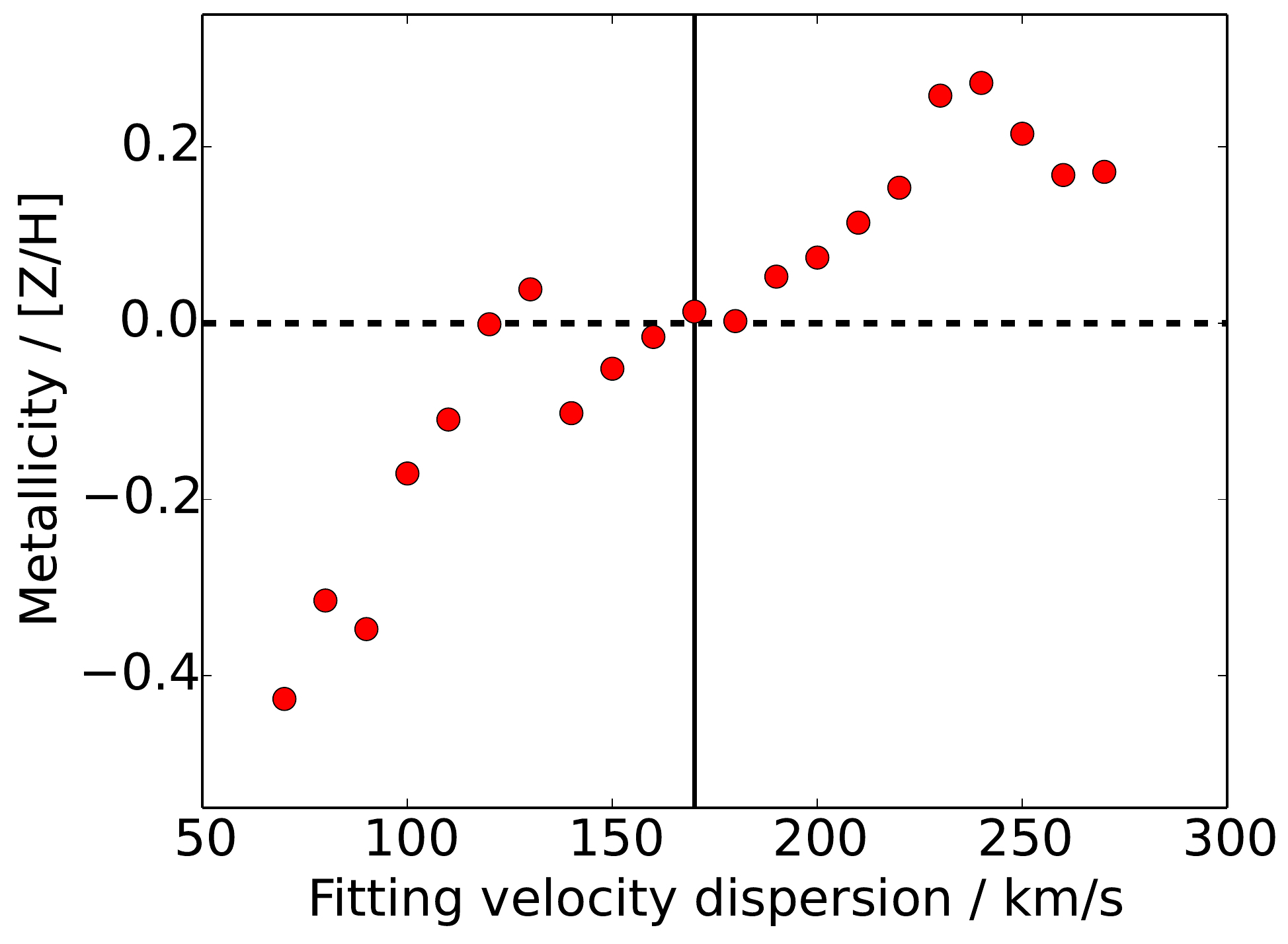}
\end{subfigure}
\caption[Recovery of age and metallicity as a function of offset in velocity dispersion.]{Recovery of age and metallicity of a 10 Gyr, solar metallicity SSP mock galaxy with 170 km/s velocity dispersion, using a range of input model velocity dispersions.}
\label{vdisp_test}
\end{figure}
\\
To this end we constructed a `mock' galaxy spectrum consisting of a 10 Gyr, solar metallicity SSP with a velocity dispersion of 170 km/s from MILES-based M11 models. We then run \FF\ using 21 sets of the same M11 models, but assume velocity dispersions from 70 km/s to 270 km/s, in 10 km/s intervals. Figure \ref{vdisp_test} shows the corresponding age and metallicity recovered when using each of these sets of models. Both panels of the Figure show an underestimate in the age and metallicity recovered when using a model velocity dispersion lower than the true one from the data, and an overestimate in the age and metallicity recovered when using a model velocity dispersion higher than the true velocity dispersion of the data. The effect in age is somewhat small in a $\pm$ 50 km/s range around the correct value, but beyond this varies by about 0.1 dex for every 20 km/s offset, going some way to support the argument of \cite{2007MNRAS.378.1550P}. However, the effect on metallicity is much more pronounced, with no clear stable region of correct metallicity recovery and instead a $\sim$ 0.1 dex displacement for every 30 km/s offset from the true input value. This effect on metallicity was also found in \citet{2008MNRAS.385.1998K} and \citet{2011MNRAS.415..709S}, where it was described as the `metallicity-velocity dispersion degeneracy'.
\\
\\ 
From these tests, we conclude that taking into account the true velocity dispersion is required to within 10 km/s velocity dispersion accuracy in order to avoid a $\sim$0.03 dex systematic error in metallicity, although this requirement is less strict for accurate age determination. This result makes intuitive sense, as one of the main effects of velocity dispersion is broadening metallic absorption lines.

\subsection{Comparing Multiple Solutions of a Fit}\label{choosingsolution}
Models are fit to data by comparing fluxes at the same wavelength. To test the goodness of each of the model fits to the data at each stage of our fitting procedure we use a chi-squared test, given by:
\begin{equation}\label{chi_test}
\	\chi^2 ( model_i | data ) = \sum\limits_{\lambda} \frac{( F_{data}(\lambda) - F_{model_i}(\lambda))^2}{\sigma(\lambda)^2},
\end{equation}
where $F$~represents the flux of the data or of model, and $\sigma$ represents the error at the wavelength point $\lambda$. Using our method we obtain a range of fits with chi-squared values that are usually close to each other (typically within a few percent of the minimum chi-squared value), which are represented by the bottom set of fits in the schematic of Figure \ref{examplefit}. For our set of models we typically have of order 1000 of these fits, all expressed as a linear combination of SSPs, as in equation \ref{basicfitformula}. To compare these fits in a statistically meaningful way for individual galaxies we must compute the cumulative likelihood function of the chi-squared probability distribution, given by:
\begin{equation}\label{chi_prob}
 \  P ( X = \chi^2_o ) = \int_{\chi^2_o}^{\infty} \frac{1}{2^{(k/2)}\Gamma(\frac{k}{2})}X^{(k/2)-1} e^{-X/2} \mathrm{d}X,
\end{equation}
where $\Gamma$ is the Gamma function, $\chi^2_o$ is the value of chi-squared we are using to find the probability density. $k$ is the degrees of freedom, which can be expressed as $k = N-\nu-1$, where $N$ is the number of (independent) observations and $\nu$ is the number of fitting parameters. Hence for our method, $N$ in the number of data flux points that we are fitting model fluxes to (of order 1000 for typical resolution and wavelength coverage of SDSS observations), and $\nu$ is the number of SSPs, $n_{SSP}$, used in linear combination to obtain our fit (which will vary according to the model library and any prior on which SSPs to include, cfr. Table~1).
\\
\\
This method relies on two assumptions. Firstly, that the fluxes and their errors are independent. This is not true for most spectroscopic surveys, such as SDSS, as fluxes from nearby pixels and their errors are somewhat correlated. Fortunately, this effect is small, on the order of the full-width at half-maximum of the instrument plus detector (see \citep{2010PASP..122..248B}, who also show how modern calibration methods could reduce this effect further). This means that the $\chi^2$ values computed from model fitting are approximately statistically correct. In any case, in \FF\ we will only output the probabilities relative to the best fit, which will be affected much less than the absolute $\chi^2$ values. Therefore, we can safely neglect this effect, especially because we use a spectrum covering a wide wavelength range. Secondly, we assume that our method explores the parameter space close to the position of the minimum $\chi^2$~value well, and this can be readily demonstrated by calculating the probabilities of similar fits\footnote{See David. M. Wilkinson PhD thesis, University of Portsmouth, UK}. 
\\
\\
The flux of the final solution of the spectral fit of the galaxy is the sum of all the fluxes of the solutions weighted by their likelihoods as:
\begin{equation}\label{sum_fits} 
 \ F(\lambda) = \sum\limits_{i}^{\mathrm{fits}}  \frac{P(\chi^2 (i)) F_i(\lambda)}{\sum_i P(\chi^2 (i))},
\end{equation}\label{prob_eqn}
where $F_i(\lambda)$ is the flux of an individual solution as described in Equation \ref{basicfitformula}. This can then be plotted as a complete spectral fit. 

\subsection{Galaxy Properties and Confidence Intervals.}\label{obtaining}
We provide best-fit physical properties and their confidence intervals for each object. For each SED analysed, we recover the likelihood distributions of galaxy properties. The best fit for each property is the peak of its distribution, with confidence intervals determined by the range of values within a given likelihood value. For example, 68\% confidence intervals in age are extracted by finding the minimum and maximum values of age that have at least 68\% likelihood, relative to the best solution. Consider for example, the lower plots of Figure \ref{example_dr7}, in which we show likelihood as a function of age. The 68\% confidence interval in this case encloses all ages with that likelihood or higher, which in this case is about 9.7 to 10.0 in $\log(t/yr)$. We provide the best fit and its 68\%, 95\%, and 99.7\% confidence intervals for all stellar population properties as default in \FF's output.
\\
\\
We note that the estimated error is based on a single SED rather than a Monte Carlo simulation based on many realisations of the SED, which is much more computationally expensive. Our method provides a fast way of measuring the spectral degeneracies between the stellar population properties based on the input data SED errors, in addition to the best fit set of properties based on this single SED. The exception to this method of evaluating model degeneracies is the dust calculation, hence the E(B-V) values, as computed in Section \ref{dust}, are not folded into these error estimates. In the next Section, we explain which types of average galaxy properties we determine and make available.

\subsubsection{Light- and Mass-Weighted Properties}
In the process of finding the contributions of stellar generations to an overall galaxy spectral fit, we normalise data and model {\it fluxes} before fitting and only fit for the spectral shape and features. Therefore, the contributions of SSP models initially obtained after a \FF\ fit are ``flux-weighted" or `light-weighted", which we shall identify as $w_{i}^{L}$\footnote{We adopt {\it light} as this is generally used, but it is important to understand that light here refers to the SED shape, and not to the amount of erg/sec.}, where i is the $i$'th SSP contribution. This is a common procedure among fitting methods (see e.g. STARLIGHT; \citet{2005MNRAS.358..363C} and STECKMAP; \cite{2006MNRAS.365...46O}). Light-weighted contributions are converted back into mass-weighted contributions after fitting as we know the relative {\it fluxes} of the models (in units of luminosity per stellar mass) compared to the data, which are the normalisation factors of models to data. 
\\
\\
We derive both mass and light-weighted properties since they can both be useful and complement each other well, identifying different processes more clearly. For example, recent star formation will dramatically reduce the light-weighted stellar age due to the high luminosity of massive stars, even when it pertains to a very small mass component. Conversely, small or negligible differences in light and mass-weighted properties will testify a small age difference between the various stellar generations, ultimately approaching a single-burst of star formation. Moderate differences can therefore be interpreted as more or less extended episode(s) of star formation. 
\\
\\
Mass-weighted properties are calculated from the light-weighted properties by the normalisation factors that were used initially to match models onto data. Data fluxes are measured in units proportional to [erg/s/\AA$\textrm{/cm}^2$], whereas model fluxes are actually luminosities hence are measured in [erg/s/\AA$\textrm{/M}_\odot$], and scaled to $1~M_{\odot}$\footnote{Scaling to 1 solar mass is true for the standard output of the Maraston's models, other models may use different units.}. Hence the contribution of each of the SSP fits in terms of stellar mass will be calculated using the normalisation factors of models onto data obtained before fitting (see Figure \ref{flowchart}). We save the values of the normalisation in \FF\ for each SSP in the model library, which we shall henceforth call ``N$_{M/D}$'':
\begin{equation}\label{modelnormal}
 \  N_{M/D}^{SSP i} = \frac{\Sigma_\lambda L [SSP i]}{\Sigma_\lambda \phi [Data]},
\end{equation}
where $L [SSP i]$ is the $i-$~SSP (model) luminosity and $\phi [Data]$ is the observed flux. The units of N$_{M/D}$ are therefore [cm$^2$ / M$_\odot$]\footnote{For example, the specific units for SDSS are [$10^{17}$cm$^2$ / M$_\odot$].}. Note that the summation corresponds to adding up fluxes or luminosities for all available wavelength data points between a defined upper and lower limit. Where possible, these wavelength limits should also be consistent when comparing models such that any systematic effect of changing the wavelength range on the stellar mass determination is reduced (see \cite{2012MNRAS.422.3285P}). The mass-weighted contribution of each SSP will then be:
\begin{equation}\label{massweights}
 \  w_{i}^{M} = \frac{w_{i}^{L}}{N_{M/D}^{SSP i}}  = w_{i}^{L} \times \frac{\Sigma_\lambda \phi [Data]}{\Sigma_\lambda F[SSP i]},
\end{equation}
where the $w_{i}^{M}$ and $w_{i}^{L}$ are the mass-weights and light-weights respectively. This expression will therefore be in units of [M$_\odot$ / cm$^2$]. The flux from the data can be converted to a luminosity knowing the object's redshift hence, assuming a cosmological model, its luminosity distance D$_L$ (in centimetres), via $L = 4 \pi D_L^2 \times \phi$. Hence, we can now convert the luminosity-weighted contributions into mass-weighted contributions as:
\begin{equation}\label{masscontributions}
\begin{aligned}
  \mathrm{M}_{SSP i} &= w_{i}^{M} \times 4 \pi D_L^2
 \\ &= 4 \pi D_L^2 \times w_{i}^{L} \frac{\Sigma_\lambda \phi [Data]}{\Sigma_\lambda L [SSP i]},
 \end{aligned}
\end{equation}
which is in units of [M$_\odot$].

\subsubsection{Total Stellar Mass and Mass Contributions}
At the end of the fitting procedure we provide global properties, such as the total mass. The total mass (stellar plus gas) is simply given by the sum of all the SSP weightings:
\begin{equation}\label{totstellarmass}
\begin{aligned}
  \mathrm{M}_{stellar}^{tot} &= \Sigma_i  M_{SSP i}  
\\ &= \Sigma ( w_{i}^{M} ) \times 4 \pi D_L^2
\\ &= 4 \pi D_L^2 \sum\limits_{i} ( w_{i}^{L} \frac{\Sigma_\lambda \phi [Data]}{\Sigma_\lambda L [SSP i]}).
\end{aligned}
\end{equation}
In addition to this, we also breakdown the total mass into more detailed segments, such as the contribution from living stars, stellar remnants and the gas that was ejected via stellar mass losses. This information is provided in tables by the models used in this work (at www.maraston.eu). The contributions depend on the assumed initial mass function (see next Section) and on the analytical prescriptions relating the initial stellar mass and their product remnant. They also depend on age, and little on metallicity (see \cite{1998MNRAS.300..872M, 2005MNRAS.362..799M}). 

\subsection{Input and Output of \FF}
In order to run \FF, the following needs to be provided as input quantities:
\begin{itemize}
\item Object's spectrum\footnote{Note that if the object contains emission lines and the models do not, the spectral fit may not be able to accurately recreate the data. This effect is discussed in Appendix~\ref{App:AppendixA}. \FF\ allows for the masking of emission lines as an input keyword.}
\item Object's velocity dispersion ($\sigma$)
\item Object's redshift 
\item Input stellar population models
\item Wavelength range for the fit
\end{itemize}
The models will be downgraded to the actual object's (galaxy) velocity dispersion ($\sigma$) after being downgraded, if necessary, to the data spectral resolution (i.e. the combination of the specific instrument resolution and intrinsic object's dispersion) using a routine developed in \citet{2013MNRAS.431.1383T}. Average fitting times for a single spectrum are $\sim 1$ minute on a standard computer. The default output of \FF\ is:
\begin{itemize}
\item Input model and fitting parameters (stellar library, IMF, wavelength range, etc.), galaxy's $\sigma$ and redshift;
\item Light-weights of each SSP entering the average solution.
\item Mass-weights of each SSP entering the average solution.
\item Light and mass-weighted averaged age and errors.
\item Light and mass-weighted averaged metallicity and errors.
\item Best-fit spectrum.
\item $\chi^2$ of the average solution.
\item $E(B-V)$~(with the method described in Sec.~3.3.1) of the average solution.
\item Total stellar mass, its fractions in stellar remnants, mass in gas.
\end{itemize}
The above outputs provide a large amount of information which should be useful in robustly determining galaxies star formation histories.

\subsection{Visualisation of Spectral Fits and Physical Properties}\label{visual}
\begin{figure*}
\begin{center}
\includegraphics[width=0.48\linewidth]{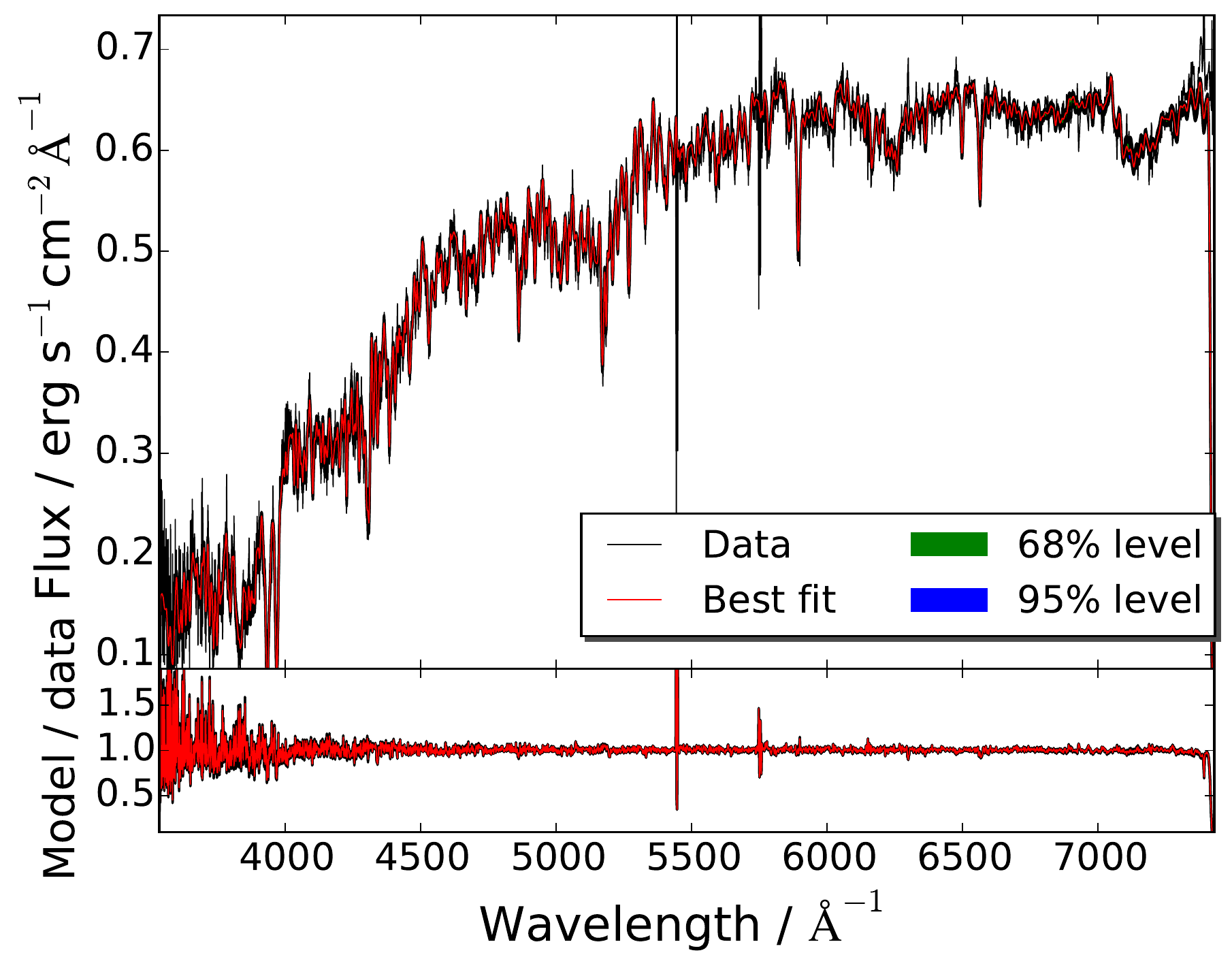}
\includegraphics[width=0.51\linewidth]{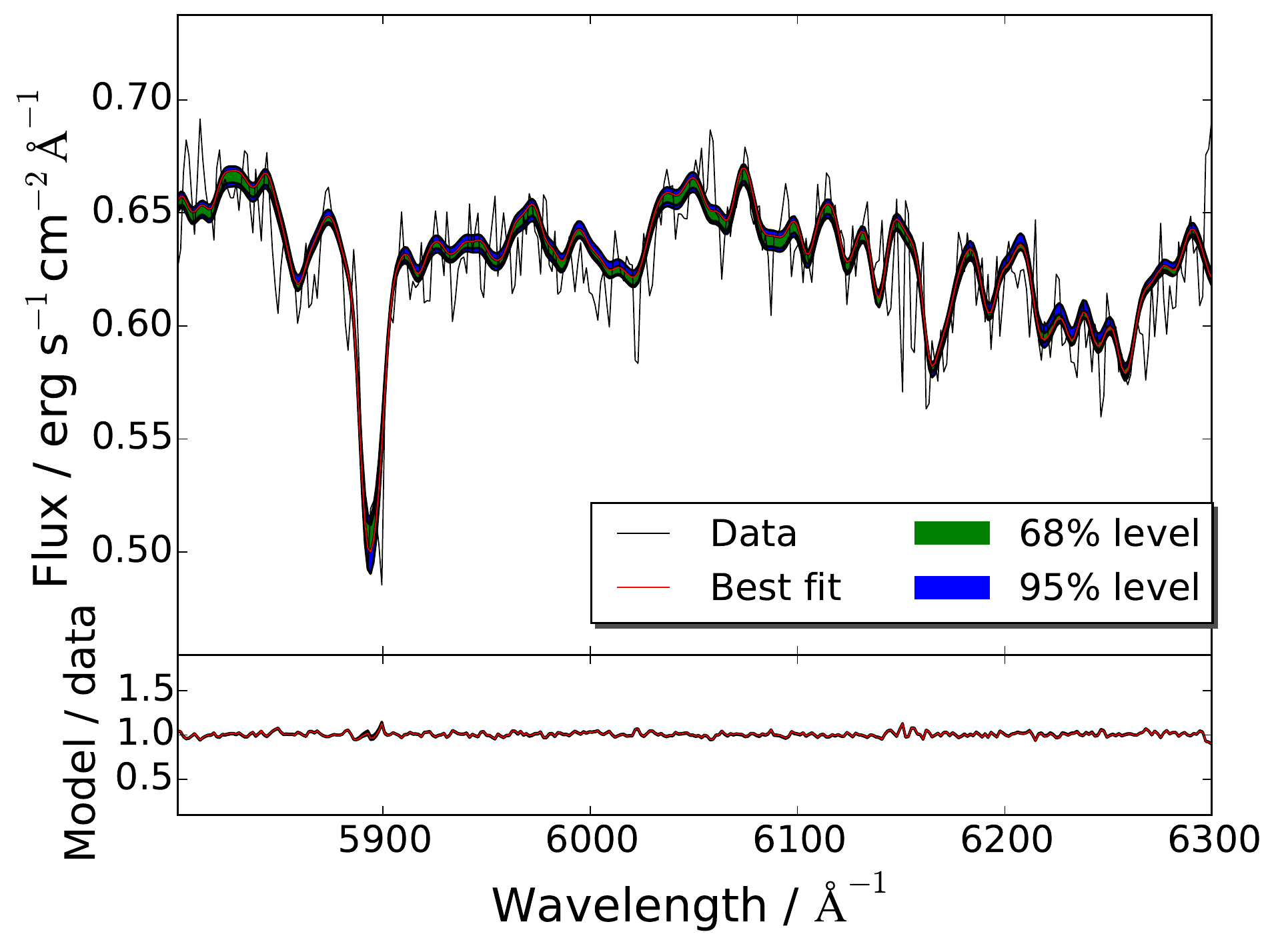}
\includegraphics[width=0.49\linewidth]{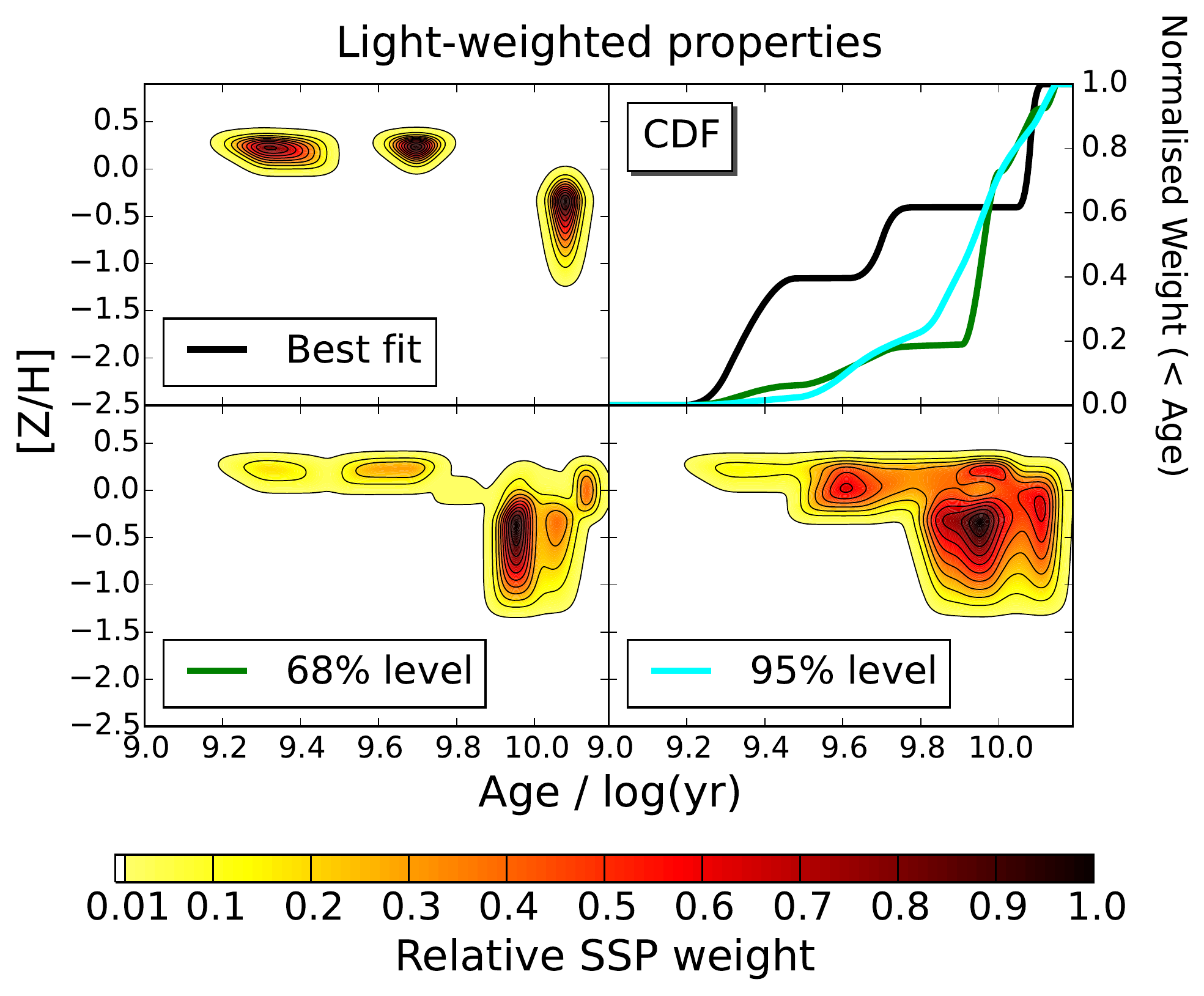}
\includegraphics[width=0.49\linewidth]{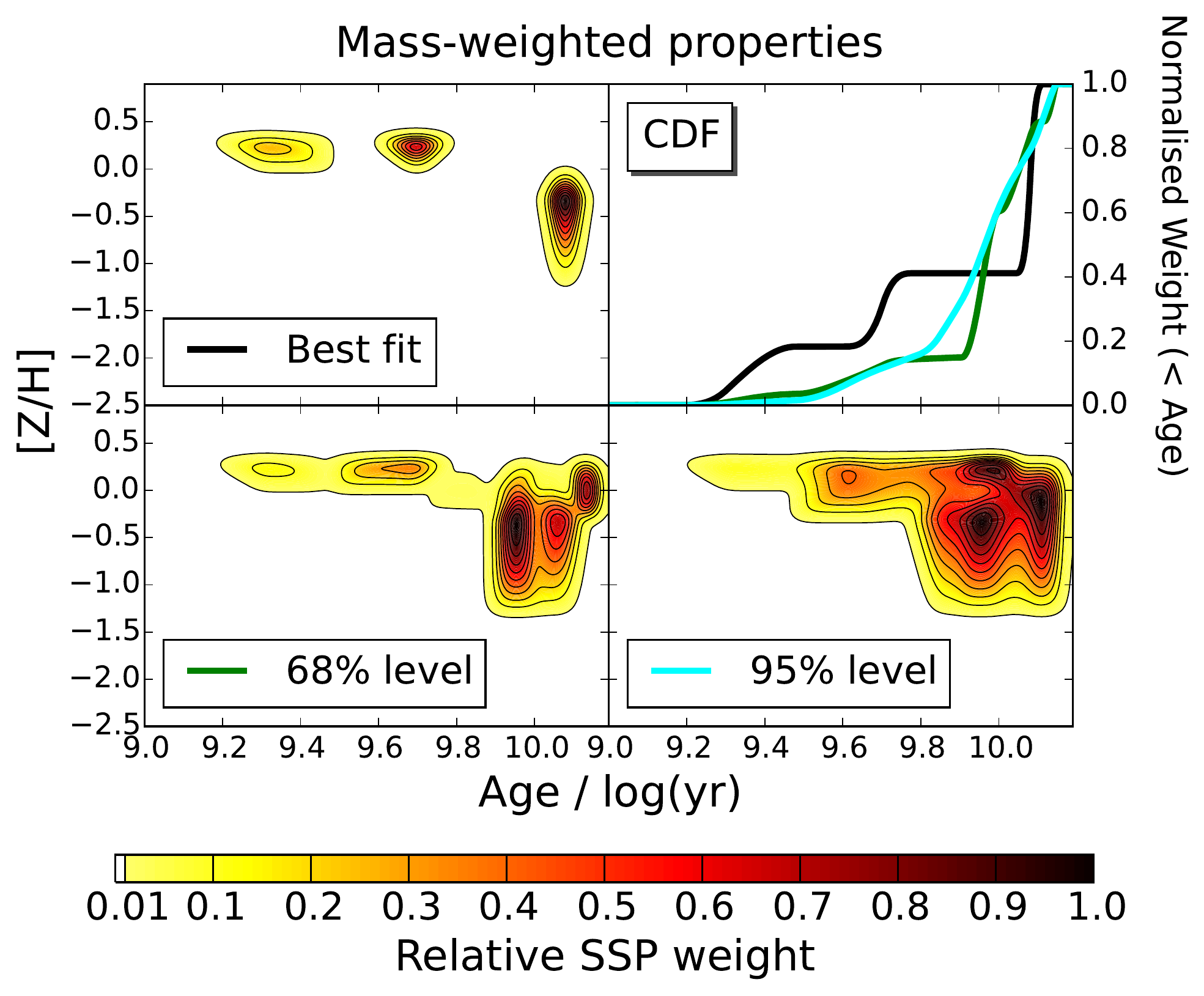}
\caption[Example fits and properties of an SDSS DR7 galaxy.]{{\it Top panels:} Example total fit (red) of a typical SDSS DR7 galaxy SED (black) with 68 percent and 95 percent confidence intervals in shaded green and blue respectively, with the residuals plotted underneath. In the left panel we show the entire spectral range, and in the right panel, to aid visualisation of the confidence intervals, we zoom in to the 5800 to 6300 \protect\AA~region. {\it Bottom panels}. Light-weighted and mass-weighted age-metallicity maps, each containing the best-fit solution, and the sum of solutions within the 68 percent and 95 percent confidence intervals. Darker contour regions correspond to higher weights of the base model SSPs found in the fits. We also apply some interpolation of the order of the sampling scale in age and metallicity of the SSPs. The age-metallicity degeneracy is visible in the spread of SSP weight contours, particularly in the island of low age, high metallicity solutions away from the dominant ones around half-solar metallicity. This is especially visible in the spread of solutions at higher confidence intervals. In the top right-sub-panels of the contours we show the cumulative distribution function of stellar age for each of the solution sets by marginalising the contour plots over metallicity. All plots in this figure are based on \FF's fits using M11-MILES models with a Kroupa IMF on the same SDSS galaxy.}
\label{example_dr7}
\end{center}
\end{figure*}
Figure \ref{example_dr7} visualises the full spectral fitting result. The final fit is a composite model containing the galaxy properties we need to know to perform galaxy evolution studies. Given the large number of contributing solutions, we need an effective way to extract them and visualise them. In the left-hand panel we show the total fit, i.e. the weighted sum of all contributions (red line) over-plotted to the empirical spectrum (black line), for a SDSS typical galaxy (see Section \ref{sdss}). The total fit includes all possible solutions as in Equation \ref{sum_fits}, in this case $\sim~2000$. The right panel is a zoomed-in version in the 5800 to 6300 \protect\AA~region. In the bottom panels we show the light-weighted and mass-weighted age-metallicity maps, each containing the best-fit solution, and the sum of solutions within the 68 percent and 95 percent confidence intervals. Darker contour regions correspond to higher weights of the base model SSPs found in the fits. 
\\
\\
The age-metallicity degeneracy is visible in the spread of SSP weight contours, particularly in the island of low age, high metallicity solutions away from the dominant ones around half-solar metallicity. This is especially visible in the spread of solutions as one considers higher confidence intervals. In the top right-sub-panels of the contours we show the cumulative distribution function of stellar age for each of the solution sets by marginalising the contour plots over metallicity. The main difference between light and mass-weighted contributions is the emphasis of the former towards the youngest population ages, which provide substantial light even when in negligible mass proportions.

\section{{Comparison to other fitting codes}}\label{comparecodes}
In this Section, we compare the methodology of \FF\ to that of other popular SED fitting codes. We reserve a comparison of results to Section \ref{sdss}. As mentioned in the Introduction, there are other fitting codes that could be described in this Section, however our focus here is on widely-used codes which have publicly available results for SDSS galaxies.

\subsection{STARLIGHT}
STARLIGHT \citep{2005MNRAS.358..363C} is a $\chi^2$-minimisation full spectral fitting code that, as \FF, uses a base of SSPs as its input. However, rather than iteratively adding SSP contributions directly, STARLIGHT explores the parameter space by finding an approximation to the minimum $\chi^2$ solution, further fine-tunes the result, and then projects the base SSPs into coarser components. Since we save a large range of fits and weight them by their final likelihood, we effectively smooth out our star formation histories across all of the fits, and so instead of requiring this base projection in order to get a realistic and stable combined fit we effectively sum over all possible solutions. \cite{2005MNRAS.358..363C} also test their procedures with mock galaxies, which are tuned to match SDSS data and so reach low S/N as ours do. They test the recovery of light and mass-weighted age and metallicity, total stellar mass, stellar velocity dispersion, and dust extinction. They report a broadly good recovery, but find it difficult to resolve the individual stellar population components, compared to our approach which works well down to a S/N of $5$. We compare with the results of STARLIGHT for SDSS DR7 in Section \ref{sdss}.
\subsection{STECKMAP}
STECKMAP \citep{2006MNRAS.365...46O, 2006MNRAS.365...74O} is a matrix inversion code also using SSPs as their base and including a penalty-based method for avoiding over-fitting in a similar way to our use of the BIC. STECKMAP retrieves likelihoods similar to \FF, but instead of combining a range of fits, performs smoothness over its solutions until robustness is achieved. This method is capable of achieving very precise recovery of sub-populations for high signal-to-noise data. The ability of the code to recover parameters successfully was shown in \citet{2006MNRAS.365...46O}, where a detailed analysis of single and double burst mock galaxies was conducted. Although the strengths of STECKMAP lie in its ability to recover parameters accurately from high S/N data, there have been studies using the code at SDSS-like signal-to-noise (see \citet{2014A&A...570A...6S}). However, since most of their work is much more tuned for high S/N objects, and different stellar population models are used, a deeper comparison between \FF\ and STECKMAP would prove difficult to make.
\subsection{VESPA}
VESPA \citep{2007MNRAS.381.1252T, 2009ApJS..185....1T} is an iterative $\chi^2$-minimisation full spectral fitting code, whose stellar population model input includes both the Bruzual \& Charlot (2003) and the Maraston (2005) and the M11 models. Fitting solutions are retrieved through the use of age bins that are larger than our single bursts. The age bins use a combination of continuous star formation rate models and exponentially-declining star formation rate models. They also test dual-burst models but find them to give inferior results in general. The resolution of the age bins varies depending on the number of parameters required to fit below the noise level, and thus the results strongly depend on the signal-to-noise of the data. This process is used to avoid over-fitting in a similar way to \FF's use of the BIC. The result of convergence in \FF\ is to return linear combination of SSPs, and include them in many other fits (weighted by likelihood) summed together to build a star formation history, whereas VESPA will return a combination of flux contributions corresponding to its age bins. Additionally they convolve a constant $\sigma = 170$ $\textrm{km s}^{-1}$ stellar velocity dispersion with the  \citet{2003MNRAS.344.1000B} models, compared to our approach of fitting for $\sigma$ separately. In Section \ref{vdisp} we show the effect such an approximation has on the derived age and metallicity across a range of simulated velocity dispersion objects. 
\\
\\ 
\citet{2007MNRAS.381.1252T} tested VESPA using mocks obtained with exponentially-declining star formation history models ($\tau$ models) with $\tau = 0.3~\textrm{Gyr}$. They test the effect of using two different wavelength ranges (1000 - 9500 and 3200 - 9500 \AA), two different S/N cases (20 and 50), and a range of dust values. It is important to note that for this value of $\tau$, our parameter recovery in \FF\ is similarly very good, even for low S/N of 5. We compare our results to those of VESPA for SDSS DR7 in Section \ref{sdss}.

\section{Testing the Fitting Method}\label{calibration}
To test the validity and stability of the results obtained from \FF\ we run a comprehensive set of tests on two types of data; mock galaxies (with both simple and complex star formation histories) and real astronomical data, comprising both globular clusters and galaxies. We use these data to assess the effect of signal-to-noise, adopted wavelength range in the fitting and star formation history, and to evaluate how realistic ages, metallicities and stellar masses derived via \FF~are. We push this testing to also investigate the recovery of the input star formation history and reddening.\\ 
The testing using SDSS galaxies is placed in Section~\ref{sdss} for clarity. This latter testing allows us to compare our results with those obtained from other fitting codes in the literature and to test the effect of the input stellar library on the resulting model fit. We test both the accuracy and the precision, in terms of error, of the recovery of stellar population properties. Our work matches for full high-resolution spectral fitting the tests by \citet{2012MNRAS.422.3285P} for broad-band spectral fitting. Throughout the next Section we show a combination of both light- and mass-weighted properties depending on which is more relevant to the dataset in question. Also note that in the following we consider the wavelength range spanned by the M11-MILES models (cfr. Table 1), but we know that the extension of the SED in wavelength especially towards the near-IR matters on improving the recovery of stellar population properties (e.g. Pforr et al. 2012).

\subsection{Effect of Signal-to-Noise, dust reddening and star formation history}\label{mockgalaxies}
To test the effect of signal-to-noise, and its relation to the assumed star formation history and dust reddening, we determine the recovery of stellar population properties from two types of mock galaxies based on model spectra which we have perturbed to simulate a range of $S/N$~ratios. 

\subsubsection{Simple Stellar Populations\label{perturbedsimple}}
\begin{figure}
\begin{center}
\includegraphics[width=8.5cm]{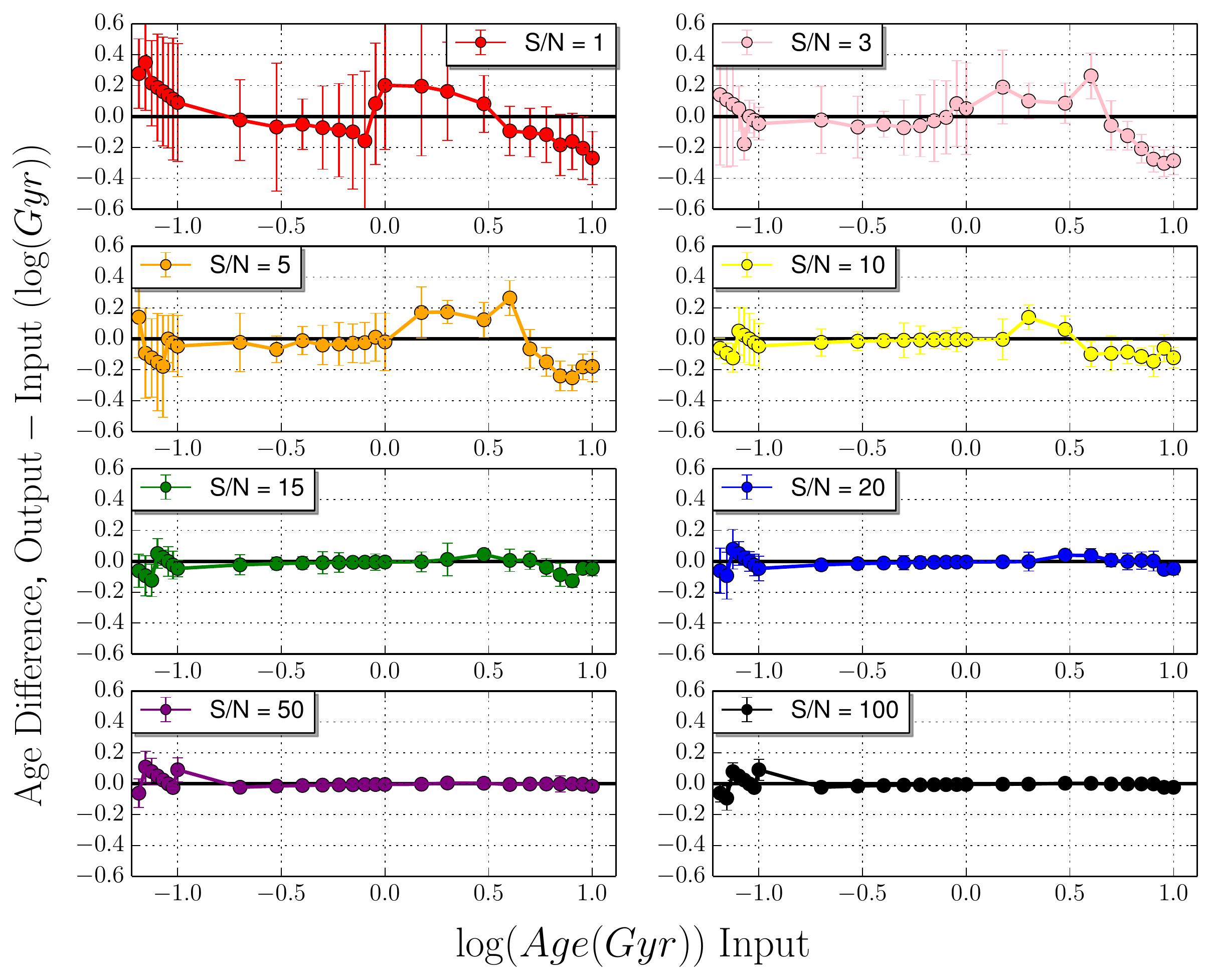}
\caption[Monte Carlo recovery of age for simple mock galaxies.]{Recovery of light-weighted ages of single-burst models perturbed as to mimic 8 different $S/N$ values. Each point is the median fitted value for the 100 Monte Carlo realisations, with errors plotted as the 68 percentile (i.e. 1 $\sigma$). Tests refer to MILES-based M11 models at the full available wavelength range available, solar metallicity, no dust.}
\label{props_mock_ssp_1}
\end{center}
\end{figure}

\begin{figure}
\begin{center}
\includegraphics[width=8.5cm]{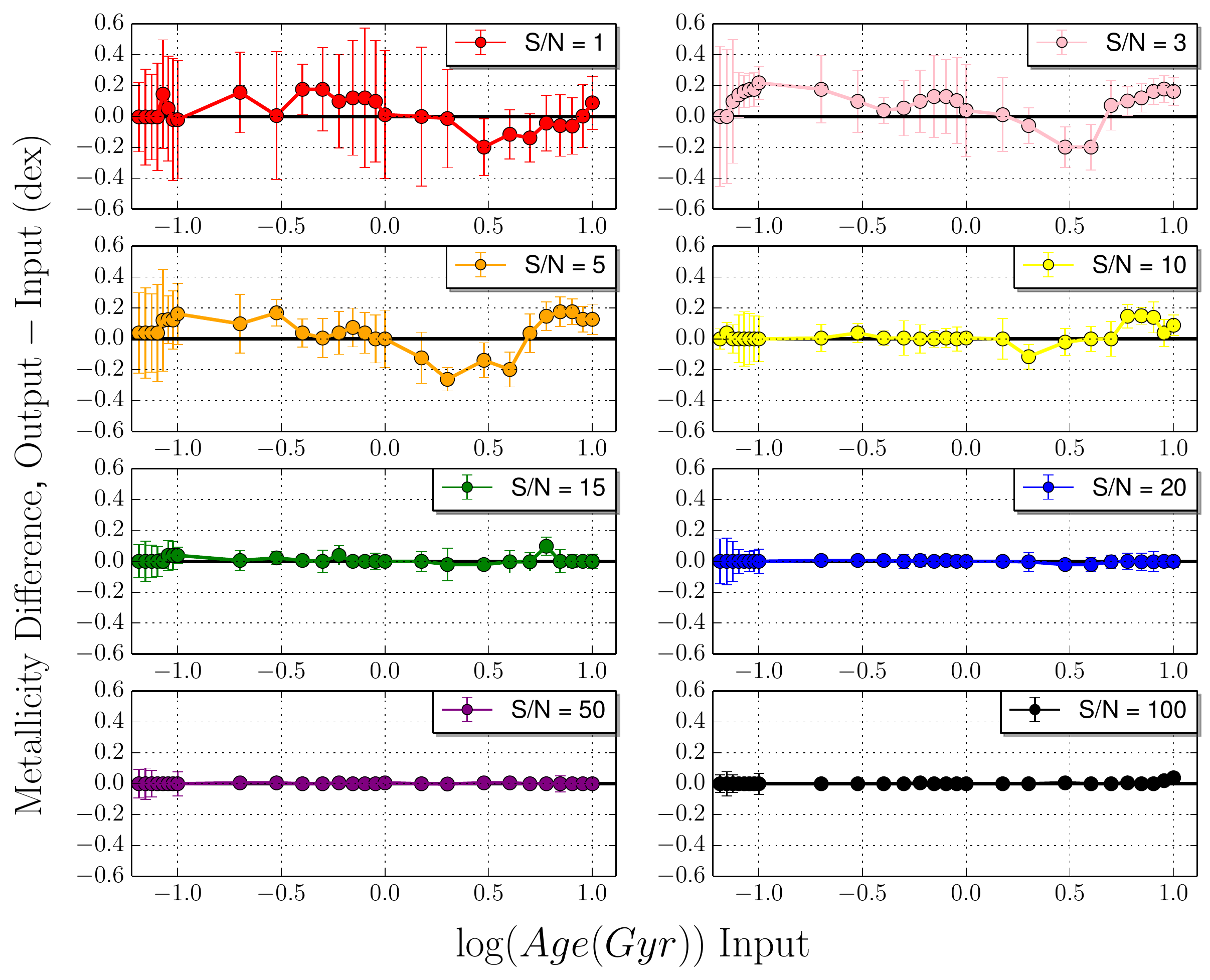}
\caption[As in Figure~\ref{props_mock_ssp_1}Monte Carlo recovery of metallicity for simple mock galaxies.]{As Figure \ref{props_mock_ssp_1}, for metallicity.}
\label{props_mock_ssp_2}
\end{center}
\end{figure}

\begin{figure}
\begin{center}
\includegraphics[width=8.5cm]{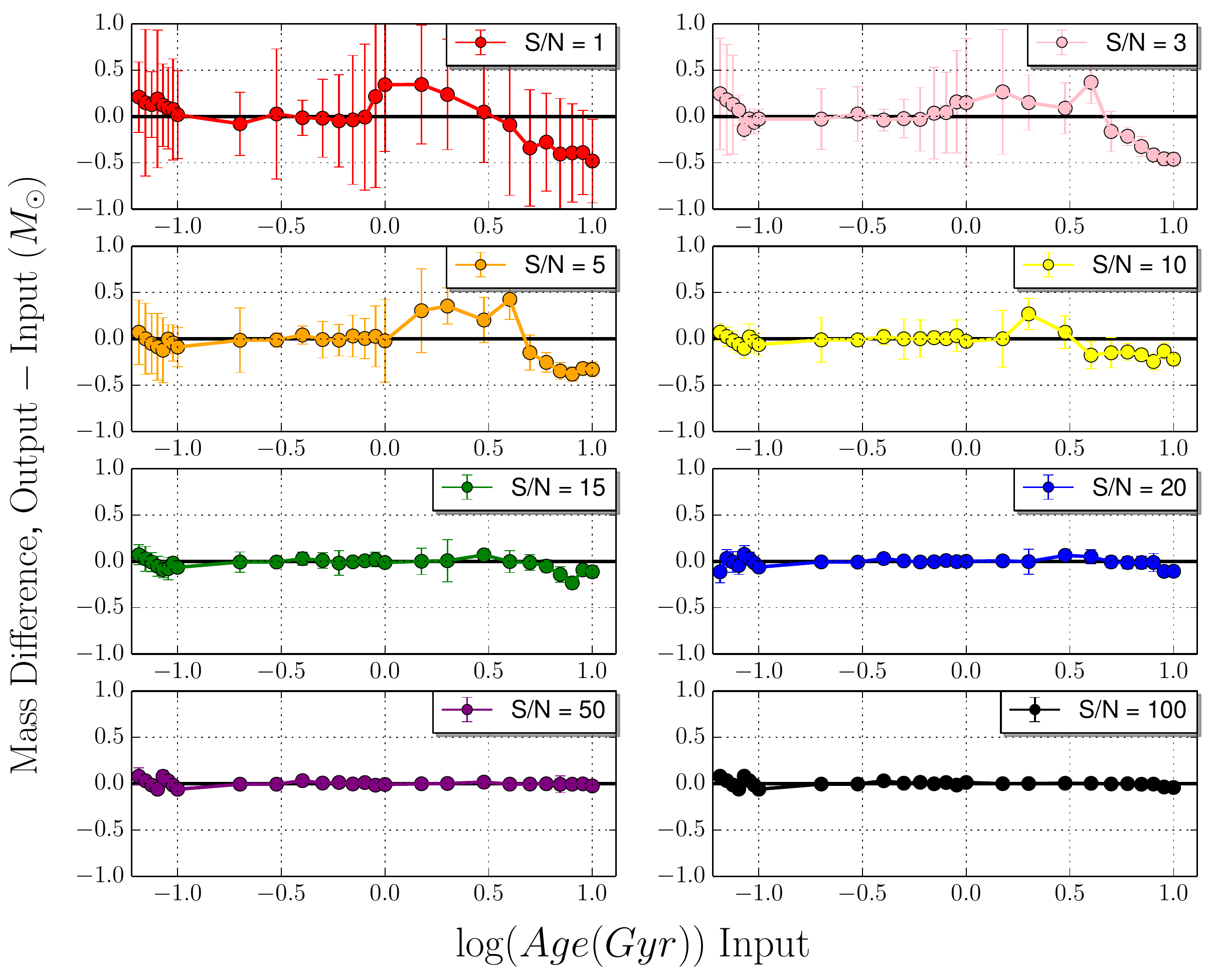}
\caption[Monte Carlo recovery of mass for simple mock galaxies.]{As Figure \ref{props_mock_ssp_1}, for stellar mass.}
\label{props_mock_ssp_3}
\end{center}
\end{figure}

In the first set of tests we use mock galaxies made from the same simple stellar population templates used to fit to them. We adopt the M11-MILES set, since it has the largest metallicity coverage, and use the full wavelength range available (see Table 1). We apply a gaussian perturbation to each flux point with a flat signal-to-noise, as described by:
\begin{equation}\label{mock_flux}
\	F_{Mock}(\lambda_i) = N\left(F_{SSP}(\lambda_i),\left(\frac{\phi_{SSP}}{S/N}\right)^2\right),
\end{equation}
where $F_{Mock}(\lambda_i)$ is the individual flux point $i$ of the mock galaxy, $F_{SSP}(\lambda_i)$ is the flux point $i$ of the input SSP, $N(x,\sigma^2)$ is the normal distribution with mean $x$ and variance $\sigma^2$, $\phi_{SSP}$ is the mean flux over the whole wavelength range of the spectrum and $S/N$ is the input signal-to-noise. We explore $S/N$~of 1, 3, 5, 10, 15, 20, 50 and 100. The 1 to 100 range covers the vast majority SDSS galaxies with higher values usually corresponding to stacked spectra. The highest signal-to-noise value case allows us to measure the intrinsic model degeneracies. For this first test we do not include dust, which will be considered in the next section when we use composite models.
\\
\\
We create a Monte Carlo simulation of 100 realisations by applying these gaussian perturbations 100 times to each SSP for each signal-to-noise bin. In Figures \ref{props_mock_ssp_1}, \ref{props_mock_ssp_2} and  \ref{props_mock_ssp_3} we plot the age, metallicity and stellar mass recovered as a function of SSP age for each S/N bin. Each point represents the average value of 100 realisations of that SSP, with the errors measured from the standard deviation of the fitted ages. The line drawn at $y$=0 visualises where the fitted property matches the input one. \\
\\
Figure \ref{props_mock_ssp_1} to Figure  \ref{props_mock_ssp_3} report the results, for age, metallicity and stellar mass, respectively. 
The recovery of the input properties is excellent for all properties, down to $S/N=5$. It remains acceptable down to $S/N=3$~at intermediate ages, while towards old ages ($t>1$~Gyr), the age-metallicity degeneracy kicks in and leads to underestimating age, overestimating metallicity and underestimating stellar mass. These results are very encouraging for our code because the code fits combinations of SSPs with dust and such a remarkable agreement for a single SSP with no dust is not trivial. We also note that when fitting the mock spectra, we always find negligible reddening, consistent with the input. However, these plots are not shown for brevity. These results will support the use of our code for low S/N populations such as those of galactic halos or high-redshift galaxies. 
\\
\\
\citet{2012MNRAS.421..314C} note that stellar masses of SDSS-III galaxies obtained by full spectral fitting of spectra with $S/N\sim5$~were larger than the stellar masses calculated for the same galaxies, but using broad-band spectro-photometry (as in \citet{2013MNRAS.435.2764M}). \citet{2012MNRAS.421..314C} show that as the S/N ratio increases, the stellar mass derived from their full spectral fitting tends to approach the one they derive from broad-band photometry fitting. Our results from full spectral fitting seem more optimistic in that even at low S/N the stellar mass is well recovered. We should note that the spectral fitting method by \citet{2012MNRAS.421..314C} employs principal component analysis and is overall different from the one we take here.
\\
\\
\subsubsection{Extended star formation histories}
\begin{figure*}
\centering
\hspace{0.3cm}
\begin{subfigure}{0.46\linewidth}
	\includegraphics[width=\linewidth]{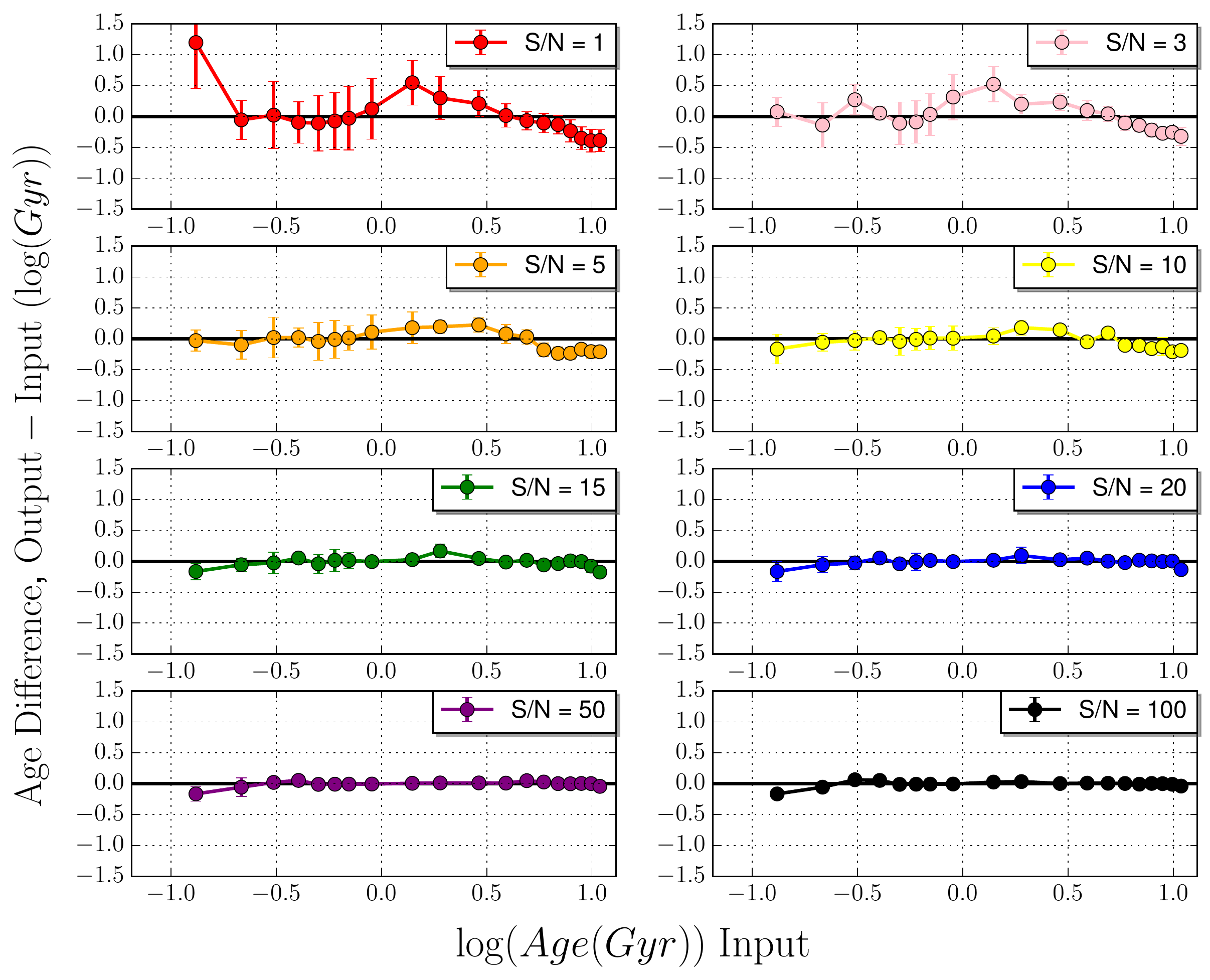}
	\caption{$\tau$ = 0.1 Gyr, age recovery.}
\end{subfigure} 
\begin{subfigure}{0.46\linewidth}
	\includegraphics[width=\linewidth]{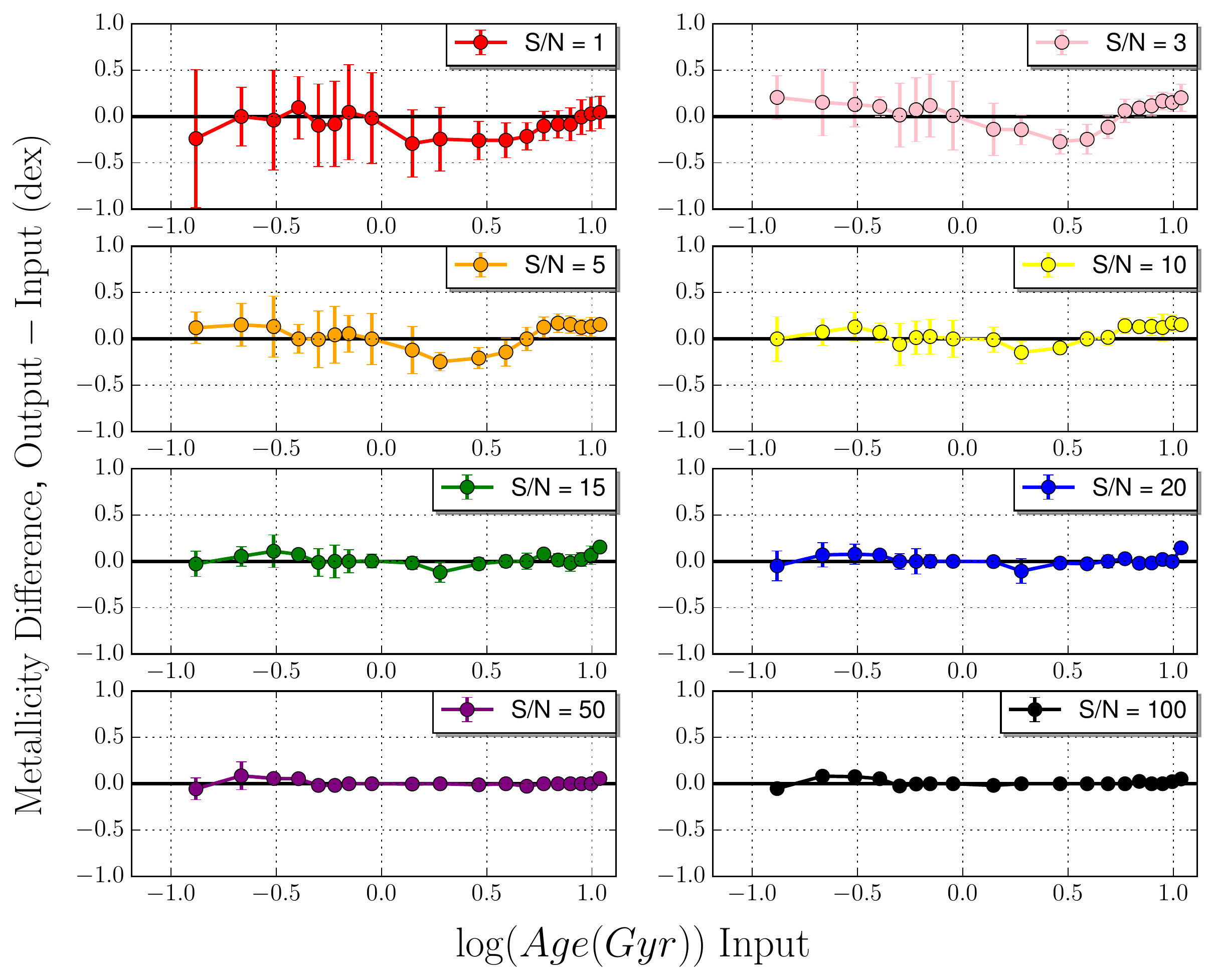}
	\caption{$\tau$ = 0.1 Gyr, metallicity recovery.}
\end{subfigure}
\begin{subfigure}{0.46\linewidth}
	\includegraphics[width=\linewidth]{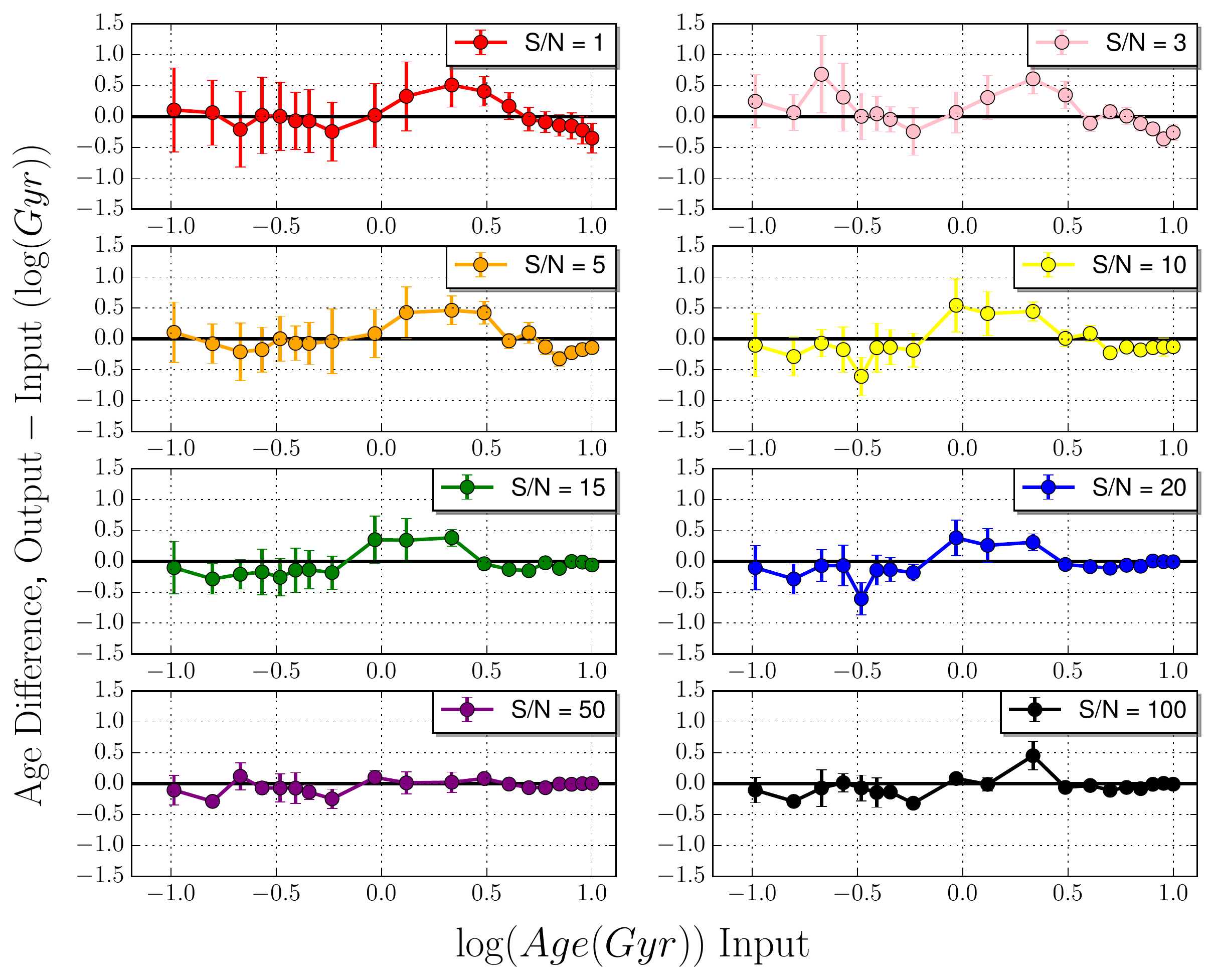}
	\caption{$\tau$ = 1 Gyr, age recovery.}
\end{subfigure}
\begin{subfigure}{0.46\linewidth}
	\includegraphics[width=\linewidth]{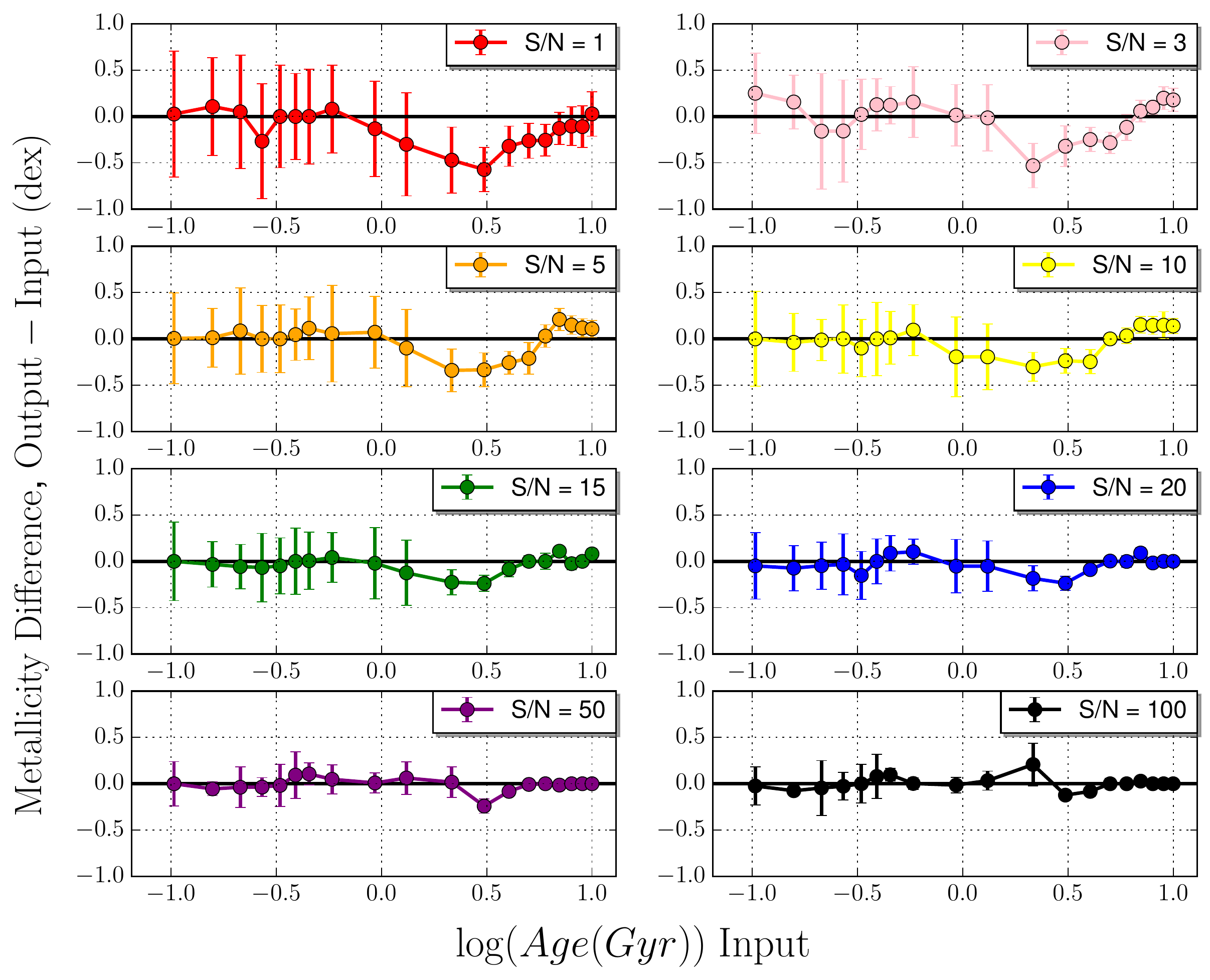}
	\caption{$\tau$ = 1 Gyr, metallicity recovery.}
\end{subfigure}
\begin{subfigure}{0.46\linewidth}
	\includegraphics[width=\linewidth]{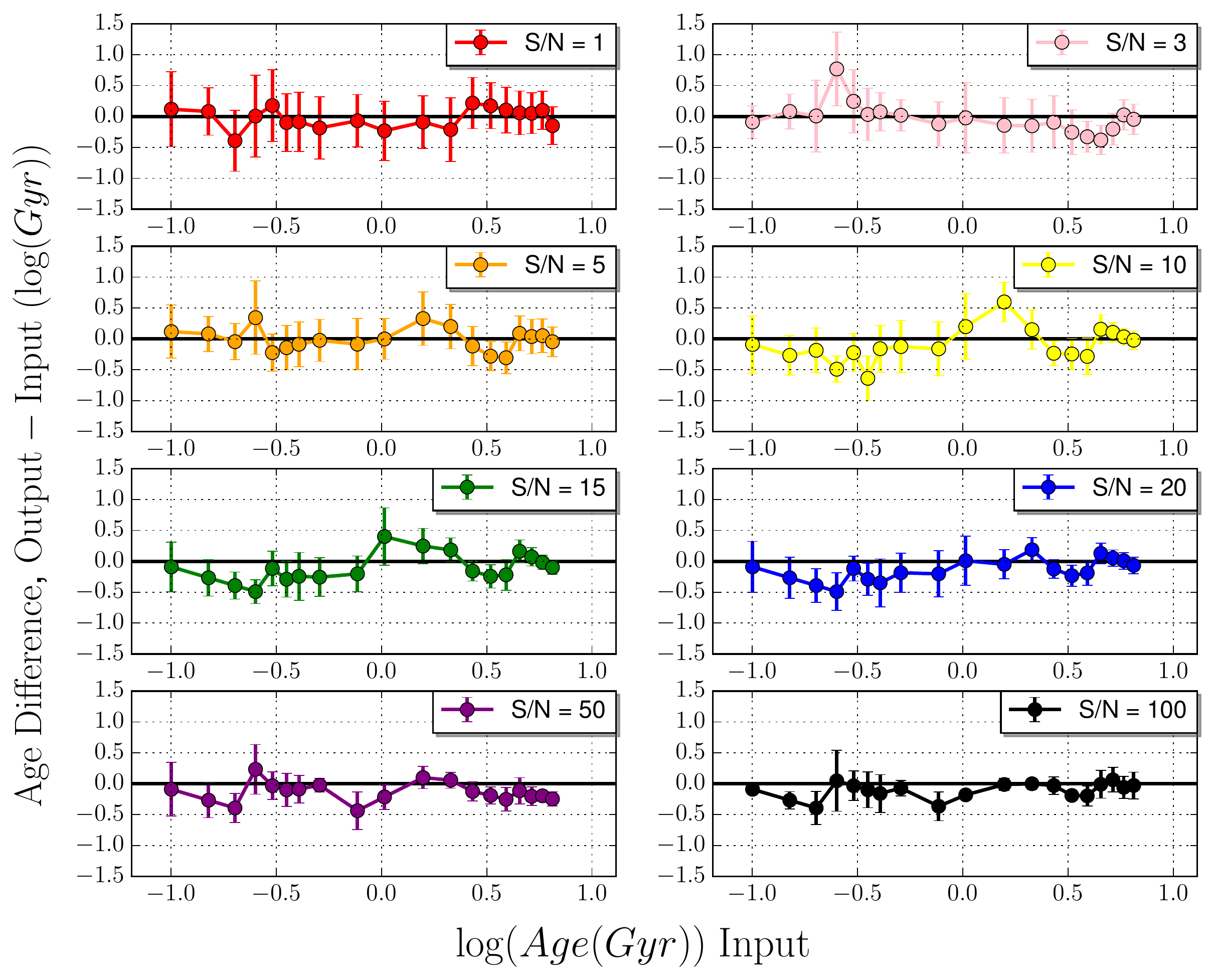}
	\caption{$\tau$ = 10 Gyr, age recovery.}
\end{subfigure}
\begin{subfigure}{0.46\linewidth}
	\includegraphics[width=\linewidth]{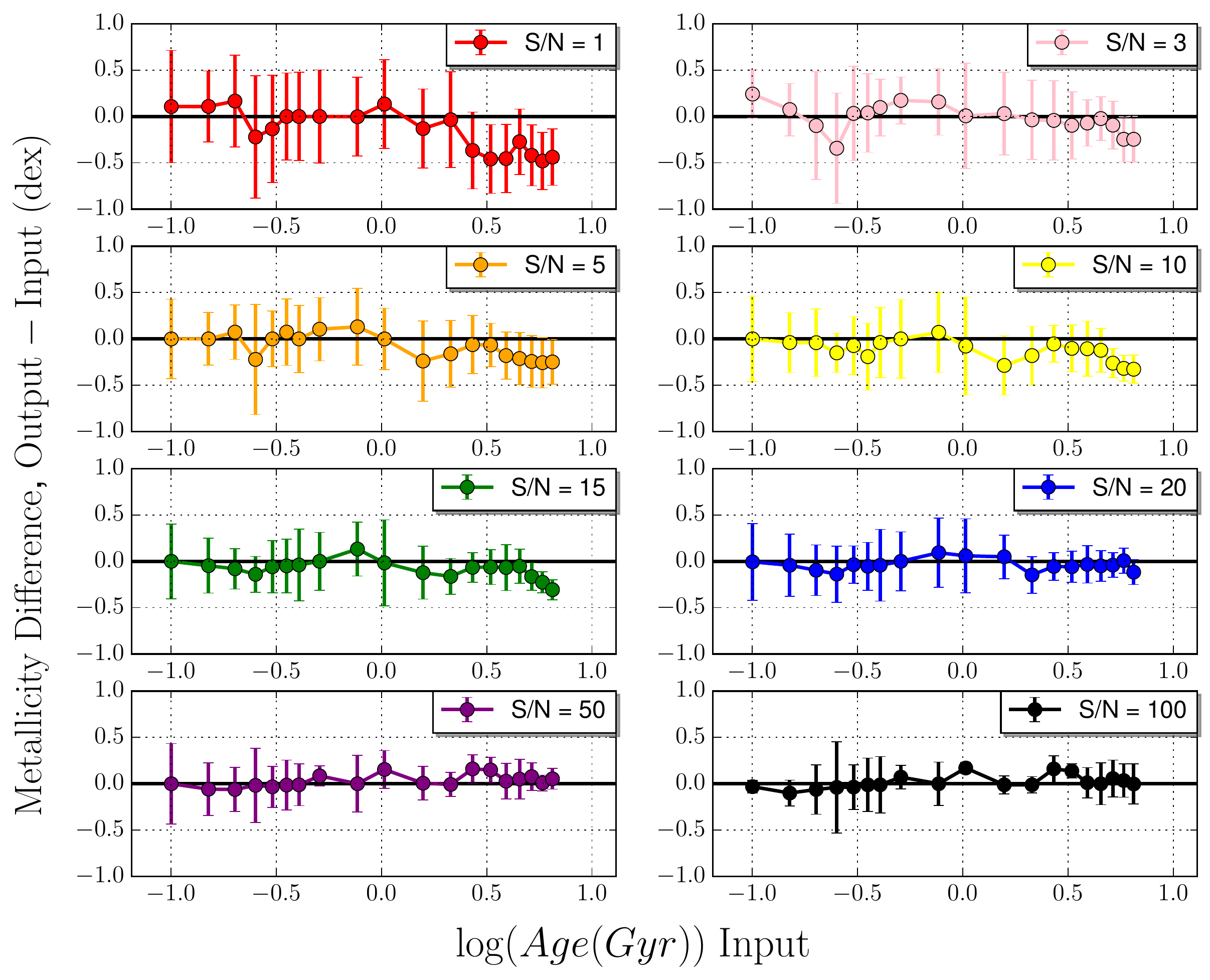}
	\caption{$\tau$ = 10 Gyr, metallicity recovery.}
\end{subfigure}
\caption[Recovery of stellar population properties for mock galaxies based on composite models.]{Recovered mass-weighted ages and metallicities (left-hand and right-hand plot series, respectively) from fitting mock spectra of composite stellar populations. Their input star formation histories have exponentially declining star formation rates given by $\Phi(t,\tau) \propto e^{-(t)/\tau}$, for $t > 0$, where $t$ is the time since the onset of star formation and $\tau$ is the star formation decay time or $e$-folding time (0.1, 1 and 10 Gyr, from top to bottom). Solutions are plotted at different of $S/N$~ratios (from 1 to 100).}
\label{cspagemetalssn}
\end{figure*}

\begin{figure*}
\centering
\begin{subfigure}{0.46\linewidth}
	\includegraphics[width=\linewidth]{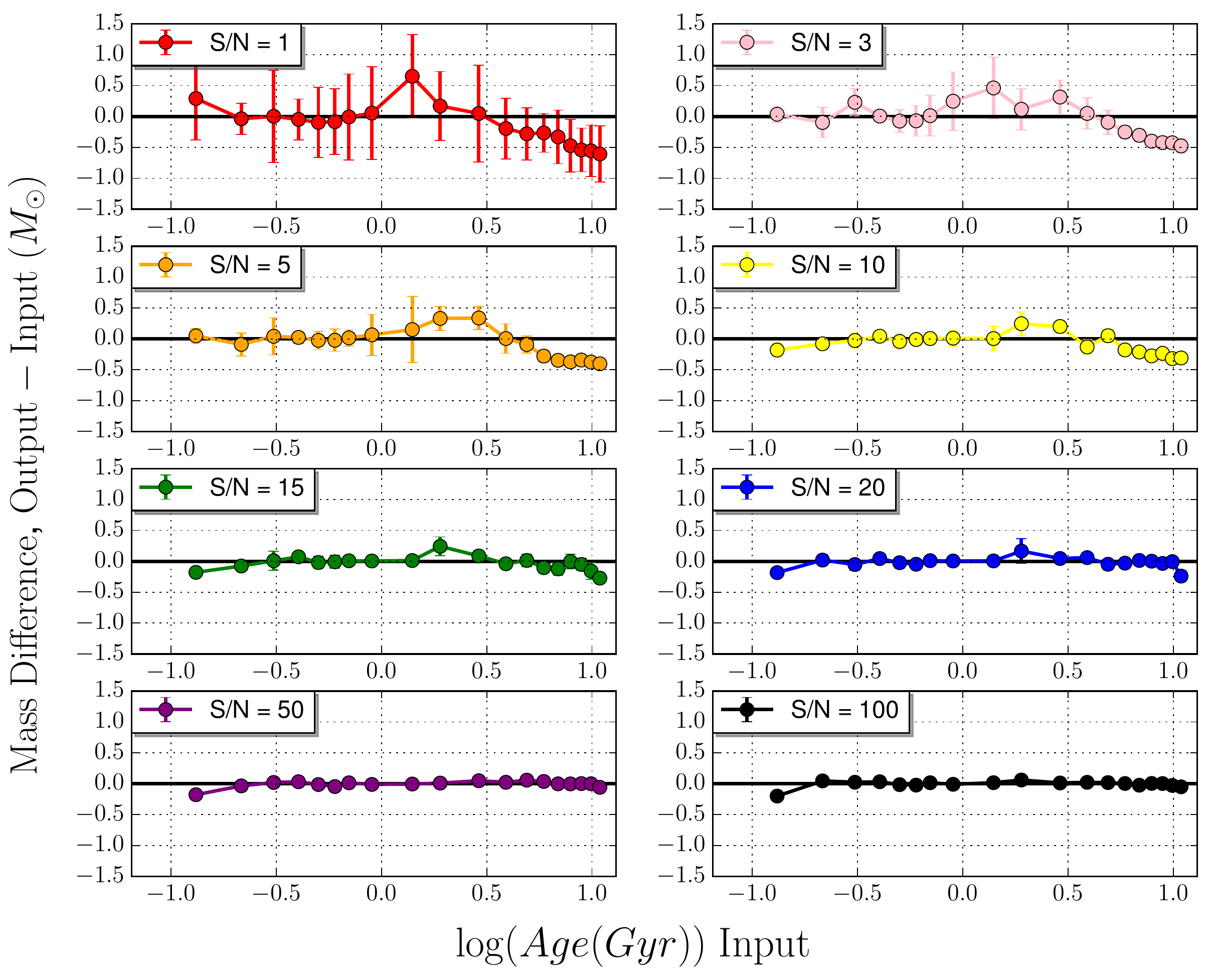}
	\caption{$\tau$ = 0.1 Gyr.}
\end{subfigure} 
\begin{subfigure}{0.46\linewidth}
	\includegraphics[width=\linewidth]{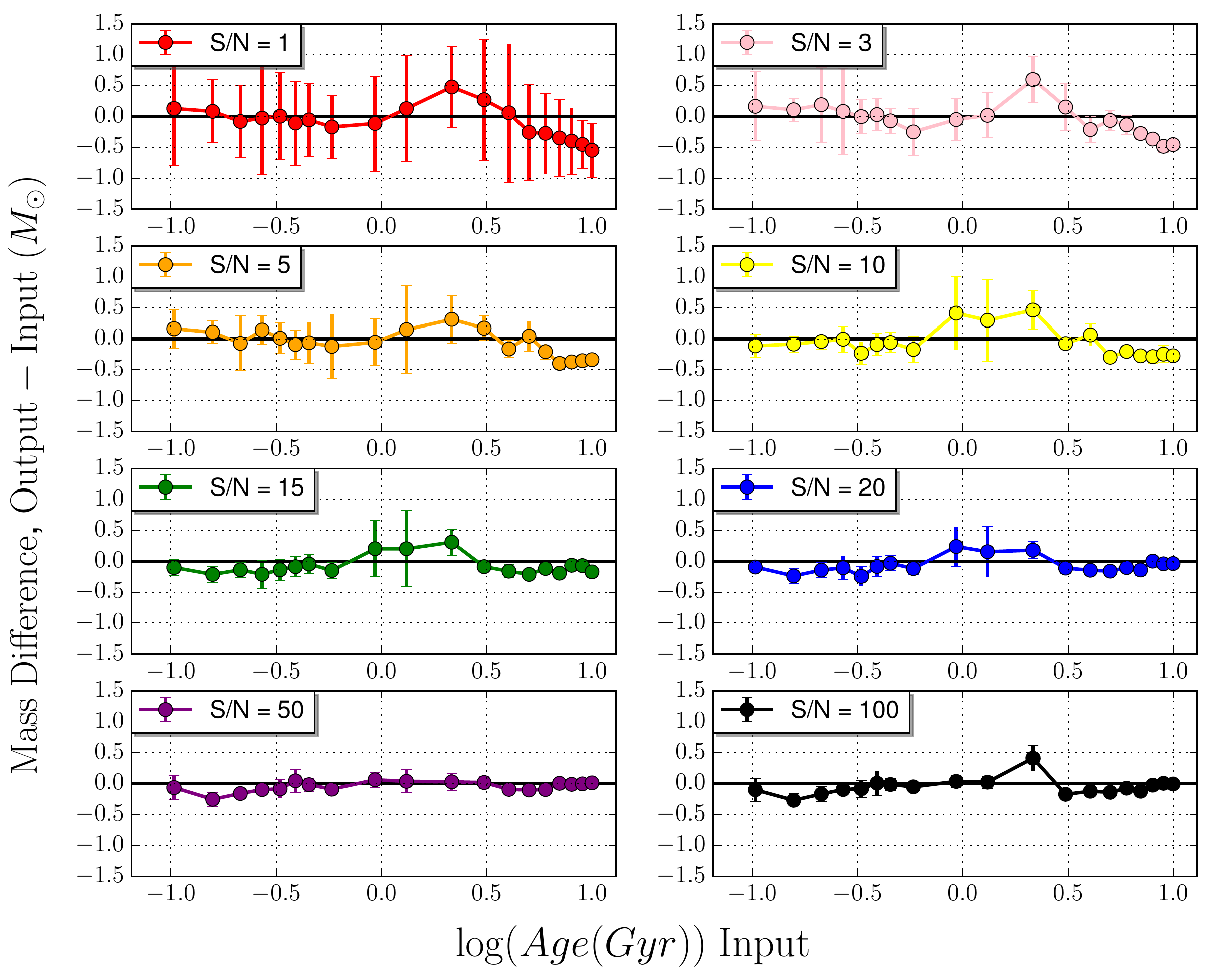}
	\caption{$\tau$ = 1 Gyr.}
\end{subfigure}
\begin{subfigure}{0.46\linewidth}
	\includegraphics[width=\linewidth]{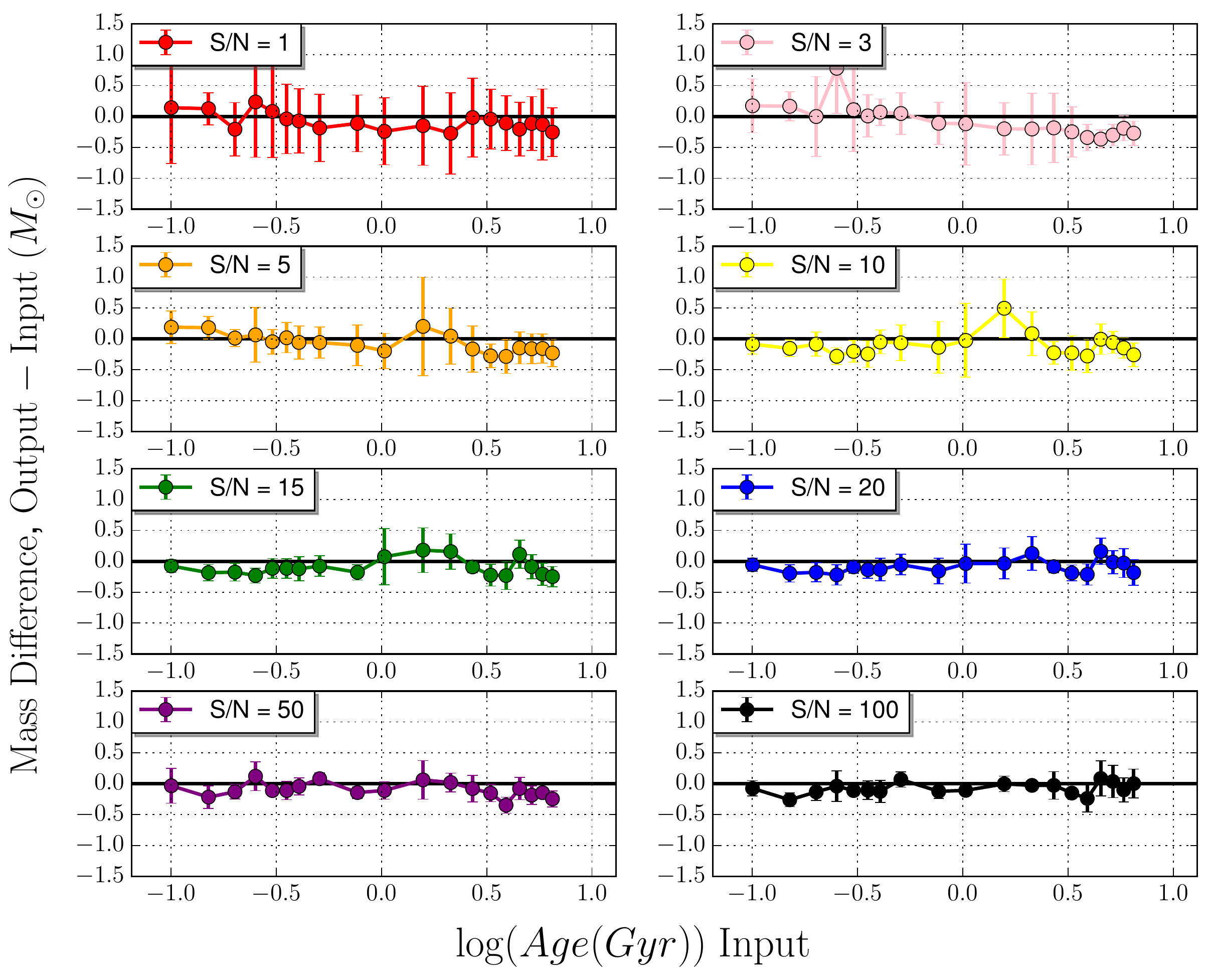}
	\caption{$\tau$ = 10 Gyr.}
\end{subfigure}
\caption[Recovery of stellar population properties for mock galaxies based on composite models.]{ As in Fig. 11 for the recovered stellar mass.}
\label{cspmasssn}
\end{figure*}

\begin{figure*}
\centering
\begin{subfigure}{0.48\linewidth}
	\includegraphics[width=\linewidth]{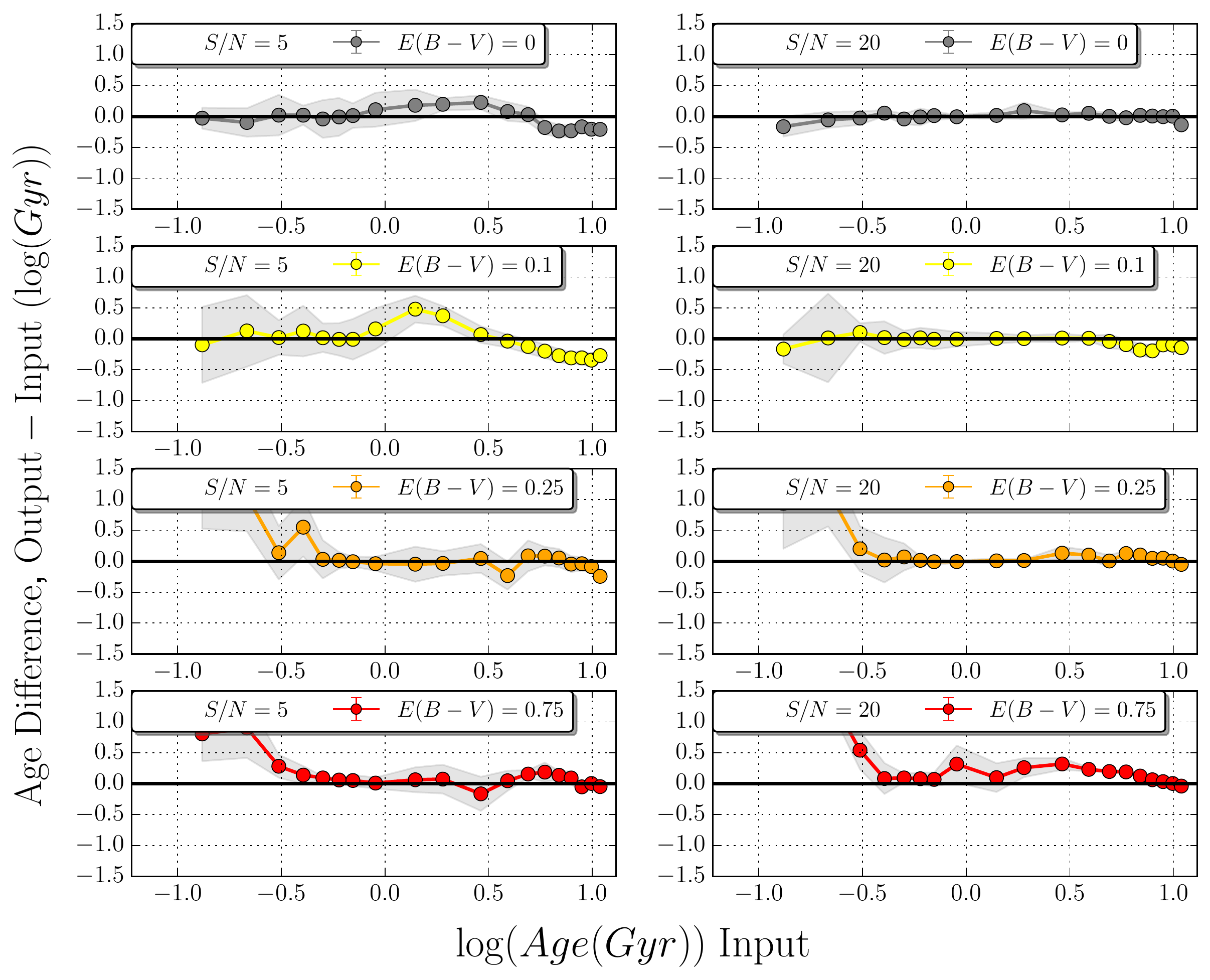}
	\caption{Recovery of mass-weighted age.}
\end{subfigure} 
\begin{subfigure}{0.48\linewidth}
	\includegraphics[width=\linewidth]{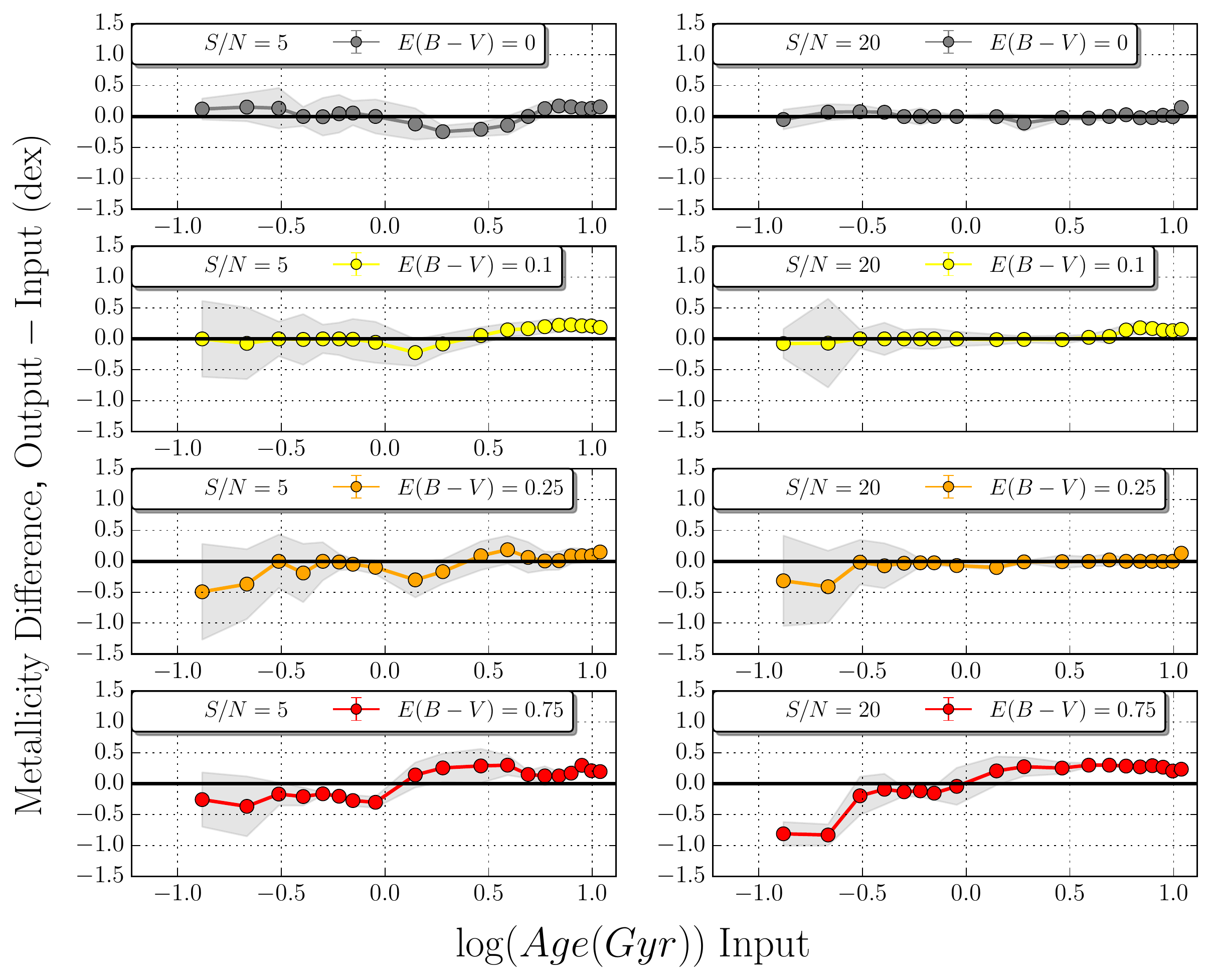}
	\caption{Recovery of mass-weighted metallicity.}
\end{subfigure}
\begin{subfigure}{0.48\linewidth}
	\includegraphics[width=\linewidth]{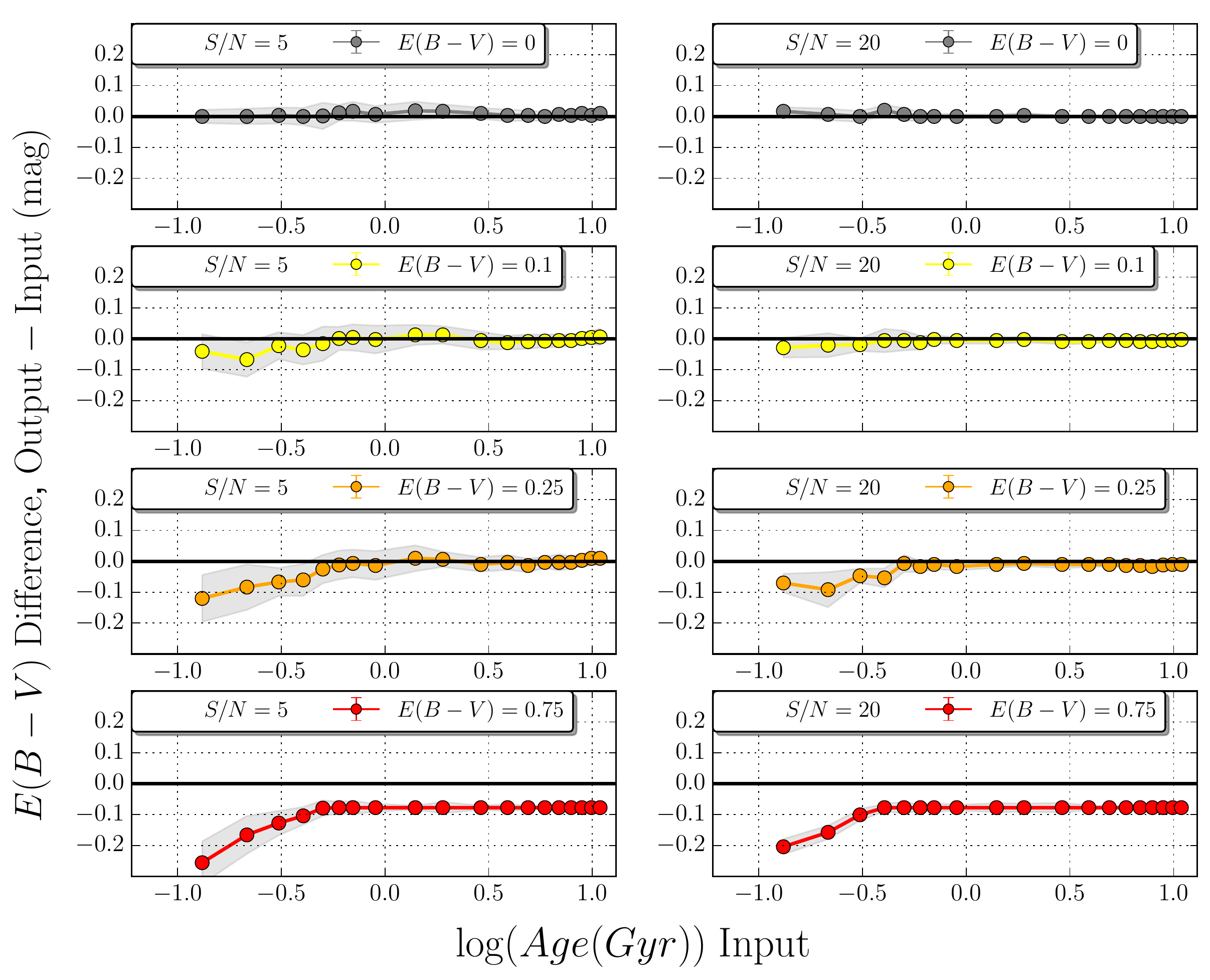}
	\caption{Recovery of reddening $E(B-V)$.}
\end{subfigure}
\begin{subfigure}{0.48\linewidth}
	\includegraphics[width=\linewidth]{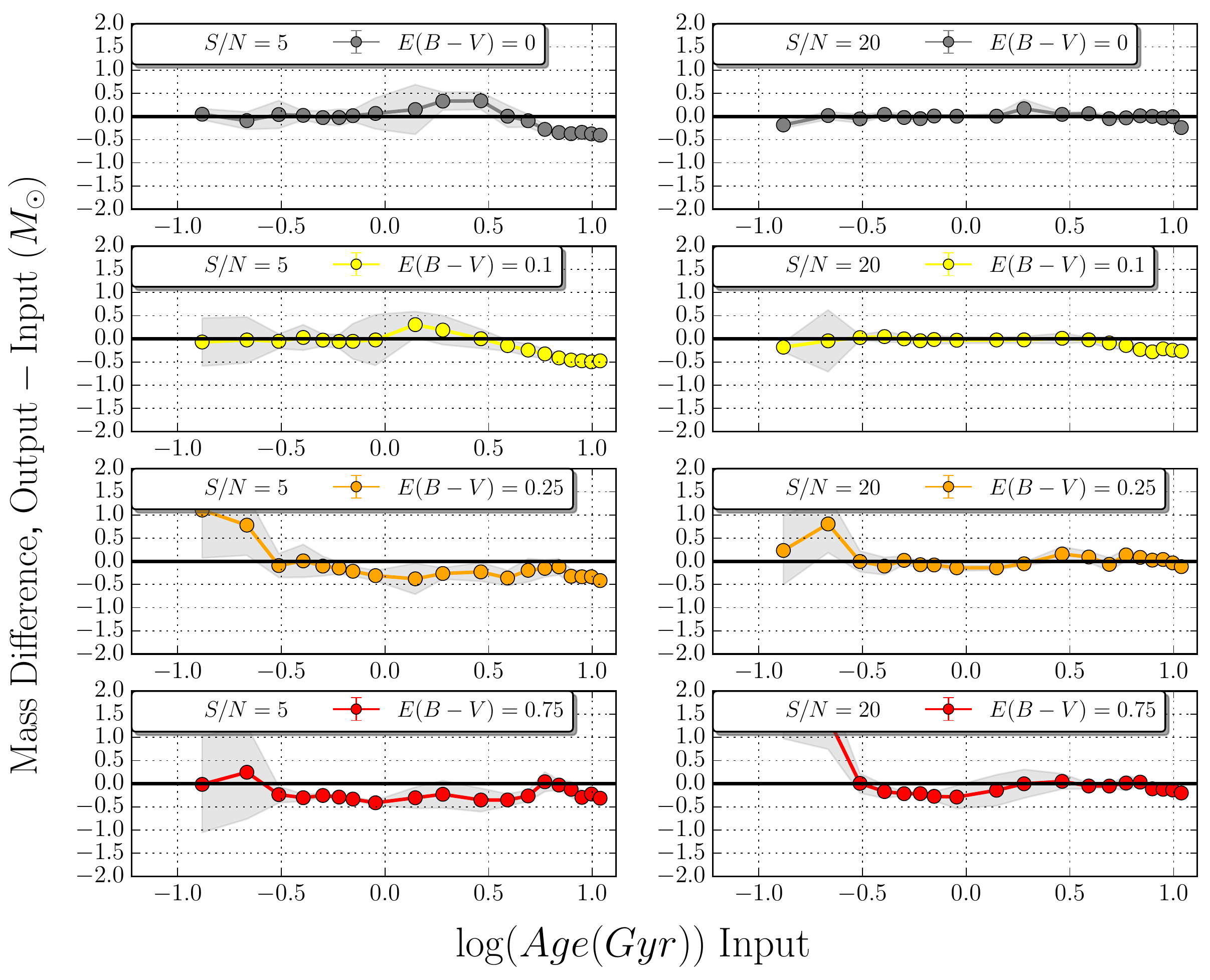}
	\caption{Recovery of stellar mass.}
\end{subfigure}
\caption[Recovery of stellar population properties for mock galaxies based on composite models.]{Recovered mass-weighted age, metallicity, reddening $E(B-V)$~and stellar mass, for input composite stellar populations including intrinsic reddening, with $E(B-V)=0, 0.1, 0.25, 0.75$, for two median signal-to-noise of 5 (LHS) and 20 (RHS). This figure refers to a $\tau=0.1$~exponentially-declining star formation history.}
\label{csp01dust}
\end{figure*}

\begin{figure*}
\centering
\begin{subfigure}{0.48\linewidth}
	\includegraphics[width=\linewidth]{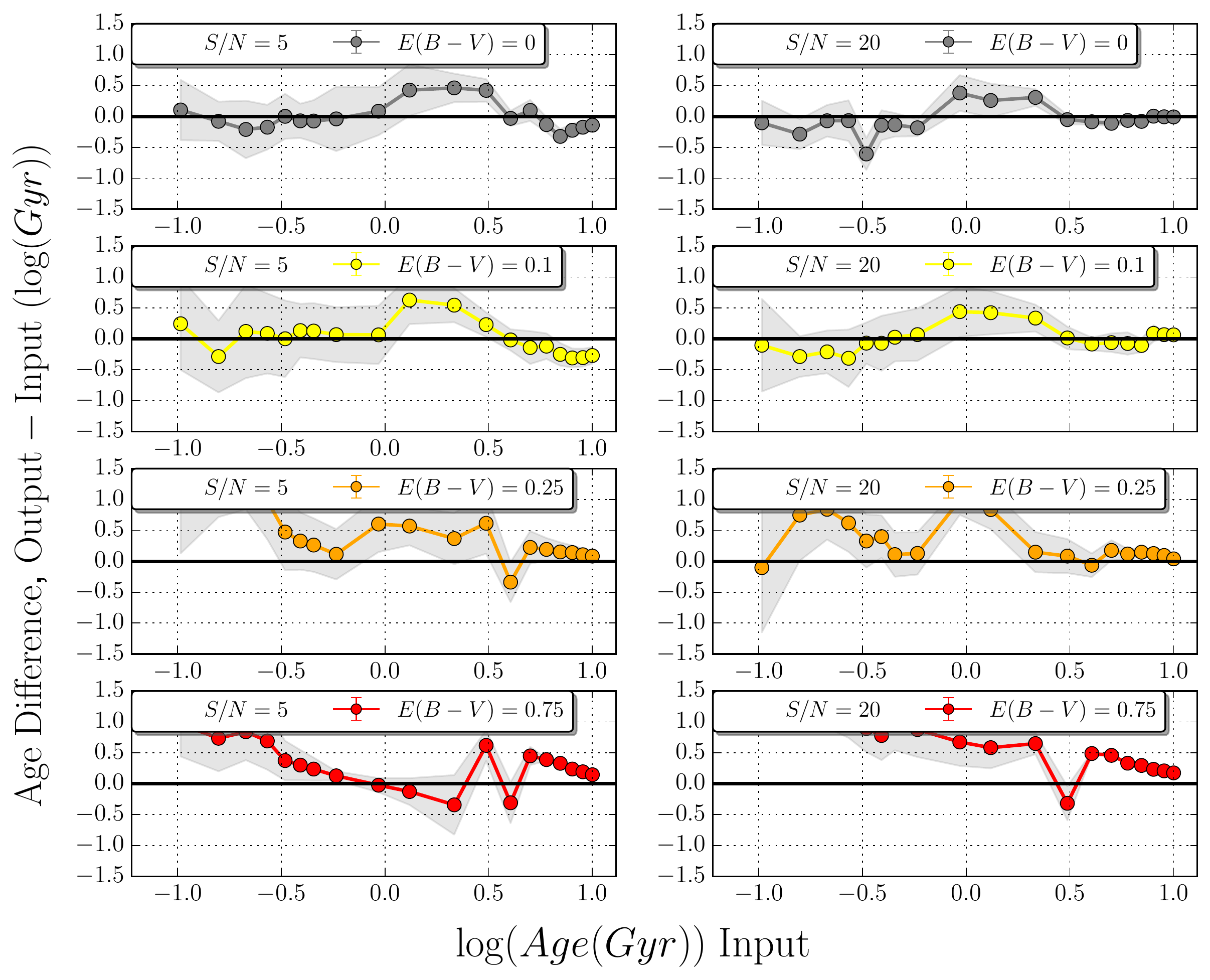}
	\caption{Recovery of mass-weighted age.}
\end{subfigure} 
\begin{subfigure}{0.48\linewidth}
	\includegraphics[width=\linewidth]{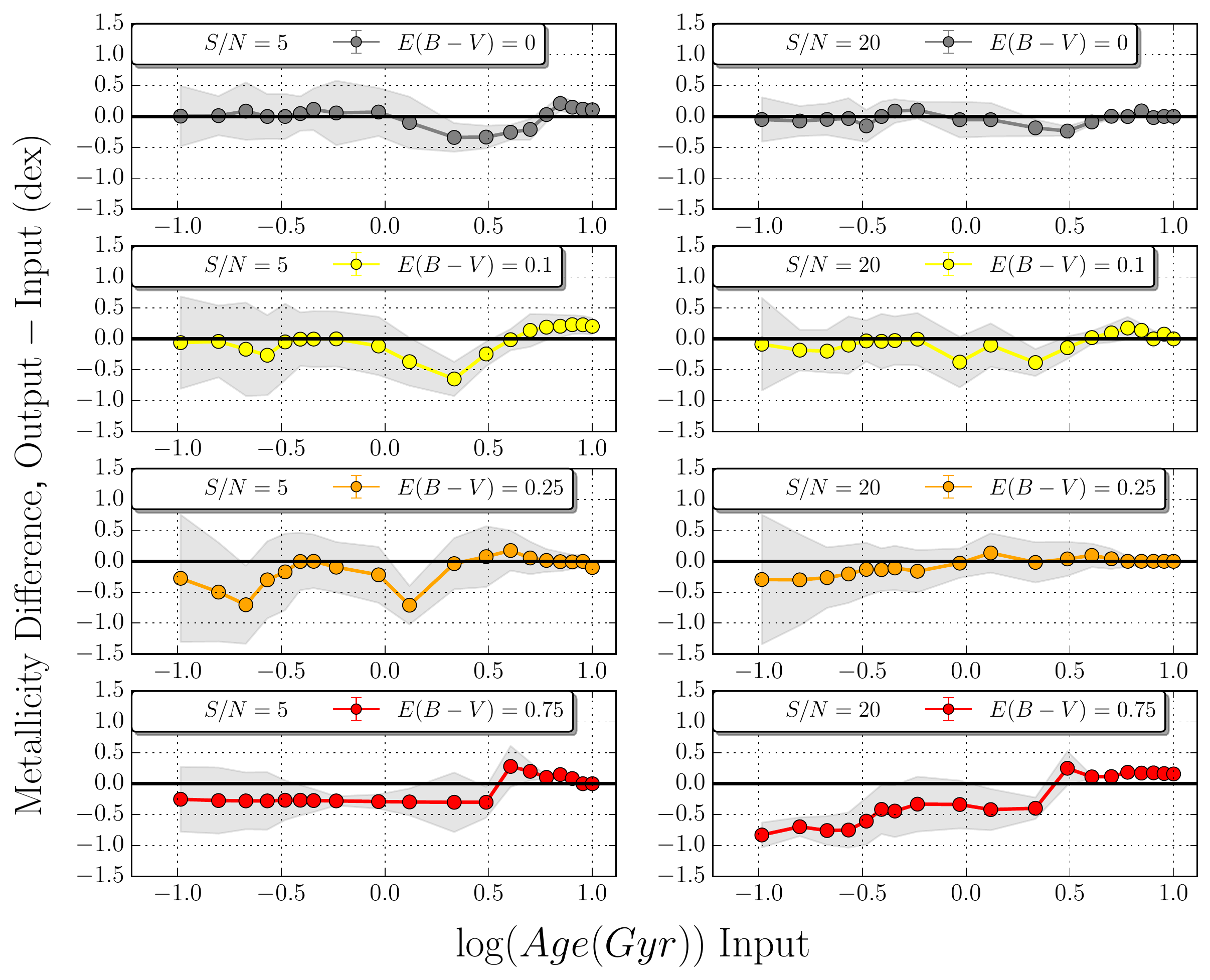}
	\caption{Recovery of mass-weighted metallicity.}
\end{subfigure}
\begin{subfigure}{0.48\linewidth}
	\includegraphics[width=\linewidth]{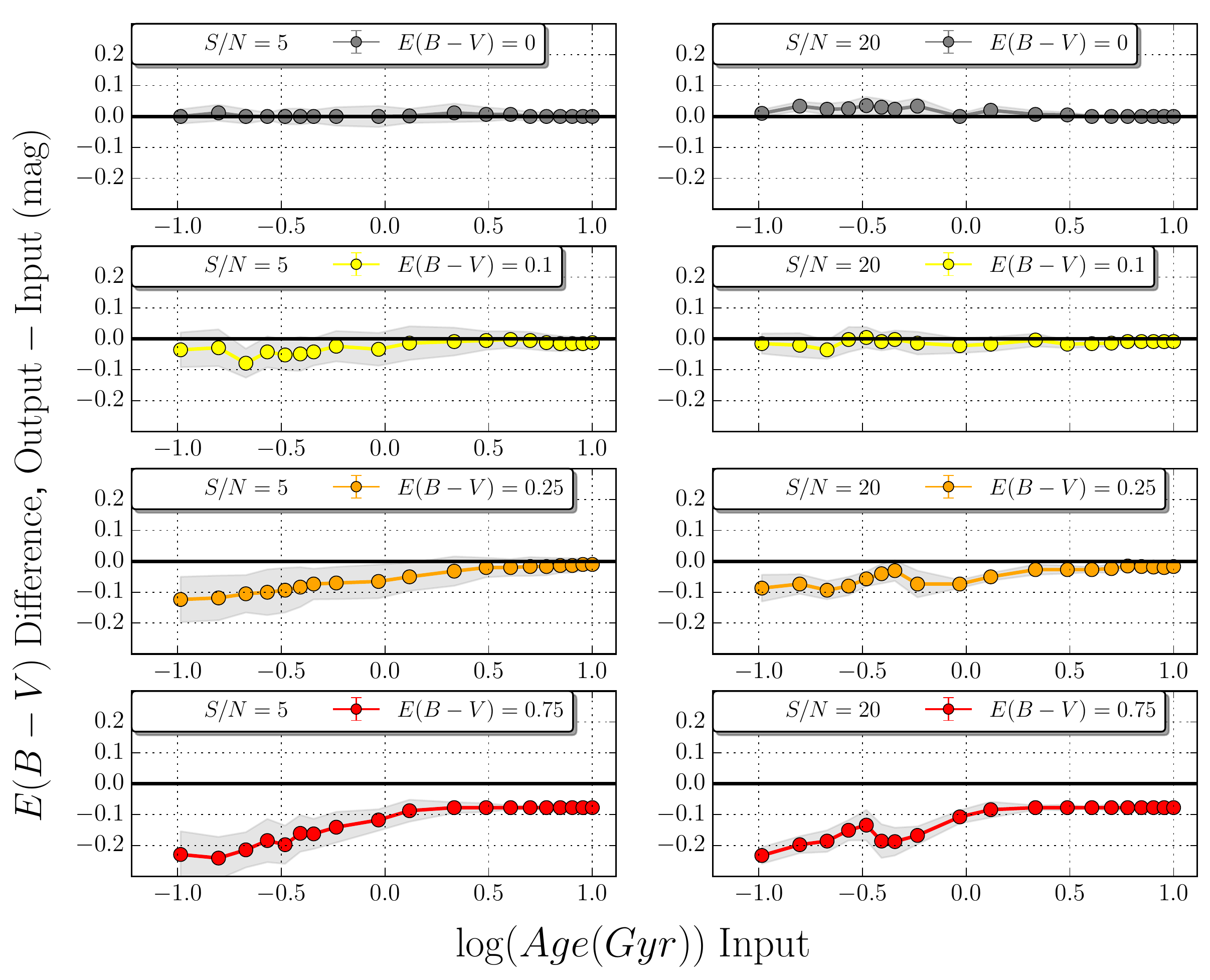}
	\caption{Recovery of reddening $E(B-V)$.}
\end{subfigure}
\begin{subfigure}{0.48\linewidth}
	\includegraphics[width=\linewidth]{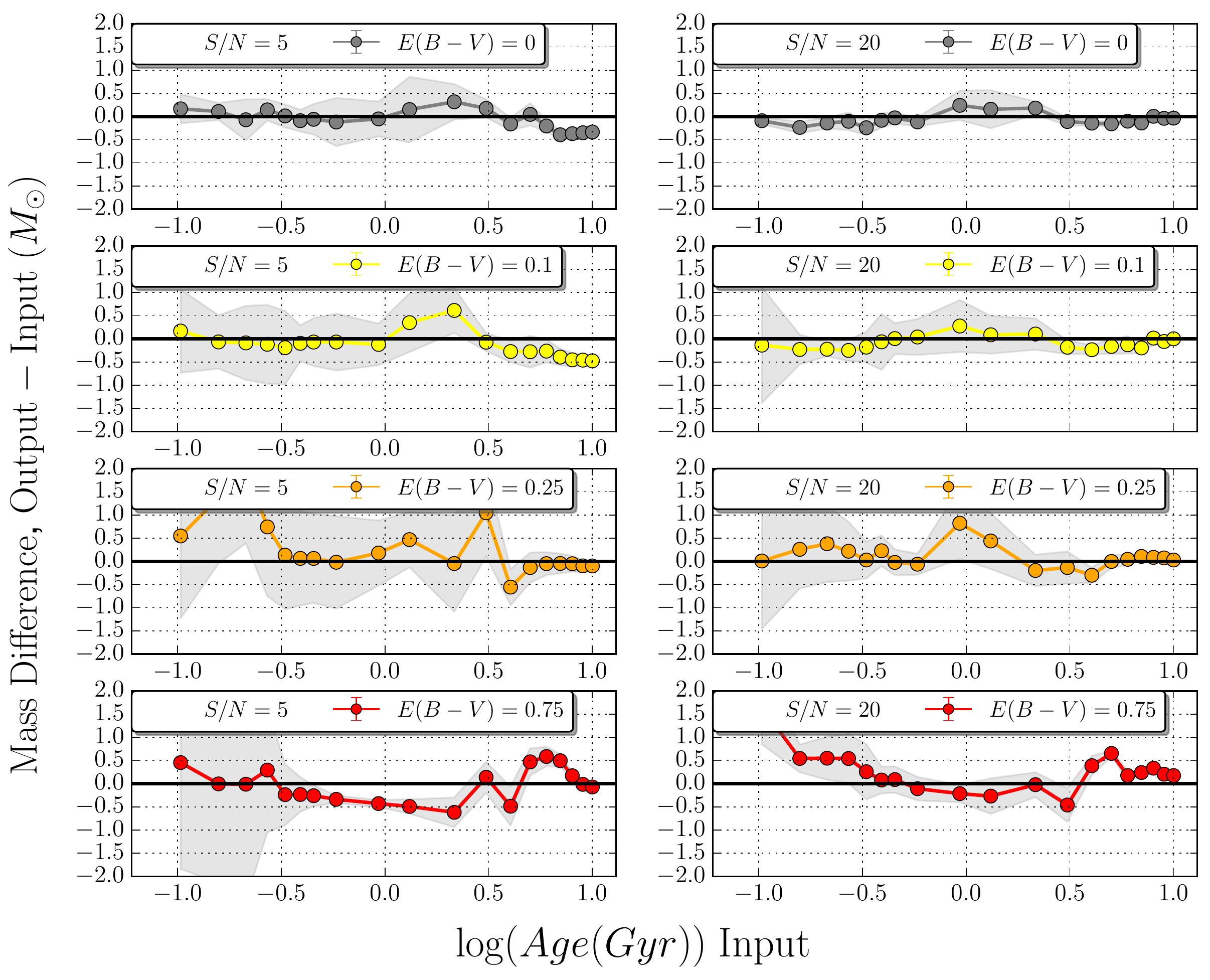}
	\caption{Recovery of stellar mass.}
\end{subfigure}
\caption[Recovery of stellar population properties for mock galaxies based on composite models.]{As in Figure~\ref{csp01dust} for a $\tau=1$ Gyr~exponentially-declining star formation history.}
\label{csp1dust}
\end{figure*}

\begin{figure*}
\centering
\begin{subfigure}{0.48\linewidth}
	\includegraphics[width=\linewidth]{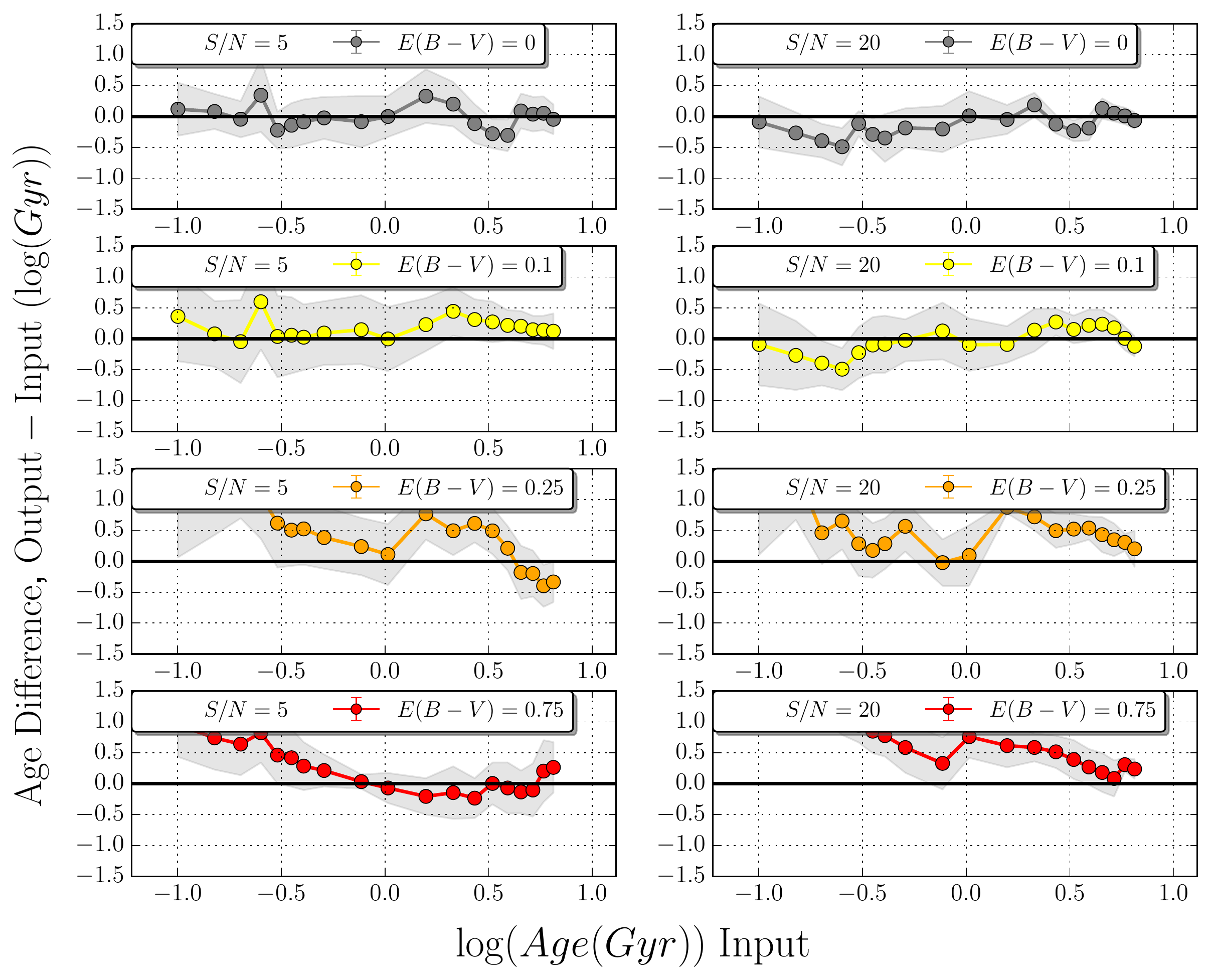}
	\caption{Recovery of mass-weighted age.}
\end{subfigure} 
\begin{subfigure}{0.48\linewidth}
	\includegraphics[width=\linewidth]{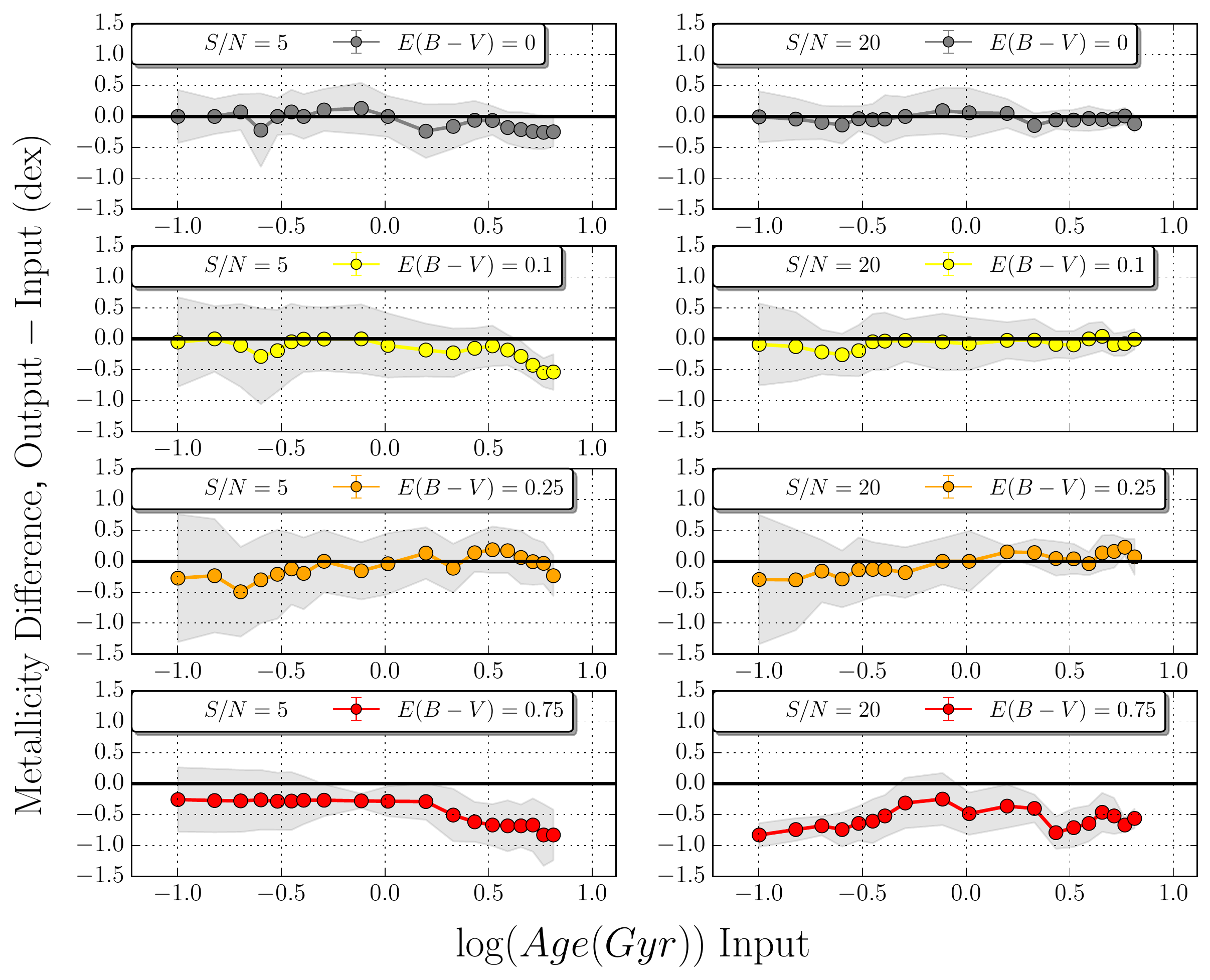}
	\caption{Recovery of mass-weighted metallicity.}
\end{subfigure}
\begin{subfigure}{0.48\linewidth}
	\includegraphics[width=\linewidth]{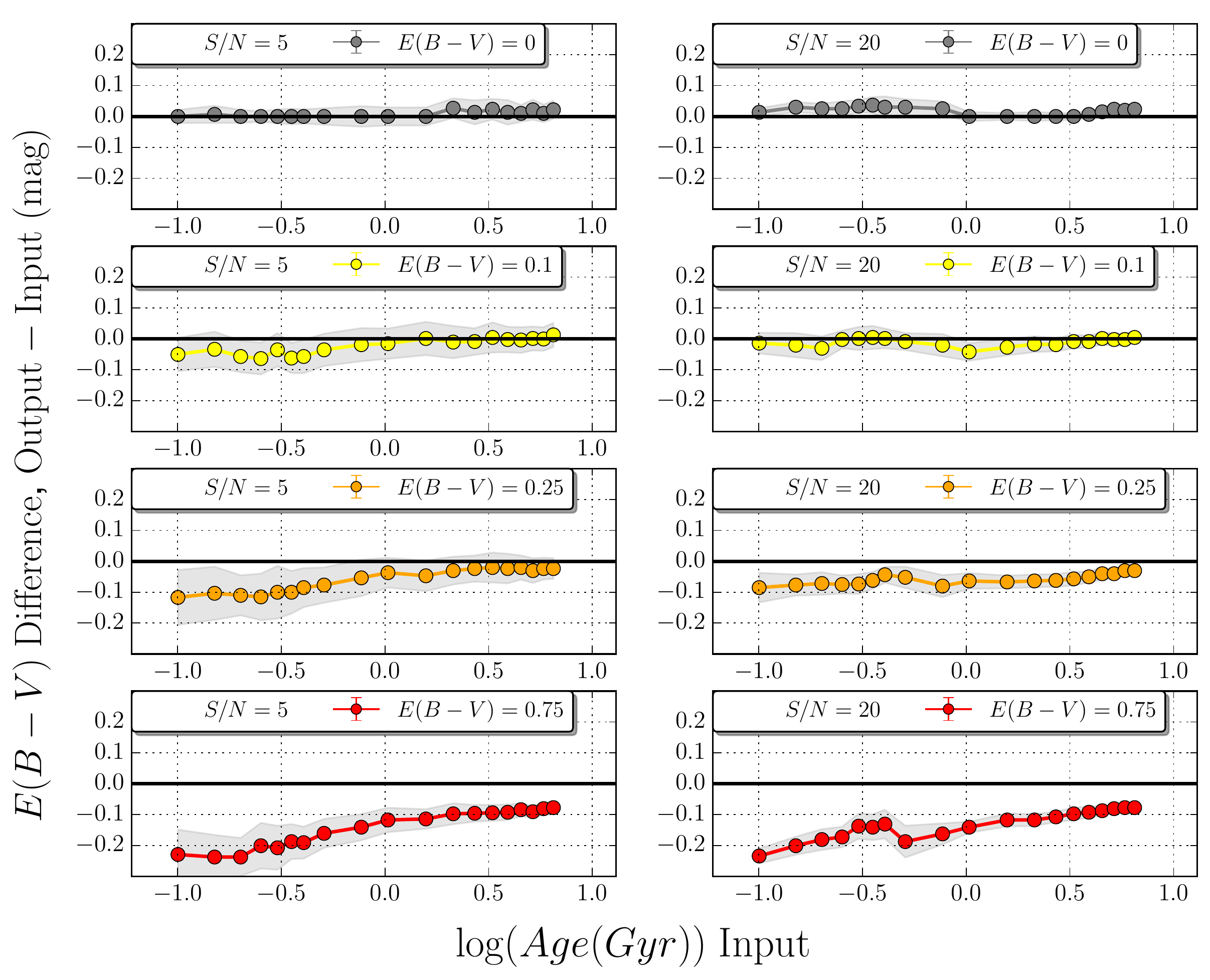}
	\caption{Recovery of reddening $E(B-V)$.}
\end{subfigure}
\begin{subfigure}{0.48\linewidth}
	\includegraphics[width=\linewidth]{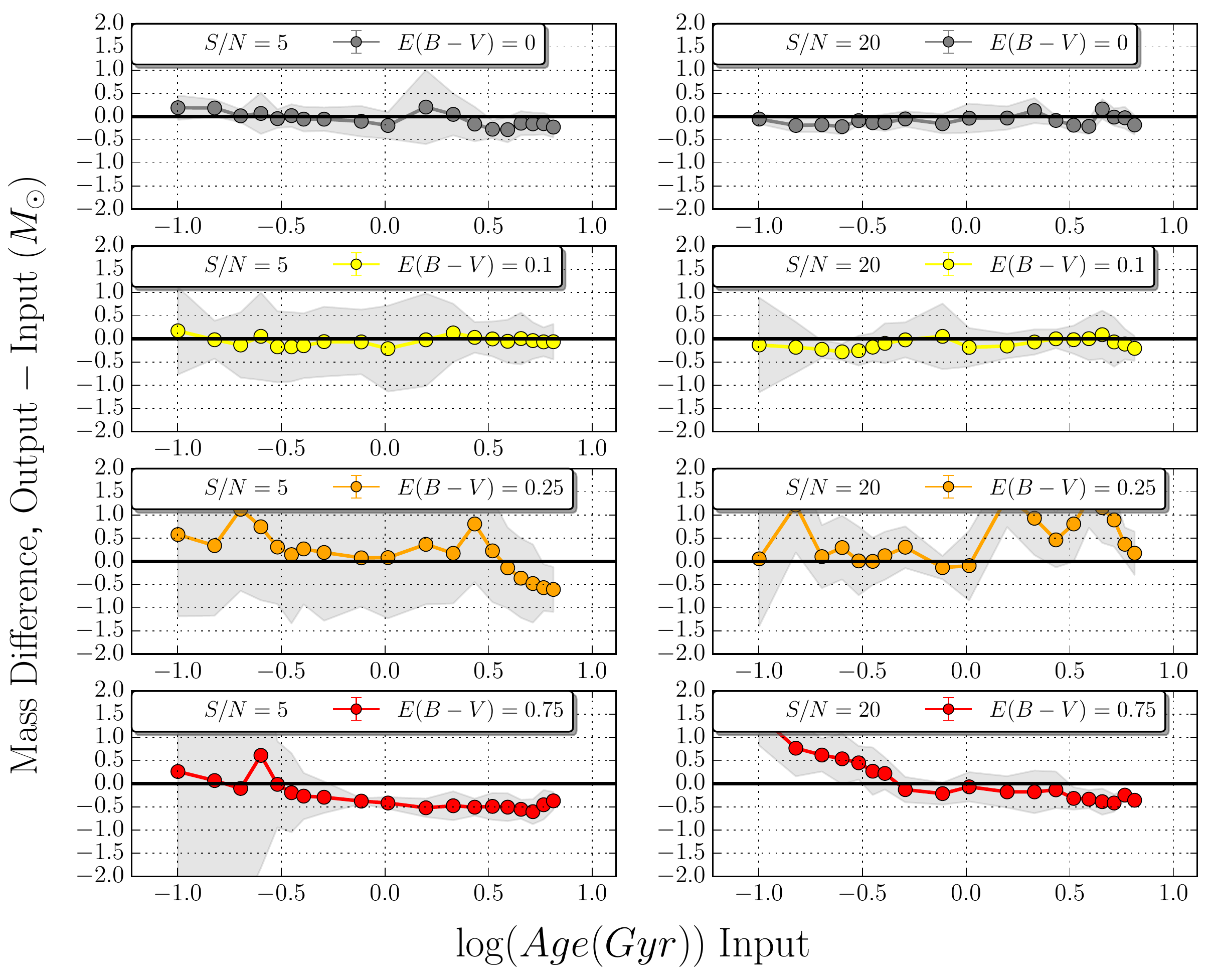}
	\caption{Recovery of stellar mass.}
\end{subfigure}
\caption[Recovery of stellar population properties for mock galaxies based on composite models.]{As in Figures~\ref{csp01dust}, \ref{csp1dust} for a $\tau=10$ Gyr~exponentially-declining star formation history.}
\label{csp10dust}
\end{figure*}

\begin{figure*}
\begin{subfigure}{\linewidth}
\centering
	\includegraphics[width=0.8\linewidth]{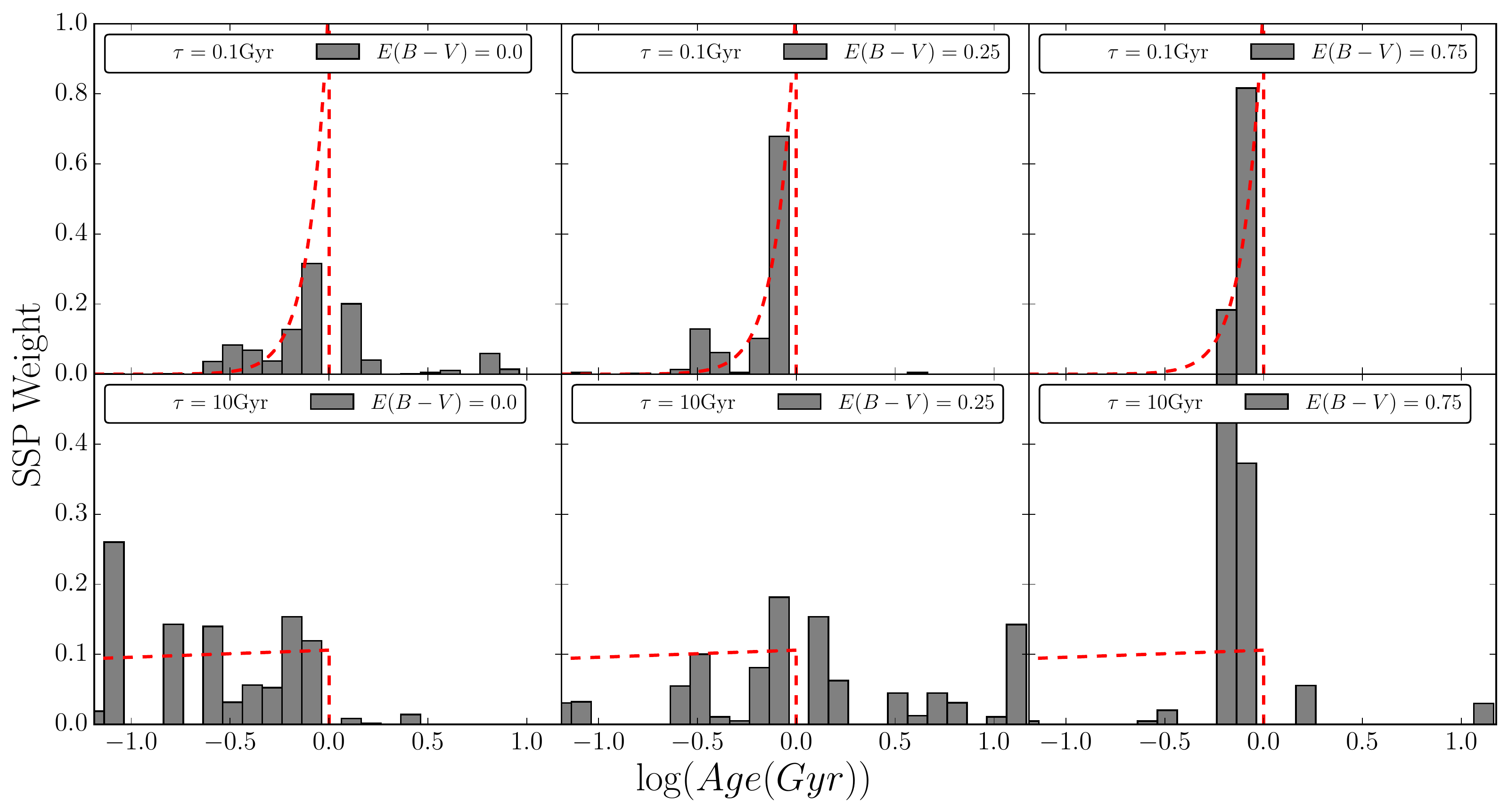}
	\caption{Simulations at $S/N=5$.}
\end{subfigure} 
\begin{subfigure}{\linewidth}
\centering
	\includegraphics[width=0.8\linewidth]{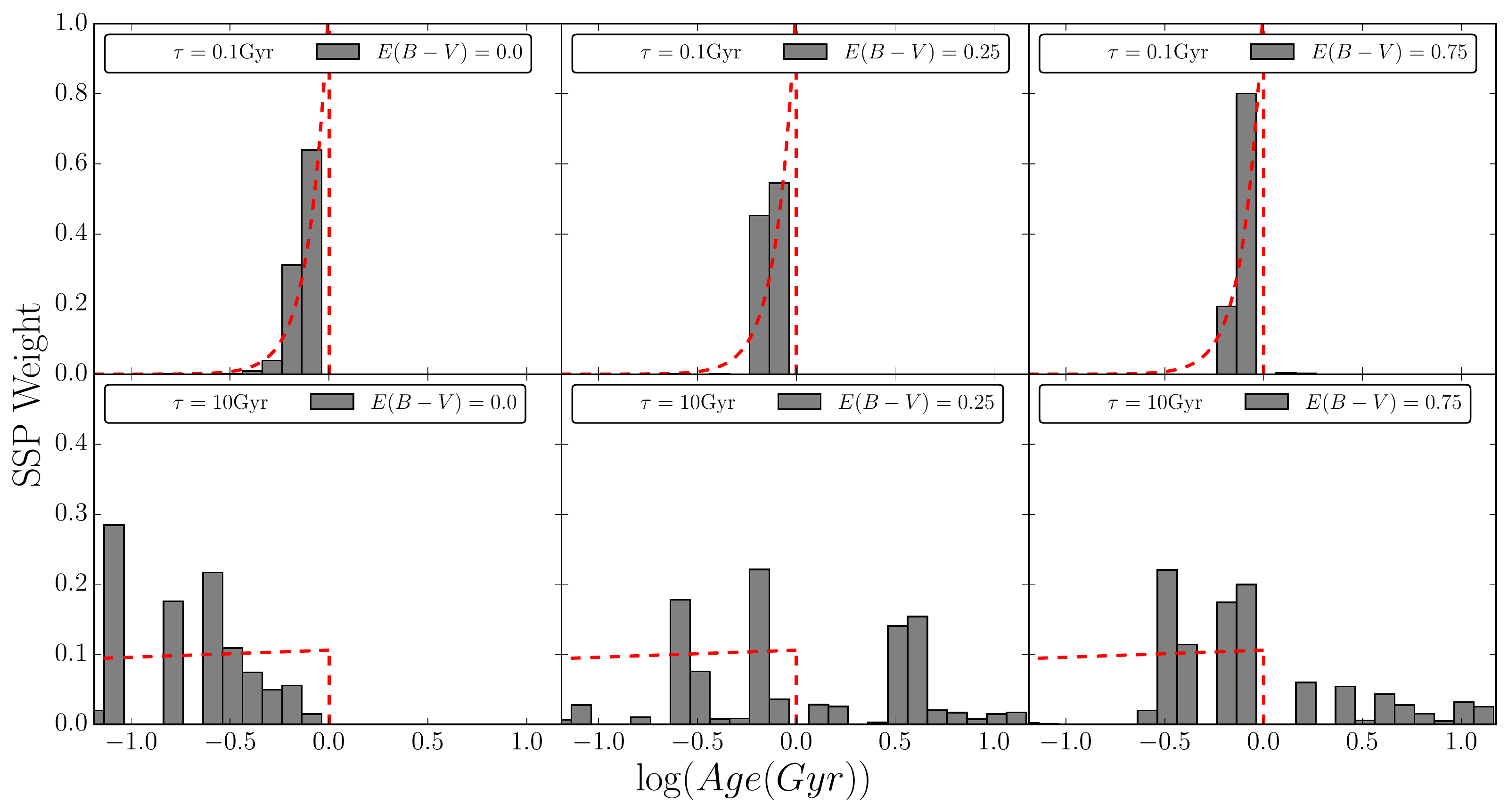}
	\caption{Simulations at $S/N=20$.}
\end{subfigure}
\begin{subfigure}{\linewidth}
\centering
	\includegraphics[width=0.8\linewidth]{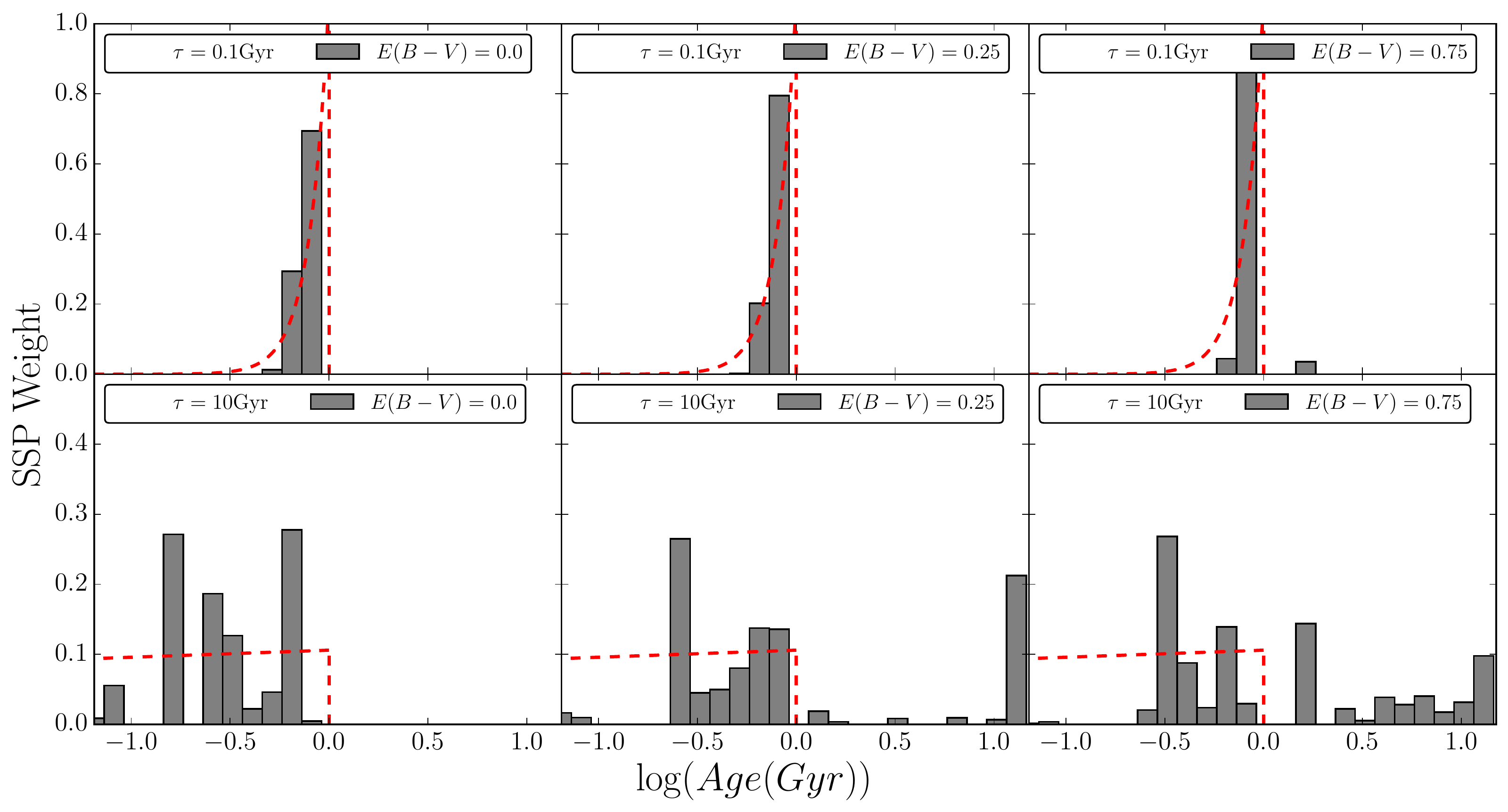}
	\caption{Simulations at $S/N=50$.}
\end{subfigure}
\caption[Recovery of stellar population properties for mock galaxies based on composite models.]{Recovery of star formation history, for input mocks with $\tau$=0.1 and 10 Gyr (upper and lower rows, respectively), 1 Gyr after the starting of star formation, as a function of reddening, increasing from left to right, for three $S/N$~ratios. Red, dashed lines show the input SFHs as smooth curves. }
\label{sfh1}
\end{figure*}

\begin{figure*}
\begin{subfigure}{\linewidth}
\centering
	\includegraphics[width=0.8\linewidth]{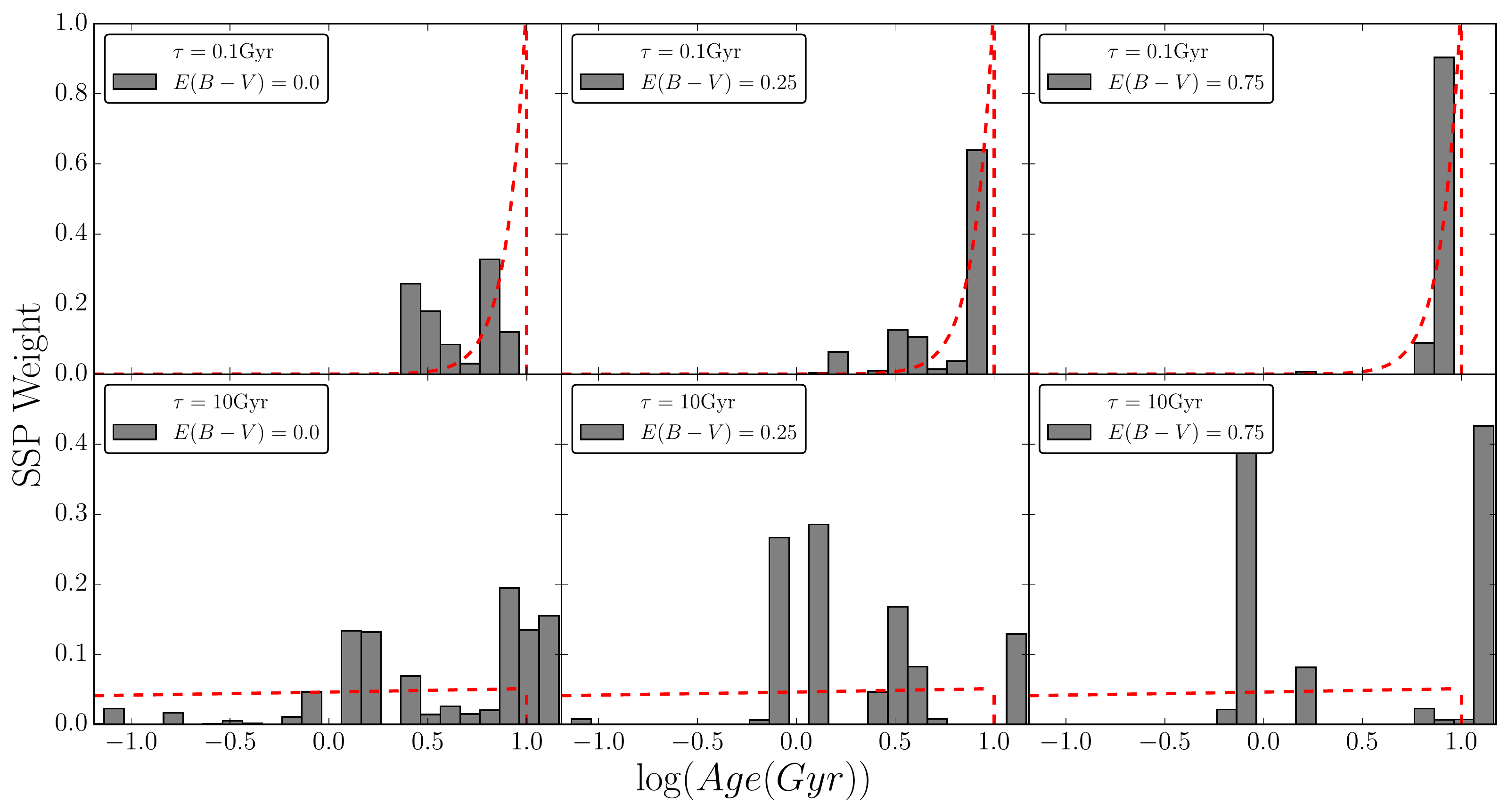}
	\caption{Simulations at $S/N=5$.}
\end{subfigure} 
\begin{subfigure}{\linewidth}
\centering
	\includegraphics[width=0.8\linewidth]{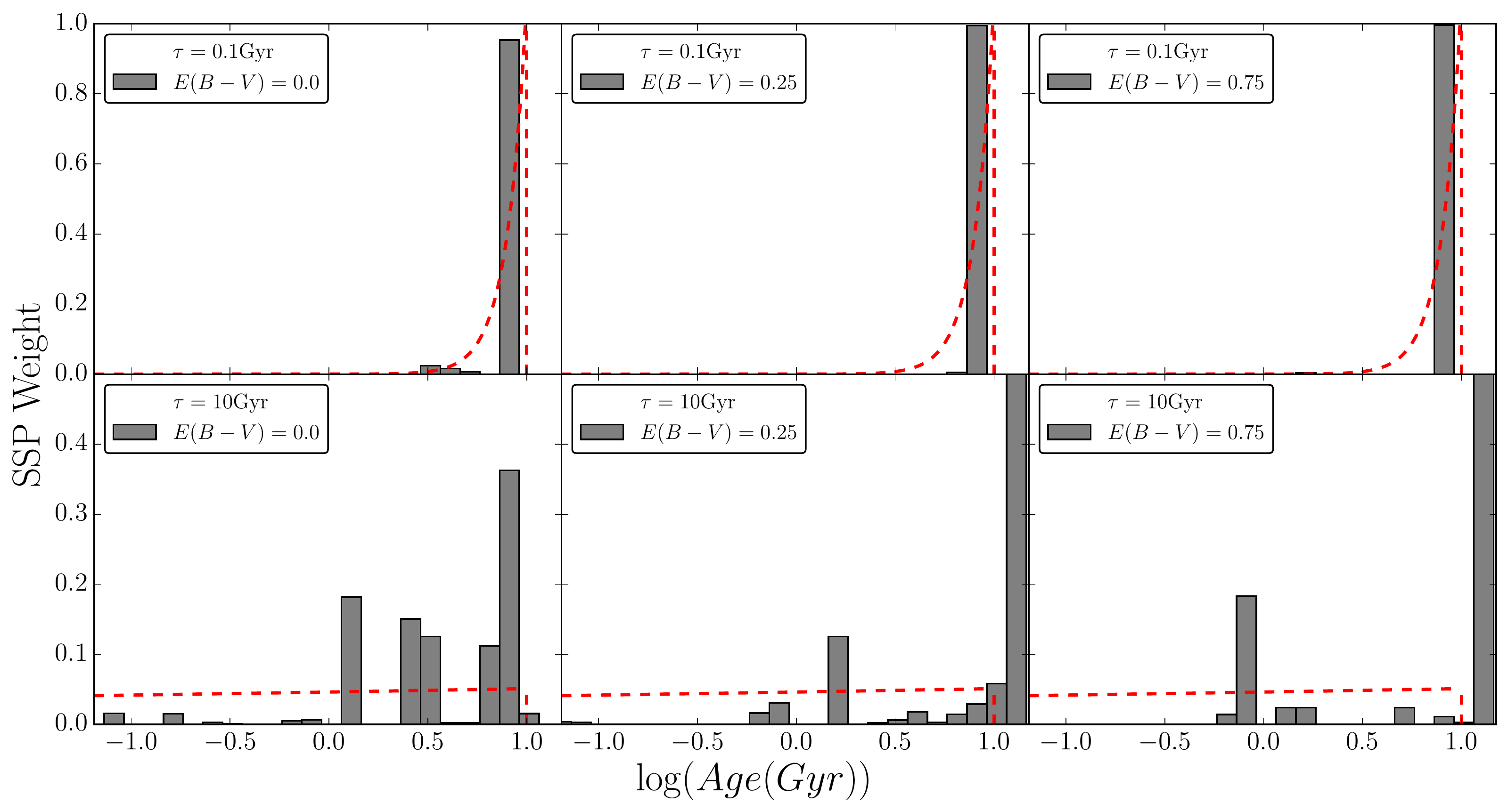}
	\caption{Simulations at $S/N=20$.}
\end{subfigure}
\begin{subfigure}{\linewidth}
\centering
	\includegraphics[width=0.8\linewidth]{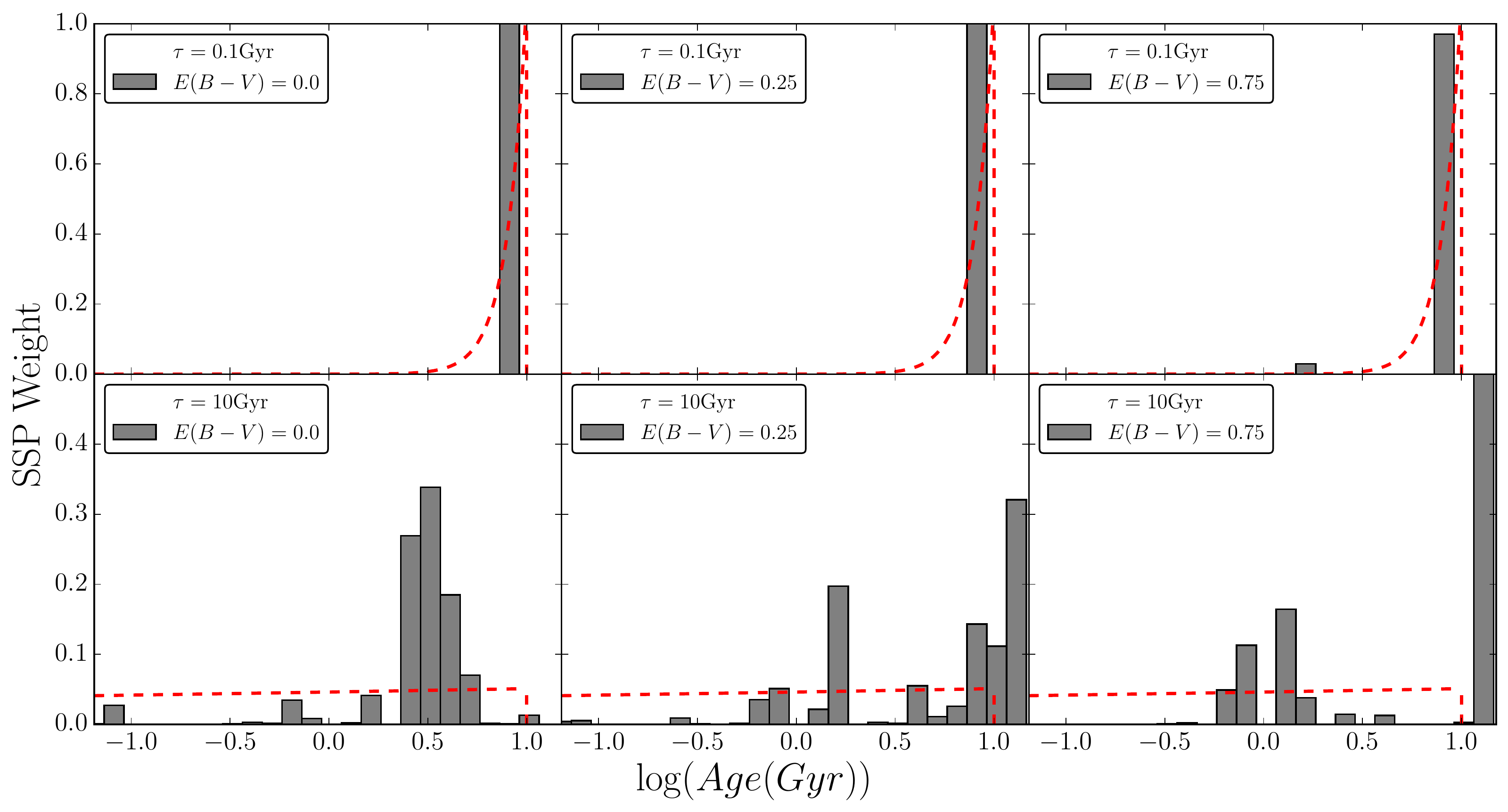}
	\caption{Simulations at $S/N=50$.}
\end{subfigure}\caption[Recovery of stellar population properties for mock galaxies based on composite models.]{As in Figure~\ref{sfh1}, but 10 Gyr after the starting of star formation.}
\label{sfh2}
\end{figure*}

In the second set of tests we use composite models obtained with a range of exponentially-declining star formation histories (known as $\tau$ models, \cite{1983ApJ...273..105B}). In $\tau$ models, the star formation rate, $\Phi(t,\tau)$, is described as $\Phi(t,\tau) \propto e^{-t/\tau}$ for $t > 0$, where $t$~is the time coordinate, with star formation beginning at $t=0$~and $\tau$~is the characteristic decay time. These models are commonly used in the literature for fitting to a variety of galaxy data (e.g. \citet{2009MNRAS.394..774L}, \citet{2010ApJ...725.1644L}) as a realistic mode of star formation. We investigate these models for every combination of: a range of decay times; $\tau = $ 0.1, 1, 10 Gyr, a range of times $t$~after star formation began (24 values spanning from $t = 0.01$ to $10$ Gyr), and a range of reddening, obtained using the Calzetti's law with dust extinction values $A_v =$  0, 0.4, 1, 3. We consider the same $S/N$~values (from 1 to 100) as in Section \ref{perturbedsimple} and we experiment with 100 Monte Carlo simulations of each single mock for the same set of $S/N$ to test the robustness of property recovery.
\\
\\
Firstly, we show the case where the mocks do not contain intrinsic reddening in Figures \ref{cspagemetalssn} and \ref{cspmasssn}. The results for the mocks with dust are in Figures \ref{csp01dust}, \ref{csp1dust}, \ref{csp10dust}, \ref{sfh1} and \ref{sfh2}. It is important to note that in all cases we use \FF\ in its full mode, e.g. we fit for reddening as well. We determine the accuracy of the fitting procedure by analysing the output in terms of age, metallicity and stellar mass, and also star formation history and dust reddening.

Figure \ref{cspagemetalssn} shows the recovered mass-weighted ages and metallicities (left-hand and right hand plot series, respectively) compared to the input mass-weighted ages, as a function of $S/N$~, from 1 to 100, for the three tau-models with $\tau = $ 0.1, 1, 10 Gyr (from top to bottom). 

We find that \FF~is able to recover remarkably well both ages and metallicities, down to low signal-to-noise ratios ($S/N\sim5$) ~and for a wide range of extended star formation histories. The recovered metallicity shows some scatter, well within 0.2 dex at $S/N > 10$. As in the case of single-burst populations, old ages at low $S/N$~are the most difficult to recover, with ages getting overestimated (around 1 Gyr), then underestimated (after $\sim$~3 Gyr), with an opposite trend displayed by  metallicity, i.e. an overestimation of age is generally accompanied by an underestimation of metallicity. No exact quantitative correspondence should be sought between the two offsets because of the additional parameter (reddening) which is considered in the fitting. 

As $\tau$~increases, we find a larger scatter in the recovered properties (larger error bars around the median). This is expected as more extended SFHs allow more room for degeneracies to creep in and hence properties become more difficult to recover.


Figure \ref{cspmasssn} shows the recovery of stellar mass, in terms of the difference between input and output stellar mass ($y$-axis), as a function of the mass-weighted age ($x$-axis), for the same models and $S/N$~ratios as in Figure~\ref{cspagemetalssn}. The recovery of stellar mass in this case of mocks without reddening is remarkably good down to low $S/N$ ratios. Trends in mass are primarily mirroring those in age, due to the age evolution of the mass-to-light ratio.
\\
\\
Figures \ref{csp01dust}, \ref{csp1dust} and \ref{csp10dust} show the recovery of mass-weighted age and metallicity, dust reddening and stellar mass, for each of the three $\tau$-models (ordered as 0.1, 1 and 10 Gyr), two values of $S/N$~(5 and 20) and four values of $E(B-V)$ (0, 0.1, 0.25 and 0.75). Note that while the no-reddening case was already shown in previous figures, we keep it here to allow a fast evaluation of reddening effects. 

We see that at the highest signal-to-noise (S/N$=20$), all properties are recovered well nearly independently of $\tau$~and reddening, with the shortest $\tau$~and the lowest reddening being the most favourable cases, as one may expect. This is nonetheless a remarkable result. At the highest $E(B-V)=0.75$, with the more extended star formation history, the accuracy of recovered properties, especially age and reddening, degrades, with for example the youngest ages being overestimated and the reddening underestimated (the well-known age-dust degeneracy). Considering that such a high-reddening is pretty rare in nature, these results are not generally a concern, but it is important to understand the limitations of the procedure.

Metallicity seems to be the most robust among all properties, although this might also be due to the fact that our mocks are calculated for a single (solar) metallicity, followed by mass, whose accuracy primarily depends on the accuracy of age. We note that reddening seems systematically underestimated, at the lowest ages. We shall investigate this event in future developments of \FF, by allowing reddening to be fit as a free parameter.  

At lower signal-to-noise~($S/N = 5$) properties are recovered well in the regime of low-reddening ($E(B-V)<0.25$), whilst at this value and above scatter up to $0.5$~dex is observed, for example with age overestimated and reddening underestimated. These effects compensate each other in most cases such that the stellar mass remains pretty well determined even in these less favourable regimes. 

Last, we consider the recovery of the star formation history (SFH) in Figures~\ref{sfh1} and \ref{sfh2}. Here we plot our reconstructed SFHs in terms of SSP weights as a function of the mass-weighted age, and we overlay the input SFH as a smooth curve (red dashed lines). Each sub-figure contains results for the shortest $\tau$ (0.1 Gyr) in the upper row, and the longest $\tau$ (10 Gyr) in the lower row, for three reddening values, increasing from left to right. Each double plot refers to a different $S/N$, namely 5, 20 and 50. 

Figure~\ref{sfh1} reports the case where the mock spectra are viewed 1 Gyr after the start of star formation, hence one should expect to find only populations younger than this limit in the ideal case. Figures~\ref{sfh2} refers to 10 Gyr after the starting of star formation, hence one should expect a larger spread in ages at the largest $\tau$.

These trends are exactly found in the plots. The recovery of the input star formation history is very good at $S/N=20$~for both short and long $\tau$ and for a range of reddening. At $S/N=50$ the recovery is basically perfect for the shortest $\tau$~independently of reddening and formation epoch. Note also the shift in the fitted ages passing from the 1 Gyr to the 10 Gyr start of star formation, as in the input star formation history. The fraction of SSPs formed are not in precise agreement with the smooth curve in the case of the long $\tau$, but the range of possible ages is generally well matched and one should also note that such a long $\tau$~is a quite complicated case.
\\
\\

In conclusions, these tests give us confidence on a good to excellent performance of \FF\ in simple or complex cases, down to a remarkably low $S/N$~and also in presence of reddening. Particularly good is the recovery of the star formation history.

\subsection{Effect of Wavelength Range}\label{mockwavelength}
Here we test the effect of the wavelength range adopted in the fitting procedure on the derived properties, focussing on age and metallicity. These tests are useful to directly probe the degeneracies of the model components and calibrate the fitting setup. We use as test populations the spectrum of the M67 star cluster, whose age is independently known, and mock galaxies as in the previous sections, whose ages and metallicity are a well defined input.

\subsubsection{M67}\label{m67_section}
\begin{figure}
\begin{center}
\begin{subfigure}{8.0cm}
	\includegraphics[width=\linewidth]{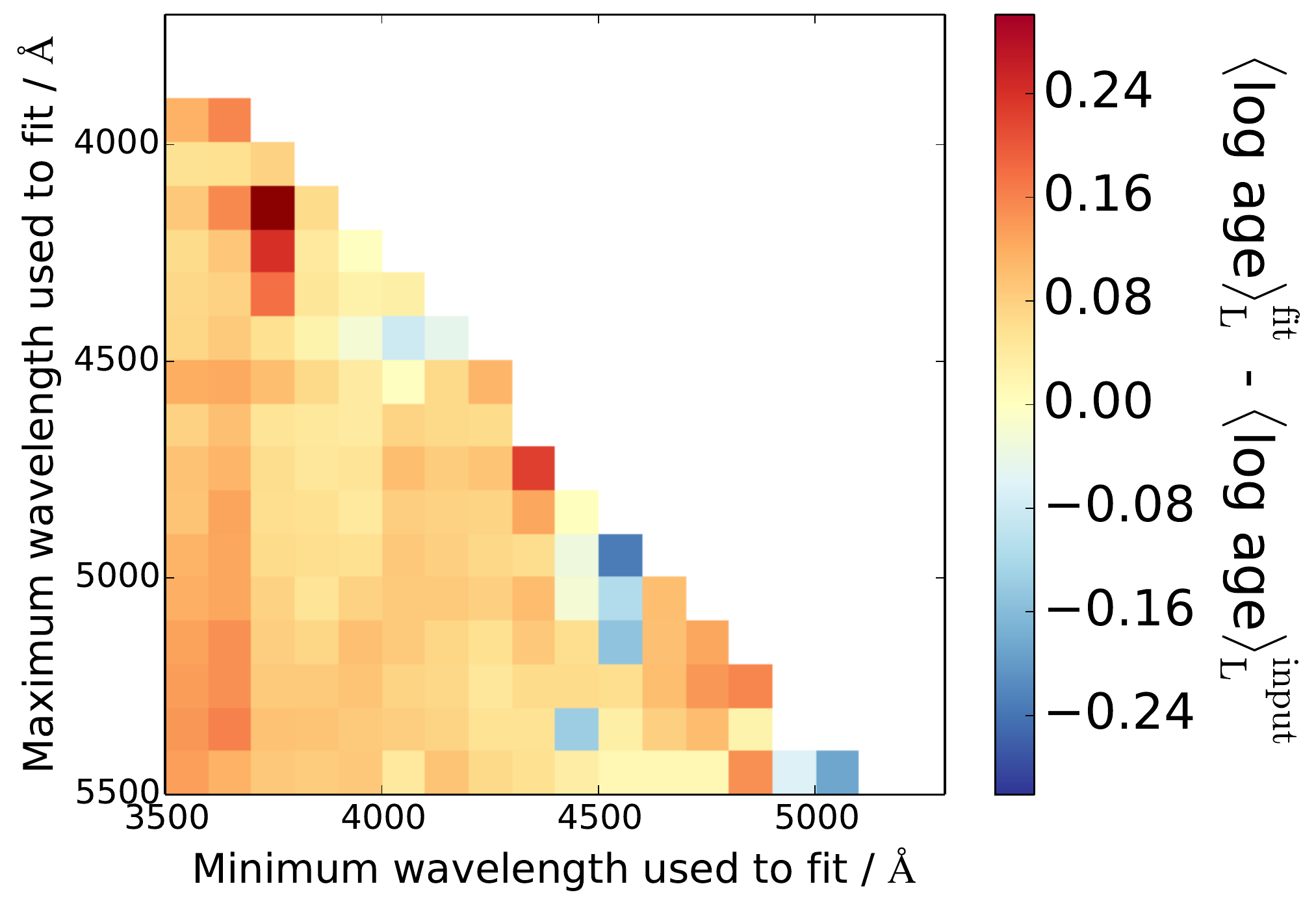}
	\caption{M11 - MILES}
\end{subfigure} 
\begin{subfigure}{8.0cm}
	\includegraphics[width=\linewidth]{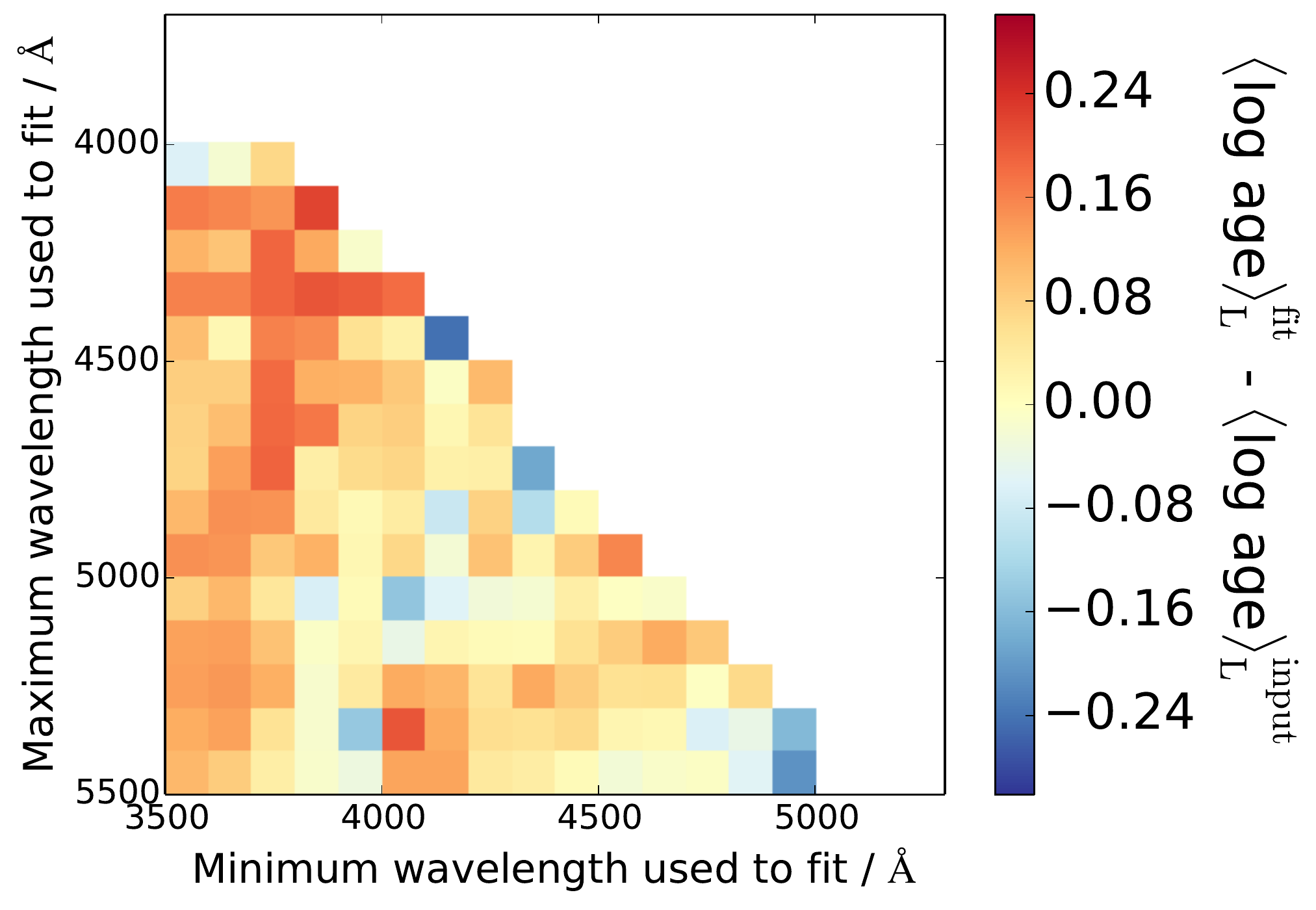}
	\caption{M11 - STELIB}
\end{subfigure} 
\begin{subfigure}{8.0cm}
	\includegraphics[width=\linewidth]{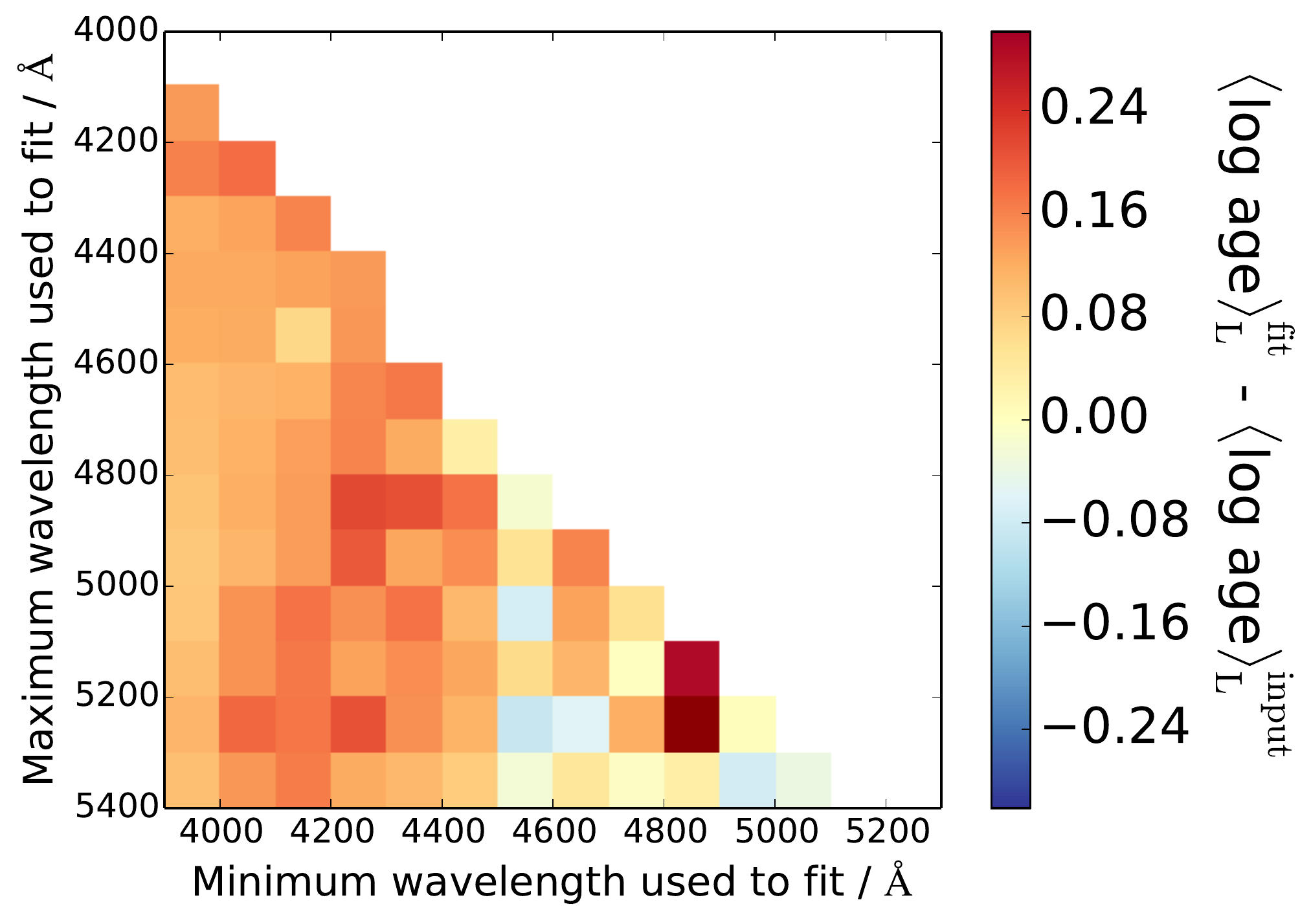}
	\caption{M11 - ELODIE}
\end{subfigure} 
\caption[Age recovery of M67 as a function of wavelength range used to fit.]{How age determinations for the star cluster M67 obtained with M11-MILES/STELIB/ELODIE-based models respectively depend on the wavelength range fitted over. Coloured squares show the average age determination for that range of wavelengths fitted across. Ages consistent with the CMD-one (4 Gyr, \protect\citet{1999ASSL..237..111C} and \protect\cite{2009ApJ...698.1872S}) are shown in colour close to yellow. The top-right of the plots are empty since those regions cover a negative, and hence non-existent, wavelength range.}
\label{agesquares}
\end{center}
\end{figure}

The wavelength range used for the best fits in Figure \ref{m67_best_fits} was the maximum used by the data (3650 - 5350 \AA) or the models. Here we test the effect on the derived age of choosing a smaller wavelength range, and we also study the effect as a function of the model input empirical stellar library. We focus on age as it is independently known from CMD-fitting. The results are shown in Figure \ref{agesquares}, where we have fitted with full freedom of SSPs and combinations of fits. CMD-consistent ages are those plotted in light yellow (cfr. vertical colour bars). \\
Looking first at the whole wavelength spanned, MILES-based models give most often the most accurate solutions, finding that for many regions of the Figure, the age determined is exactly correct, with about 0.1 dex spread in values for large wavelength ranges. STELIB-based models have more variance in the ages recovered, about 0.2 dex, but generally find accurate solutions. ELODIE-based models generally overestimate the age by about +0.2 dex.
\\
The following features related to wavelength are noted:
\begin{itemize}
\item{STELIB-based models release the most consistent fits that are also CMD-consistent when using a relatively small, blue wavelength range (around 3900 to 4300 \AA): this is the well-known D-4000 \AA~break (see below).}
\item{MILES and ELODIE-based models release CMD-consistent ages when a large wavelength range is used, but fails to do so when not taking account of the lower wavelength ranges (approximately below 4300 \AA).}
\item{All models release ages that are too high (which correspond to metal-poor solutions) when cutting out the blue part of the spectrum (up to $\sim$ 4300 \AA). In general the models are more sensitive to the minimum wavelength range set, clearly showing that the models can discriminate the age much more clearly at low wavelengths, for simple stellar populations such as in M67.}
\end{itemize}
The dominant effect on the age accuracy as a function of wavelength range is the sampling of the `D(4000)' (or the 4000\AA ~break, defined and described in \citet{1983ApJ...273..105B}), which is a spectral break generated by the onset of photospheric opacities, due to a variety of elements heavier than Helium in various stages of ionisation, short ward 4000~\AA. Hence the strength of the break depends on temperature, which can be related to the age of the population, and metallicity. Therefore including this region in the analysis explains why this is a necessary wavelength region to fit in order to accurately determine the age of M67. This shows that in order to accurately trace the age of intermediate-age stellar populations, we must use this wavelength region in any spectral fit where these stars are dominant. 

\subsubsection{Mock Galaxies}
\begin{figure*}
\begin{center}
\begin{subfigure}{7.5cm}
	\includegraphics[width=\linewidth]{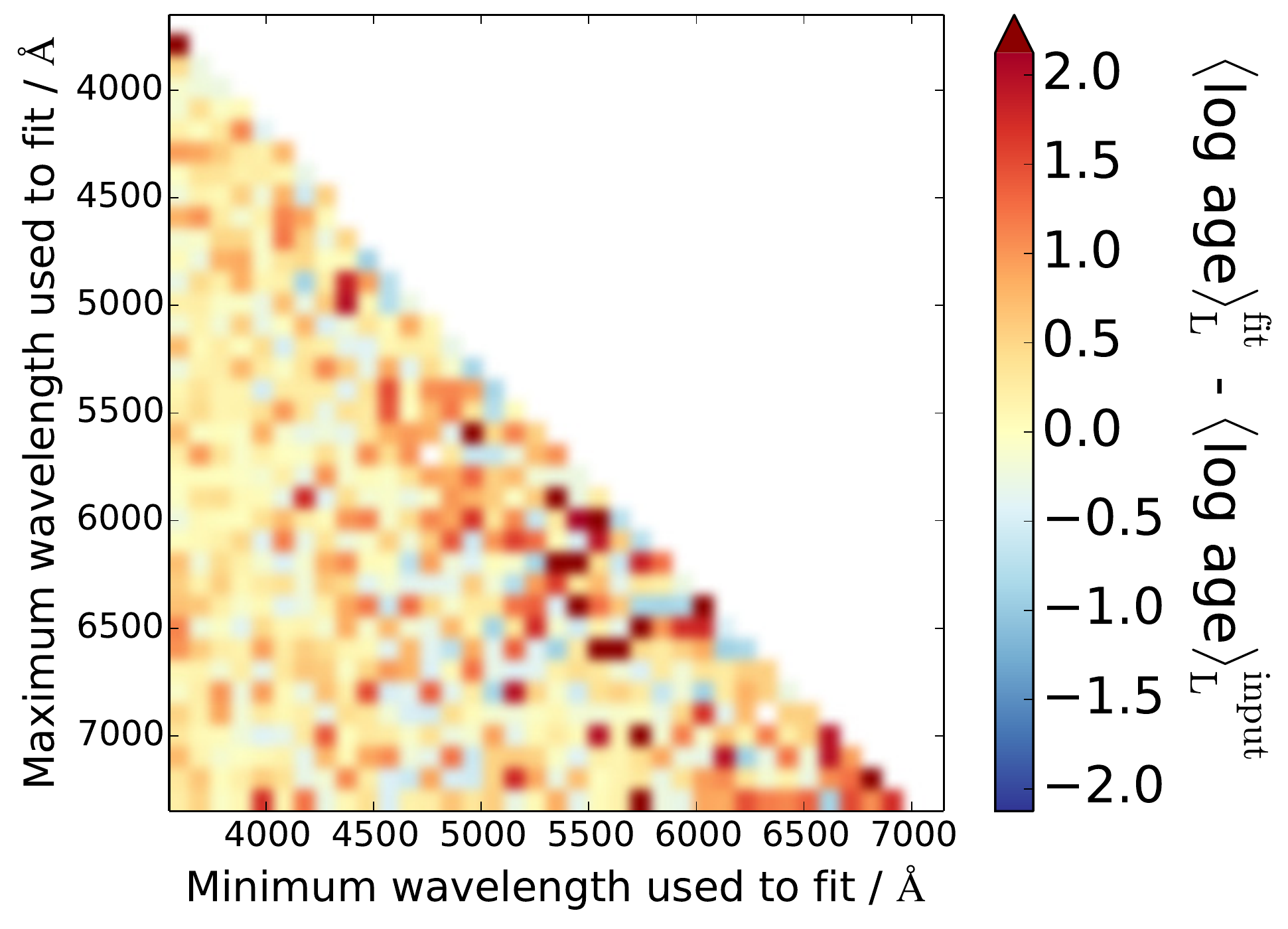}
	\caption{Age recovery of a 55 Myr mock galaxy.}
\end{subfigure} 
\begin{subfigure}{7.5cm}
	\includegraphics[width=\linewidth]{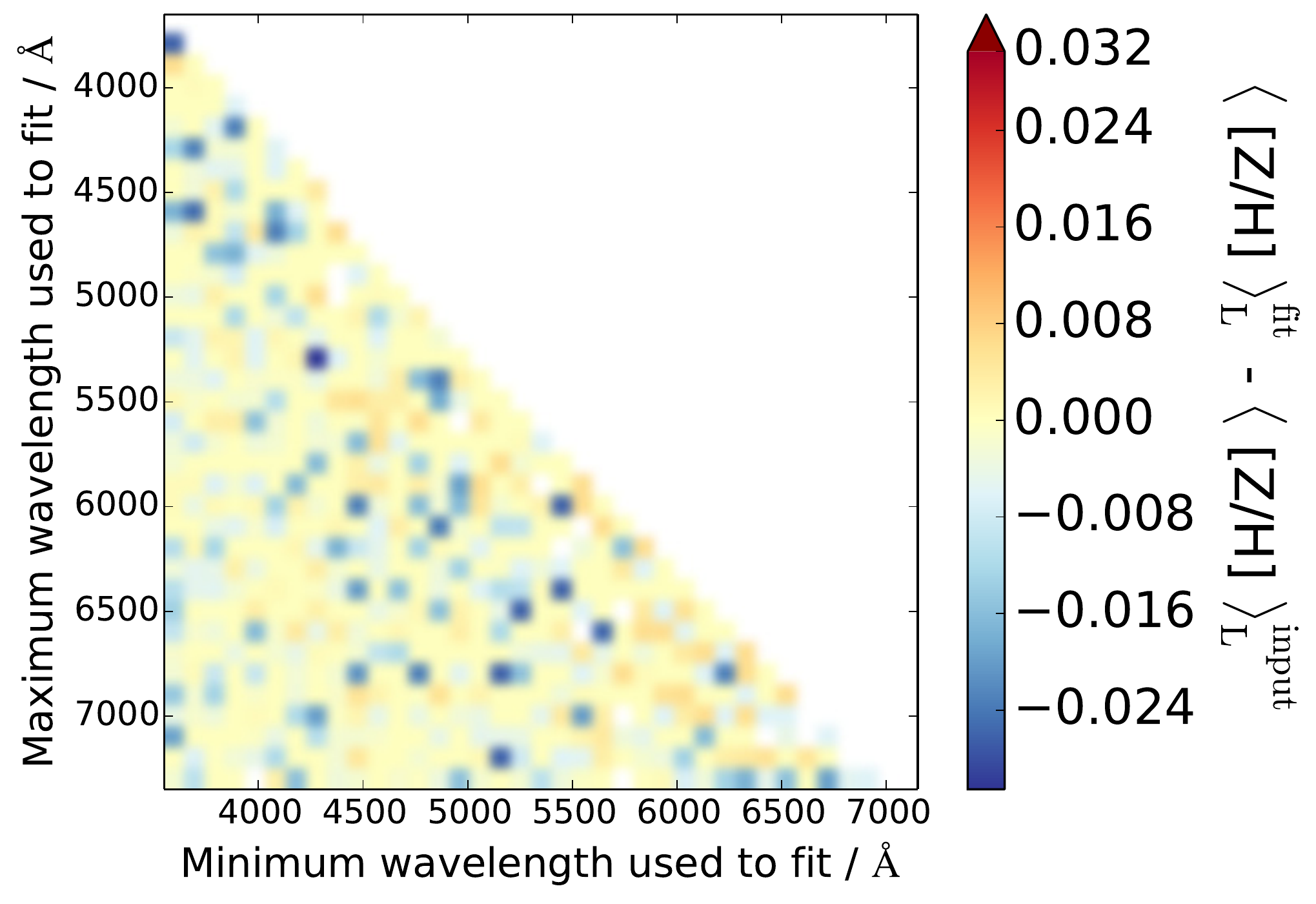}
	\caption{Metallicity recovery of a 55 Myr mock galaxy.}
\end{subfigure} 
\begin{subfigure}{7.5cm}
	\includegraphics[width=\linewidth]{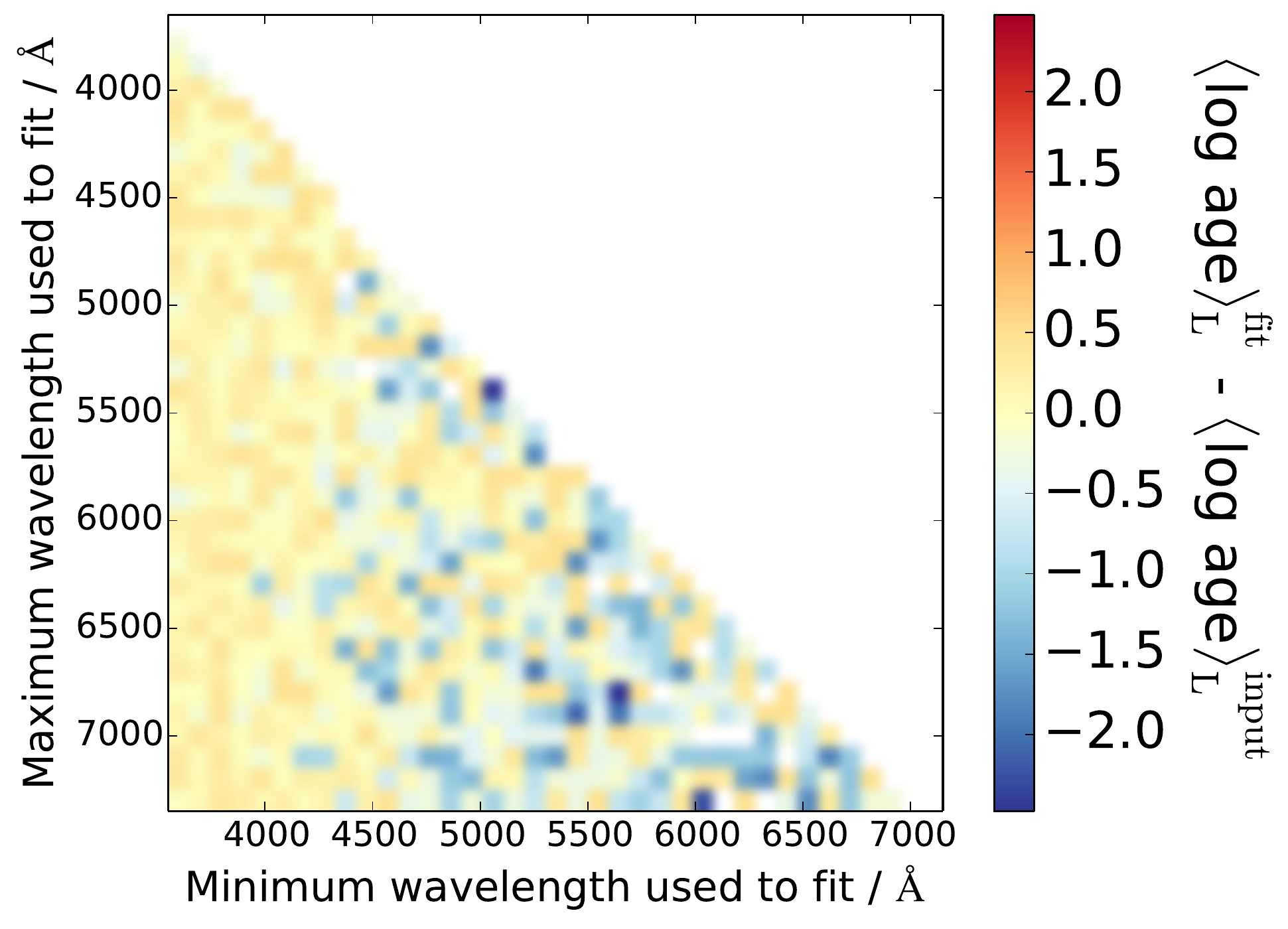}
	\caption{Age recovery of a 7 Gyr mock galaxy.}
\end{subfigure} 
\begin{subfigure}{7.5cm}
	\includegraphics[width=\linewidth]{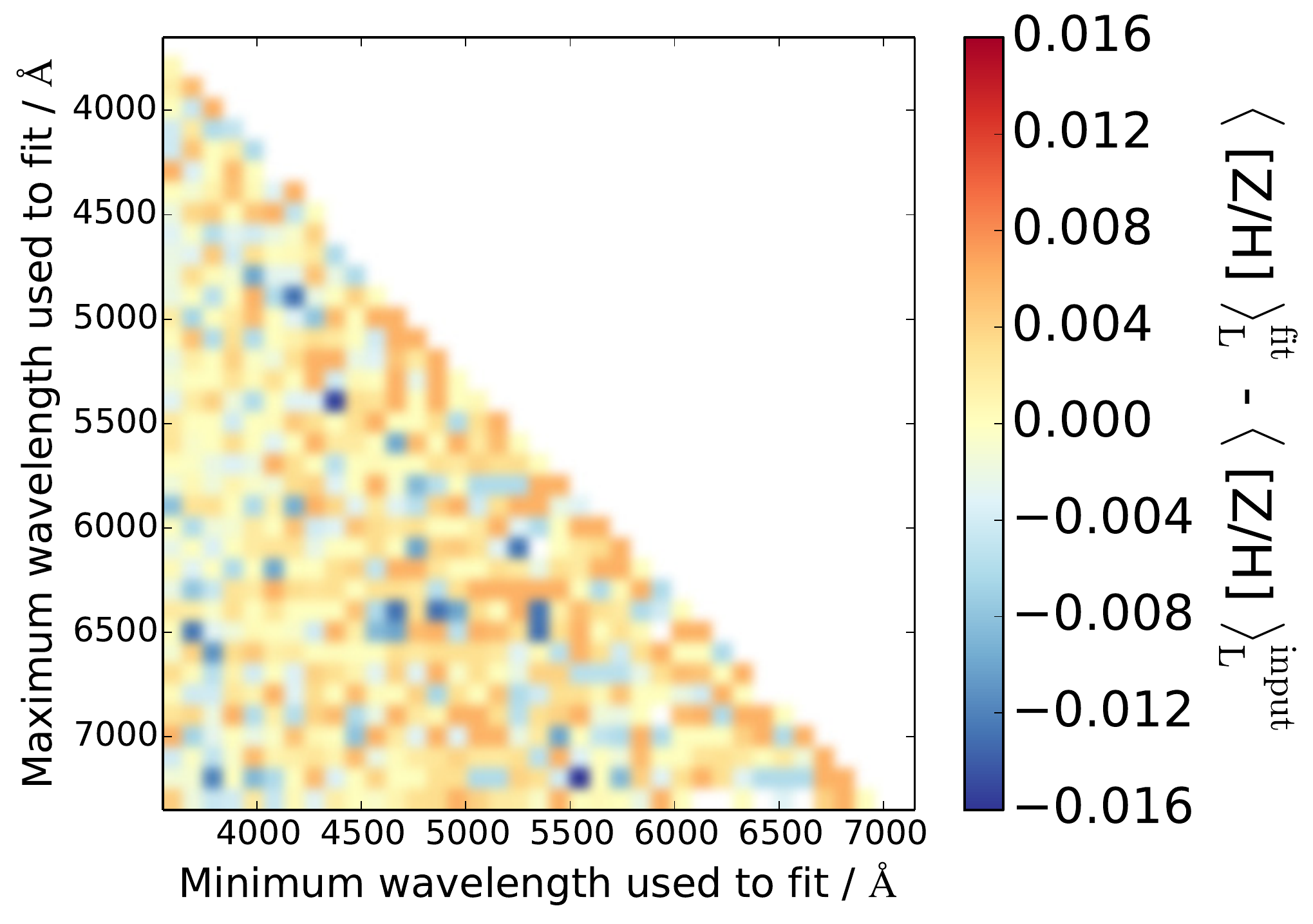}
	\caption{Metallicity recovery of a 7 Gyr mock galaxy.}
\end{subfigure} 
\caption[Age and metallicity recovery of SSP-based mock galaxies as a function of wavelength range used.]{Light-weighted age (LHS) and metallicity (RHS) recovery of SSP-based mock galaxies as a function of wavelength range used for the fitting (at 100 \protect\AA~intervals). The top panel shows results for a 55 Myr, solar metallicity SSP and the bottom panel shows results for a 7 Gyr, solar metallicity SSP. All use MILES-based models to fit to MILES-based mocks. The top-right of the plots are empty since those regions cover a negative, and hence non-existent, wavelength range.}
\label{wavelength_masses}
\end{center}
\end{figure*}

To complement the above tests on a real star cluster as M67, we carry out an investigation of the effect of varying the wavelength range used to fit to the spectra, both as a function of input model SSP age and signal to noise. For brevity, we display only the ages and metallicities obtained from fitting solar metallicity SSPs with no dust component, using a signal-to-noise of 5 and MILES-based models. Two representative cases are shown in figures \ref{wavelength_masses} for SSP mocks, with a very young (55 Myr) and an old (7 Gyr) input age, in the top and bottom panels, respectively. \\
From Figure \ref{wavelength_masses} we see two different cases of age-metallicity degeneracy. For the very young 55 Myr mock, we see that in general the recovered ages are overestimated ($\sim$ 0.3 dex), while the metallicities are very well determined, being only slightly underestimated ($\sim$ 0.01 dex). These effects worsen when the wavelength range is smaller. Conversely, the 7 Gyr mock shows that as long as one includes a wavelength fitting region below 4100 \AA, ages and metallicities are well estimated. However, once the region is removed then the solutions tend towards younger (by up to 0.5 dex), generally more metal-rich ($\sim$0.01 dex) solutions. This was the case also when fitting M67. Together, these plots show that wider wavelength ranges give more ability to overcome the effects of age-metallicity degeneracy, as also concluded in \cite{2012MNRAS.422.3285P}, and confirms that for intermediate to old age solutions, the 4000 \AA~region is essential for precise age determinations.
\subsection{Testing of Age and Metallicity Derivation with Globular Clusters}\label{calgc}
\begin{figure*}
\begin{center}
\includegraphics[width=8.5cm]{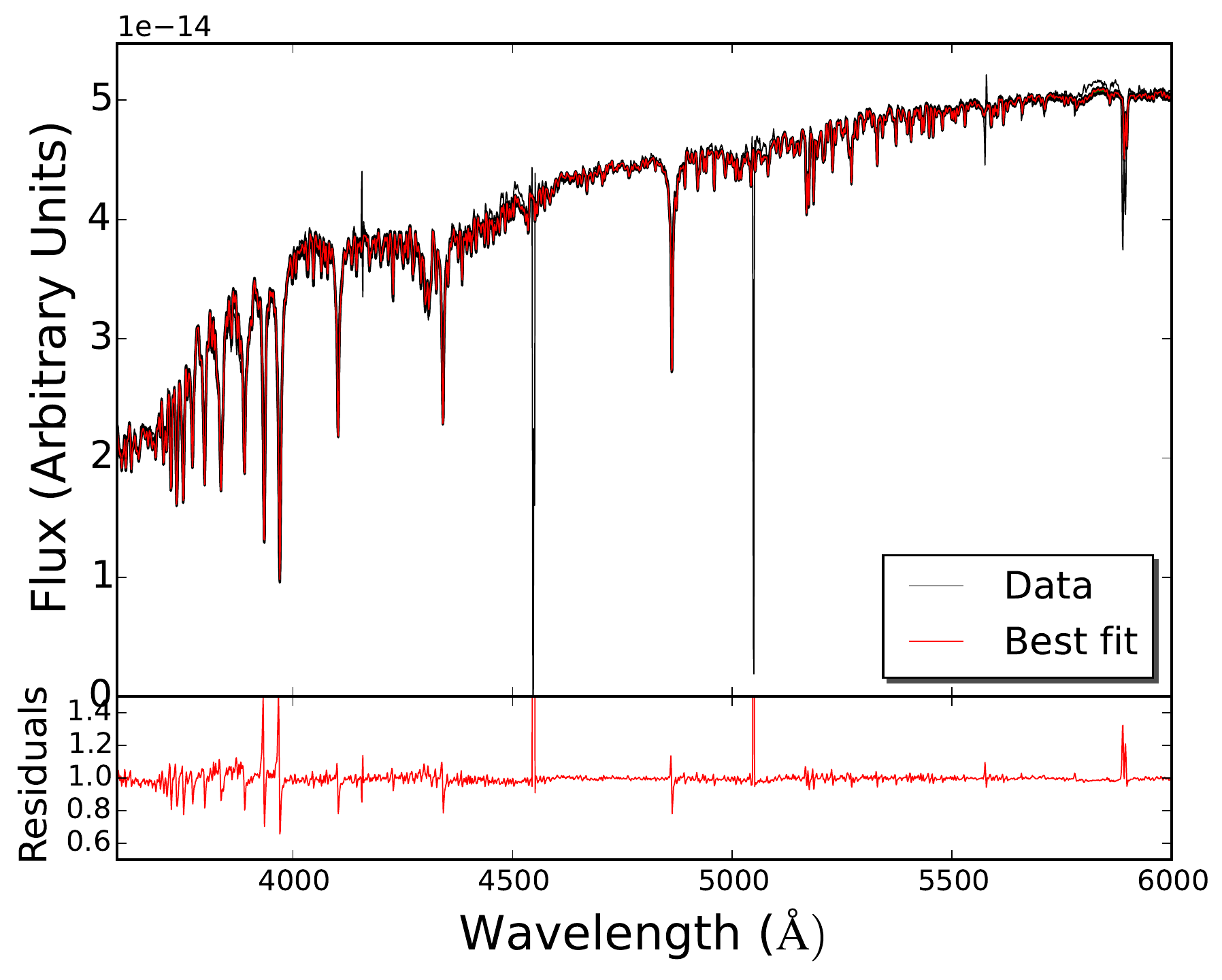}
\includegraphics[width=9.1cm]{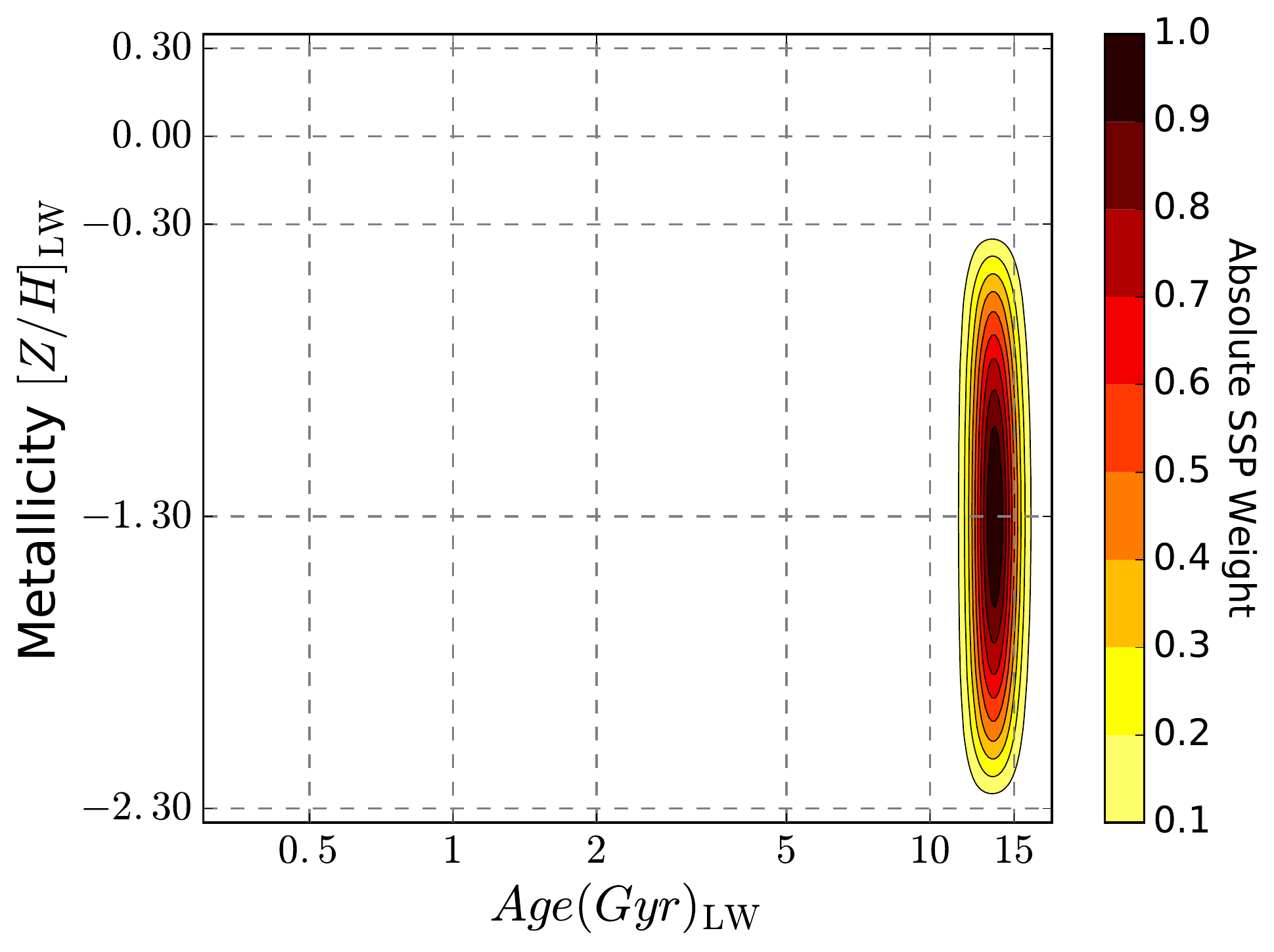}
\includegraphics[width=8.5cm]{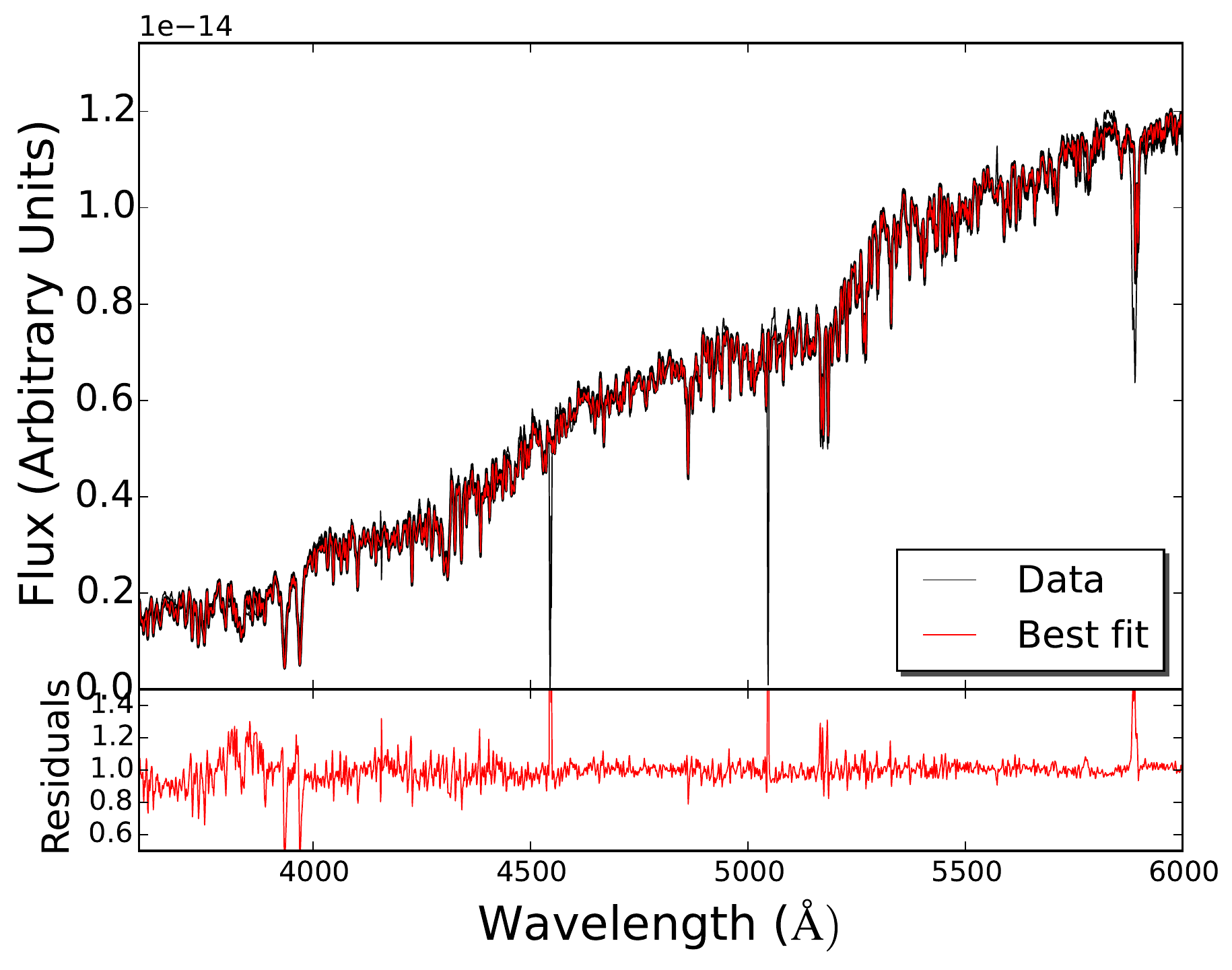}
\includegraphics[width=9.1cm]{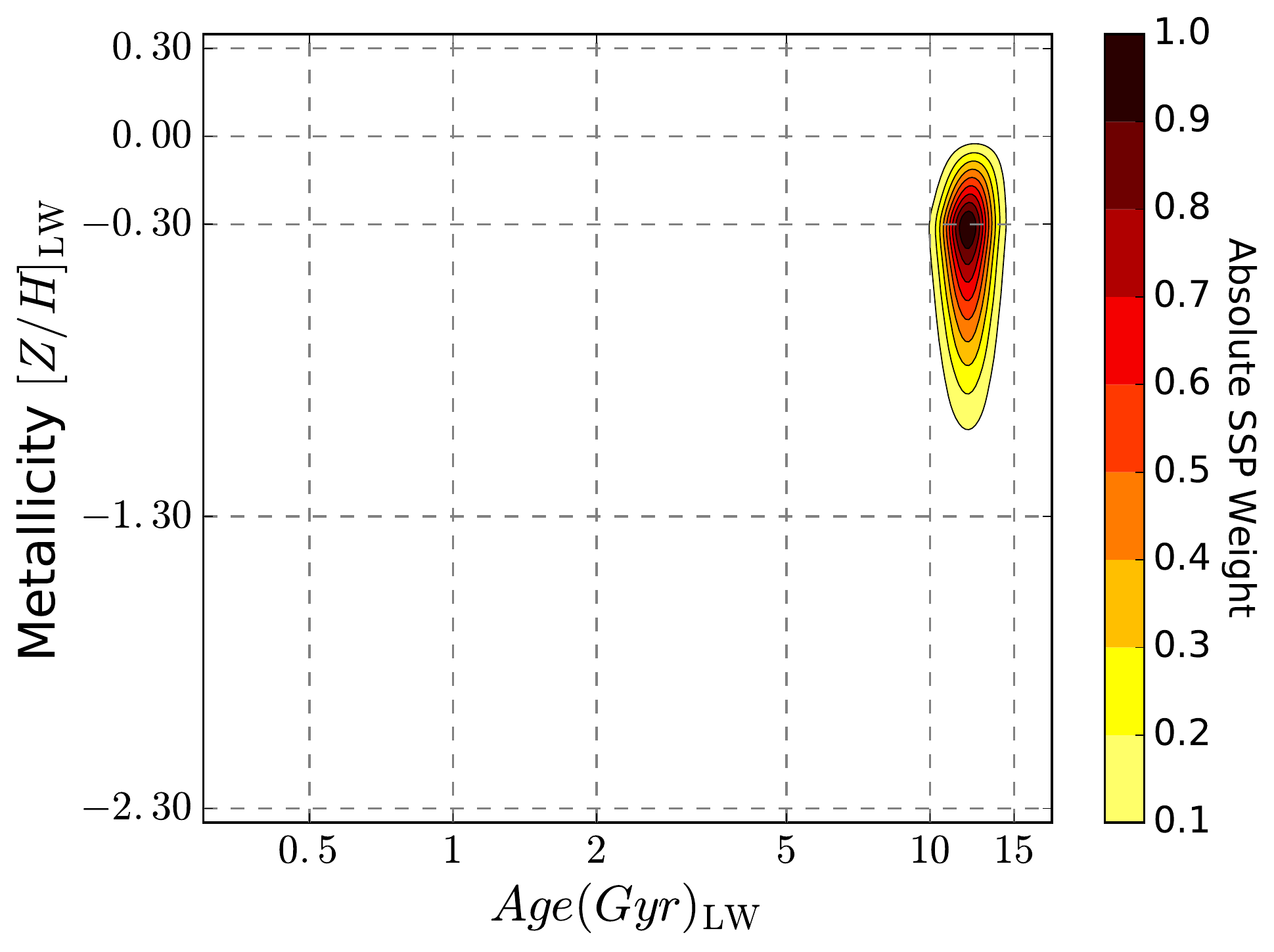}
\caption[Example fits of globular clusters.]{Fits and stellar population contours of two example globular clusters, NGC 5286 and NGC 6528 from \protect\cite{2005ApJS..160..163S}. Measurements from colour-magnitude diagram fitting show NGC 5286 is very metal-poor, whereas NGC 6528 is approximately half-solar in [Z/H]. \FF\ using M11-Miles is able to reproduce their properties remarkably well.}
\label{gc_examples}
\end{center}
\end{figure*}

\begin{figure*}
\begin{center}
\includegraphics[width=8.5cm]{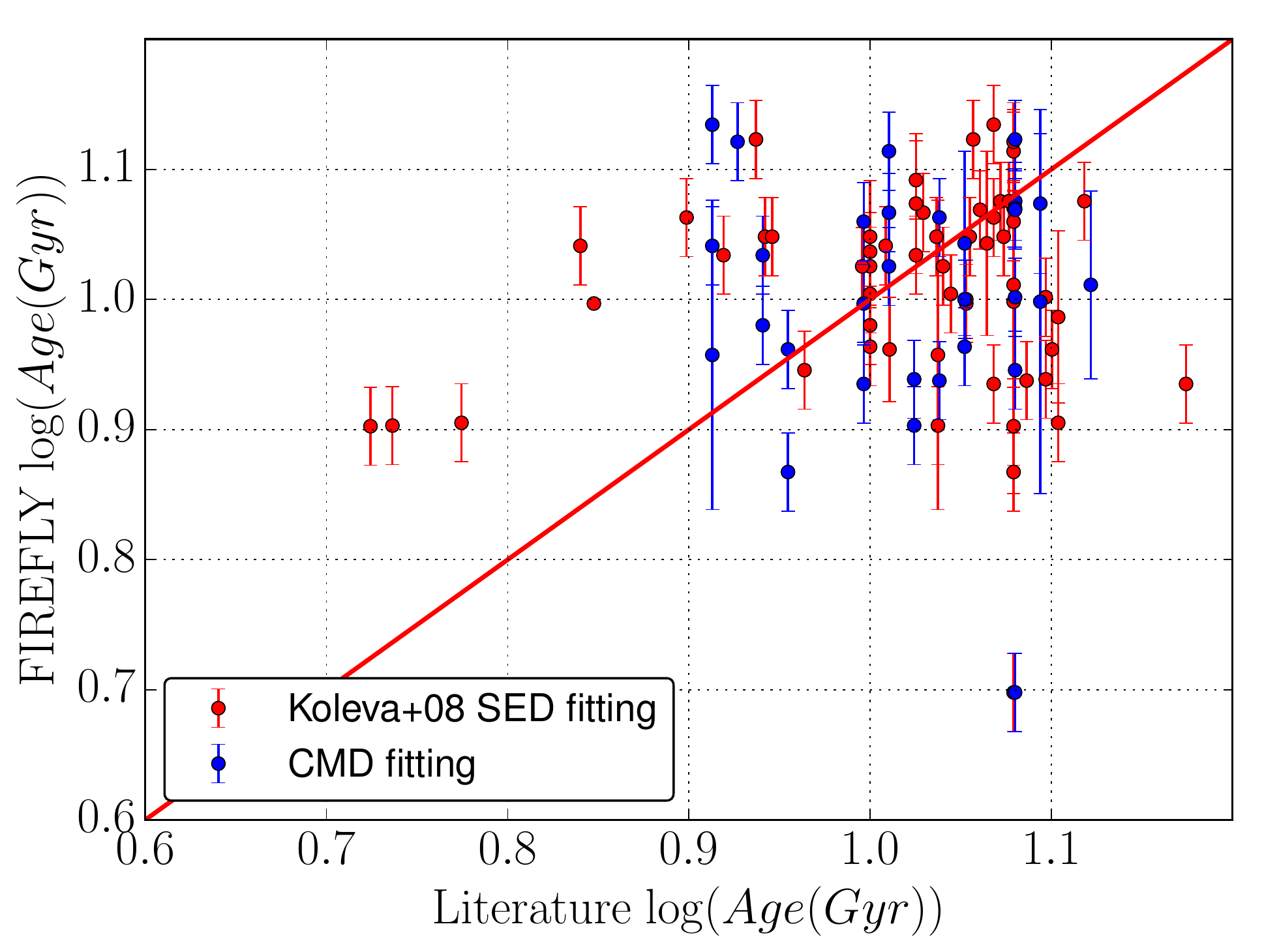}
\includegraphics[width=8.5cm]{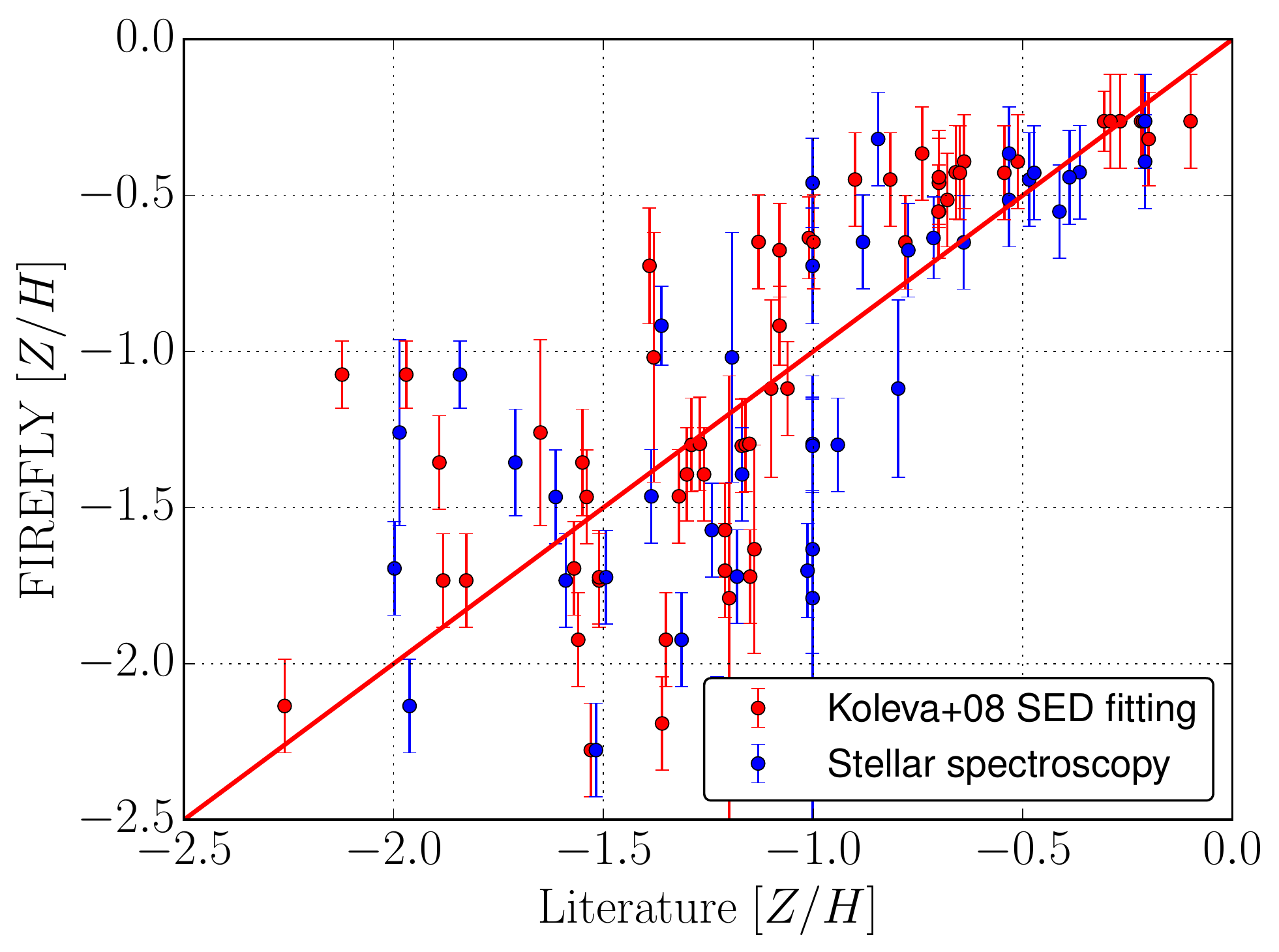}
\caption[Stellar population from \FF\ in comparison to the literature.]{Light-weighted age and metallicity derived from \FF\ with M11-MILES models compared to results obtained with another spectral fitting code (red points, from \cite{2008MNRAS.385.1998K}) or to literature values which are independent of SED fitting procedures (blue points). The latter refer to 
colour-magnitude diagram isochrone fitting in case of ages (left-hand panel), or stellar spectroscopy in case of metallicity (right-hand panel).}
\label{gc_lit}
\end{center}
\end{figure*}

As stated in the Introduction, globular clusters, especially those in the Milky Way or Magellanic Clouds, are invaluable for calibration of methods using stellar population models since we can compare our fitted ages with their independent determinations based on either fitting Colour-Magnitude diagrams (CMD-fit) for ages or stellar spectroscopy for chemical abundances~\citep{1988ARA&A..26..199R}\footnote{It should be noted, as discussed in M11, that the CMD fit depends somewhat on the tracks adopted for the fitting, which in turn are also a basic ingredient of stellar population models.}. Star clusters provide the `simplest' form of stellar populations in nature, i.e. approximately a single coeval and mono-metallicity population of stars, something we can assess with \FF. 
In this Section, we use globular cluster data from \cite{2005ApJS..160..163S}, as also used in M11. In Figure \ref{gc_examples} we show two examples of \FF's fits to a metal-poor and metal-rich globular cluster, respectively. \\
\\
We see that the metal poor globular cluster NGC 5286 is well-represented by nearly a single group of SSPs at high age and low metallicity, whereas NGC 6528, a more metal-rich cluster, is fitted by an old component of metallicity approx half-solar.  The average values of age and total metallicity derived with the full spectral fitting with M11-MILES models are: $ t= 11.8~{\rm Gyr}, \rm{[Z/H]} = -1.80$~for NGC 5286 and $t= 11.2~\rm{Gyr}, [Z/H] = -0.33$~for NGC 6528 in remarkable agreement with the independent determinations based on CMD, namely $t\sim12~\rm {Gyr}, \rm{[Z/H]}\sim -1.73$~for NGC~5286 and $t\sim 10$ Gyr, $\rm{[Z/H]}\sim -0.2$~for NGC~6528. This is an important test that a fitting code plus a set of models has to pass. 
The dispersion in solutions might be testifying the presence of multiple generations very close in age and with a metallicity spread, although an extreme horizontal-branch not included in our library or a fraction of blue stragglers could induce subtle effects in the fitting.
One caveat here is that NGC 6528 is metal-rich and enhanced in $\alpha$~elements (e.g. \citet{2003MNRAS.339..897T}), whereas the M11 models are most likely solar-scaled at high-metallicity (see discussion in \ref{models}). We shall investigate in the future whether the hints towards multiple populations is indeed a real effect.
\\
\\
In Figure \ref{gc_lit} we compare all our light-weighted ages and metallicities to those derived in the literature, as tabulated by \cite{2008MNRAS.385.1998K}. In cases where a CMD age has been measured, we use this value, and for others we use the value derived in \cite{2008MNRAS.385.1998K} from their full spectral fitting code NBURSTS \citep{2007IAUS..241..175C} based on the Pegase-HR models \citep{2004A&A...425..881L}. Since the models and code are different, we just provide a qualitative comparison of the results. \\
\\
Very pleasingly, the ages derived from \FF\ are close to those from CMD isochrone fitting (blue symbols) within 0.1 dex, for the vast majority of cases. Those for which the two determinations are not in agreement could either host Blue-Horizontal branches (we have used M11-MILES with red HB for this test), or multiple populations. We refer to a future project for studying in detail globular clusters with \FF. 

Metallicity is compared in the right-hand panel. The blue points refer to determinations based on stellar spectroscopy. As these are generally tied to the so-called Carretta \& Gratton's scale (1997, {R. Gratton, {\it private communication}), they refer to iron abundance $[Fe/H]$ ~rather than to the total metallicity $[Z/H]$~determined via \FF. In order to make a meaningful comparison, we shifted the GCs values by $+0.3$~dex, which corrects $[Fe/H]$~to $[Z/H]$~for a $\alpha$-enhancement value around $[\alpha/Fe]\sim0.3$, using the scaling by \citet{2003MNRAS.339..897T}. \\
The metallicity recovery is generally good.
\\
\\
The CMD ages are in better agreement with \FF\ than the values derived from fitting in \cite{2008MNRAS.385.1998K}, where they seem to give lower ages than the CMD-ones. This mismatch may originate in the combination of fitting procedures and adopted models. \\
Overall, the comparisons show that \FF\ is capable of correctly matching the properties of globular clusters derived from CMD fitting and stellar spectroscopy.
\section{Testing with SDSS DR7}\label{sdss}
We focus on spectra from the Sloan Digital Sky Survey Data Releases 7 (DR7) and 9 in this presentation paper because this is a very large sample which covers a range of galaxy types and which has been widely used in the literature. Obviously, our fitting code can be applied to any spectra. For example, in Wilkinson et al. (2015) we have analysed integral field unit spectroscopy data from SDSS-IV/MANGA survey and in a forthcoming work we shall use it to fit the spectra of the most massive galaxies across a redshift range, from the SDSS-III Baryon Oscillation Spectroscopic Survey Data Release 12 (BOSS-DR12) \citep{2013AJ....145...10D}. The observed galaxy spectra that we have used in this work come from the Sloan Digital Sky Survey (SDSS) II Data Release 7 (DR7). Equipped with two multi-object spectrographs, SDSS acquired spectra of more than 0.93 million galaxies in its `Legacy' survey of a very large area of sky across a large wavelength range, covering 9380 square degrees at 3800 -- 9200 \AA~at a spectral resolution of $\sim 3$~\AA, across a redshift range of $0.0 \le z \le 0.5$. .
\subsection{Data Pre-processing}
When analysing DR7 SEDs, some important considerations need to be made to ensure good recovery of galaxy physical parameters. In \FF, these considerations become features in the code that can make the analysis more versatile and robust. We list these considerations and features here:
\begin{itemize}
\item{We always use the actual velocity dispersion (in our case as output by GANDALF and pPXF) to downgrade the resolution of our models as described in Section \ref{elines}. This routine also provides us with emission-cleaned spectra that we use to fit each of the galaxies used in this paper. As described in Section \ref{dust}, we pre-process the data SEDs for Milky Way extinction by assuming a \cite{1999PASP..111...63F} reddening law, and de-reddening the data.}
\item{All SDSS datasets include quality flags (known as the `good-pixel array') on each datapoint that signify if it is untrustworthy. For example, pixels can show extreme residuals or very low signal to noise that can arise from artefacts in the data, or high amounts of sky flux. These points are removed from the analysis and shown in the spectral fits as if that part of the wavelength space had no data. Additionally, skylines from atmospheric scattering occur at 5577, 6200, and 6363 \AA ~(in the observed frame) - these points are similarly removed with 5 \AA~masks to ensure they are not part of the spectral fits in their corresponding galaxy rest frame.}
\item{Occasionally, these flags can miss poorly processed fluxes, which would bias our fitting by weighting the chi-squared values obtained towards them. To prevent this, we use sigma-clipping of points at every measurement of chi-squared as used in many popular codes such as PPXF, \cite{2004PASP..116..138C}.}
\end{itemize}

\subsection{Results from DR7}
In this Section, we show the application of \FF\ to SDSS DR7 and construct the star formation history of full DR7 dataset by summing the contributions from all the likelihood-weighted fits. An example fit to a typical DR7 galaxy was shown in Figure \ref{example_dr7}. We use the full range of age, metallicity and wavelength coverage available from each of the MILES-, STELIB- and ELODIE-based M11 models since we want to assess the ability of each of the models to recover galaxy properties to their fullest extent. The light-weighted and mass-weighted ages vs metallicities derived for DR7 galaxies are shown in Figure \ref{dr7sfh}. We note that we have computed a full set of stellar population properties including mass and light-weighted average ages and metallicities, dust, mass, and chi-squared as a function of stellar library, but the plots shown are representative of the full set of possible plots. The calculated properties are publicly available\footnote{www.icg.port.ac.uk/firefly}.\\
\begin{figure*}
\centering
\begin{subfigure}[t]{0.47\linewidth}
	\includegraphics[width=\linewidth]{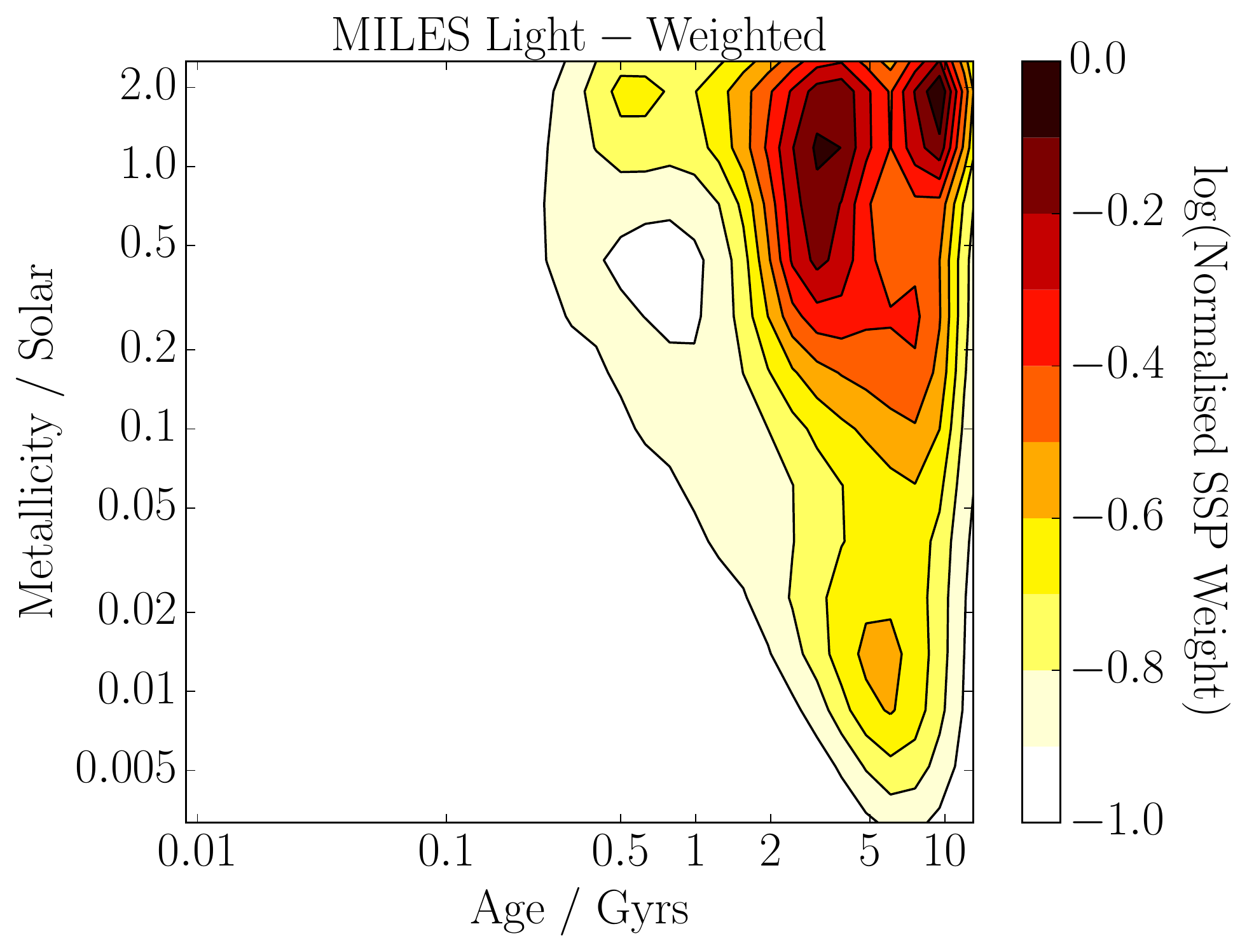}
	\caption{Light-weighted MILES-based solutions.}
\end{subfigure}
\hspace{0.5cm}
\begin{subfigure}[t]{0.47\linewidth}
	\includegraphics[width=\linewidth]{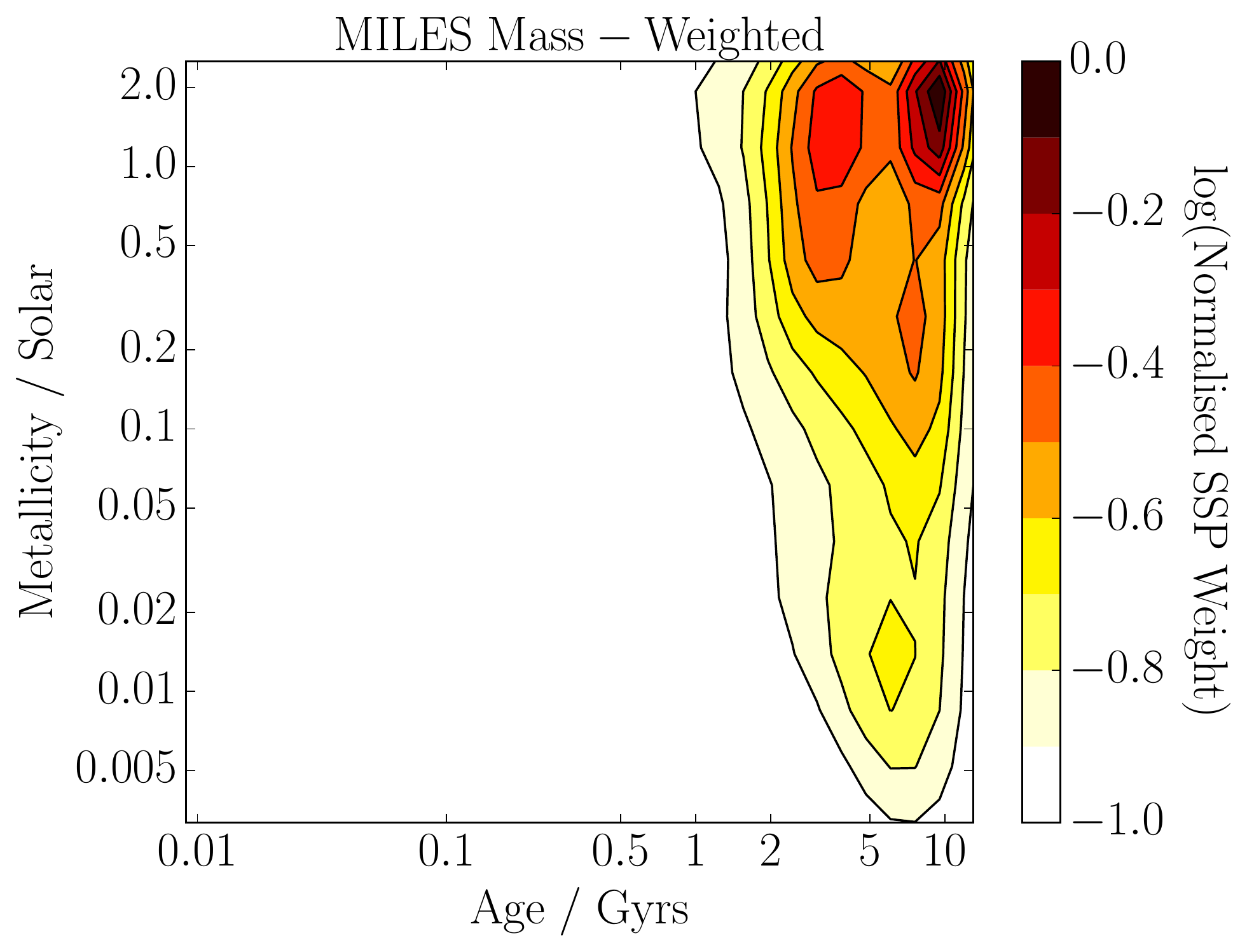}
	\caption{Mass-weighted MILES-based solutions.}
\end{subfigure}
\begin{subfigure}[t]{0.47\linewidth}
	\includegraphics[width=\linewidth]{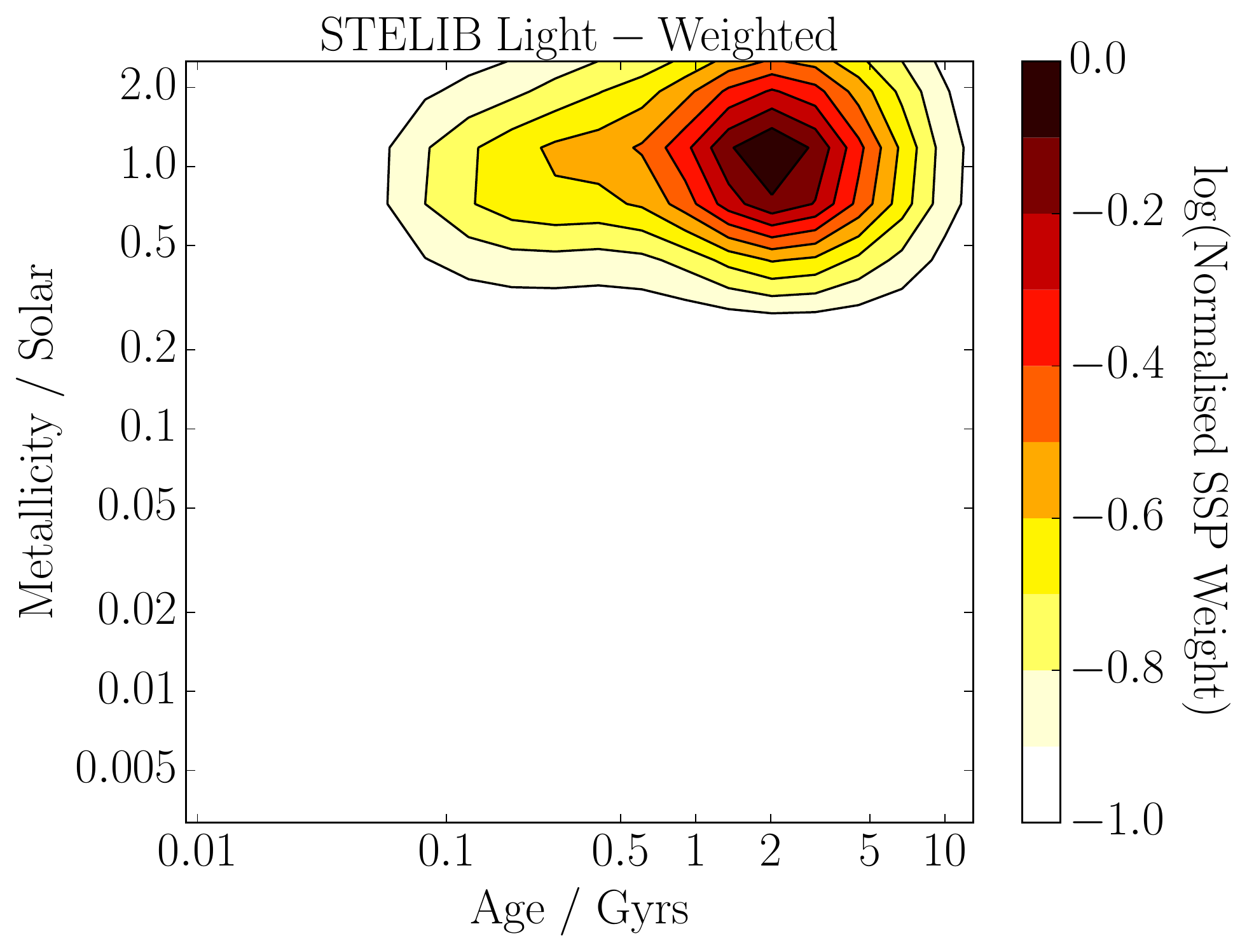}
	\caption{Light-weighted STELIB-based solutions.}
\end{subfigure}
\hspace{0.5cm}
\begin{subfigure}[t]{0.47\linewidth}
	\includegraphics[width=\linewidth]{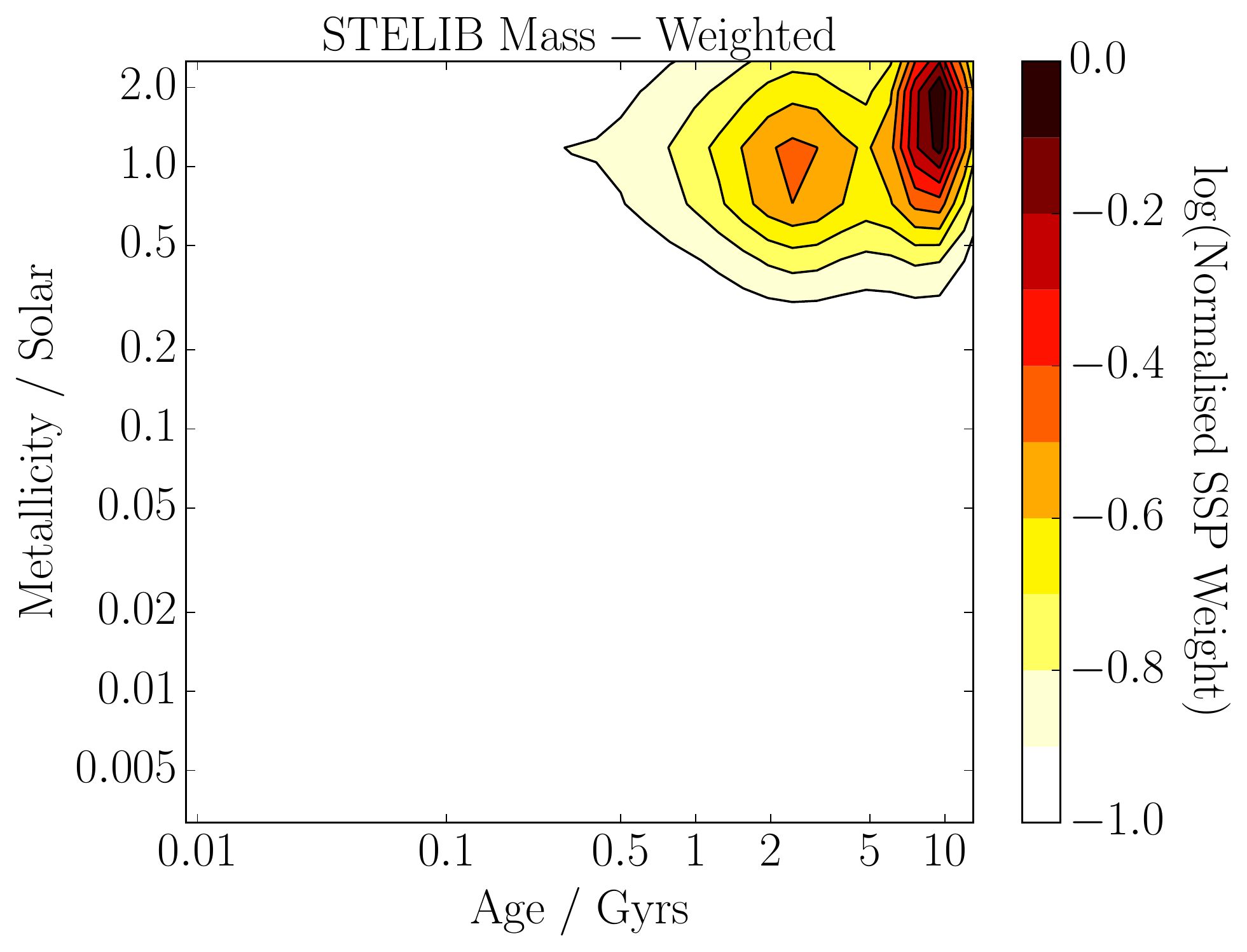}
	\caption{Mass-weighted STELIB-based solutions.}
\end{subfigure}
\begin{subfigure}[t]{0.47\linewidth}
	\includegraphics[width=\linewidth]{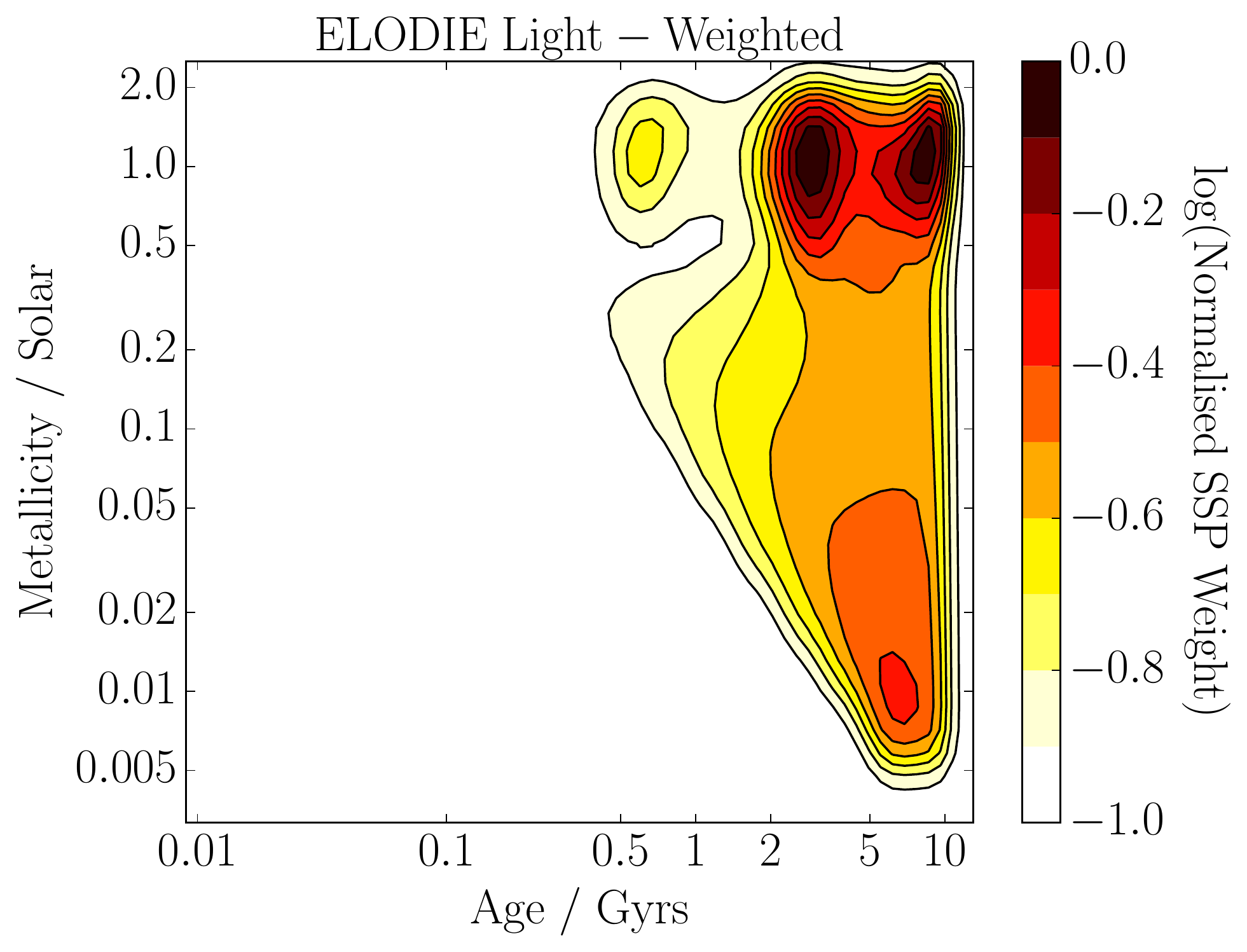}
	\caption{Light-weighted ELODIE-based solutions.}
\end{subfigure}
\hspace{0.5cm}
\begin{subfigure}[t]{0.47\linewidth}
	\includegraphics[width=\linewidth]{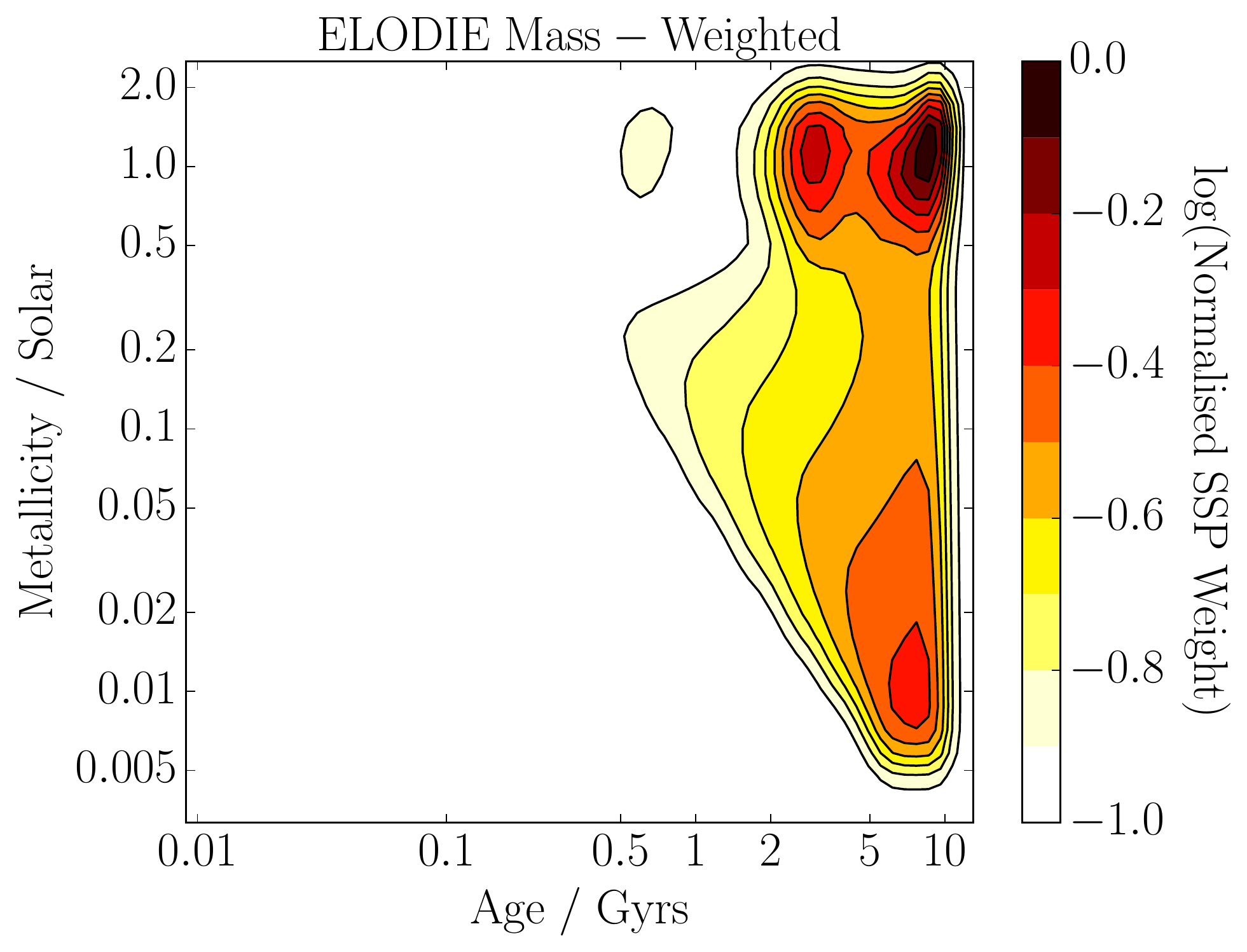}
	\caption{Mass-weighted ELODIE-based solutions.}
\end{subfigure}
\caption[Star formation histories of all DR7 galaxies.]{The likelihood-weighted sum of all SSP contributions from all galaxies in the SDSS DR7 survey, as a function of empirical stellar library model ingredient, by light (i.e. flux) (LHS) or by stellar mass (RHS). Each contour represents the fractional weight of stellar population solutions in that part of age-metallicity space.  An interpolation algorithm has been used to smooth the distributions.}
\label{dr7sfh}
\end{figure*}

We now briefly analyse the distributions of Figure \ref{dr7sfh}.
For all three models we see that DR7 galaxies are dominated in mass-weighted contributions (right-hand panels) by a major component with old age, $\sim12$ Gyr and high metallicity (solar or above), plus minor components at younger ages (for MILES and STELIB-based models) or low-metallicity (for ELODIE-based models). \\
Differences due to the input stellar library become more evident in the light-weighted contributions (left-hand panels). M11-MILES and M11-ELODIE solutions display a bimodality in light-weighted ages, with a group of galaxies with low ($\sim 3$~Gyr) ages. STELIB-based models is more consistent with a unimodal, broader distribution with somewhat intermediate-ages. Results based on this model flavour lack the extension to low metallicity because the library lacks low-metallicity models (cfr. Table~1). Probably as a consequence of a narrow span in metallicity, the STELIB-based solutions show a wider spread in age, compared to the other models whose solutions are more compressed towards older ages. MILES-based models show the greatest number of solutions with high metallicity, both at high $\sim$ 10 Gyr and intermediate $\sim$ 2 Gyr ages. STELIB-based models are similar, but to a lesser extent. ELODIE-based solutions cluster around solar metallicity. Hence, the input stellar library affect both the recovered ages and metallicities, mostly in a light-weighted sense.\\
\begin{figure}
\centering
\begin{subfigure}[b]{0.75\linewidth}
	\includegraphics[width=\linewidth]{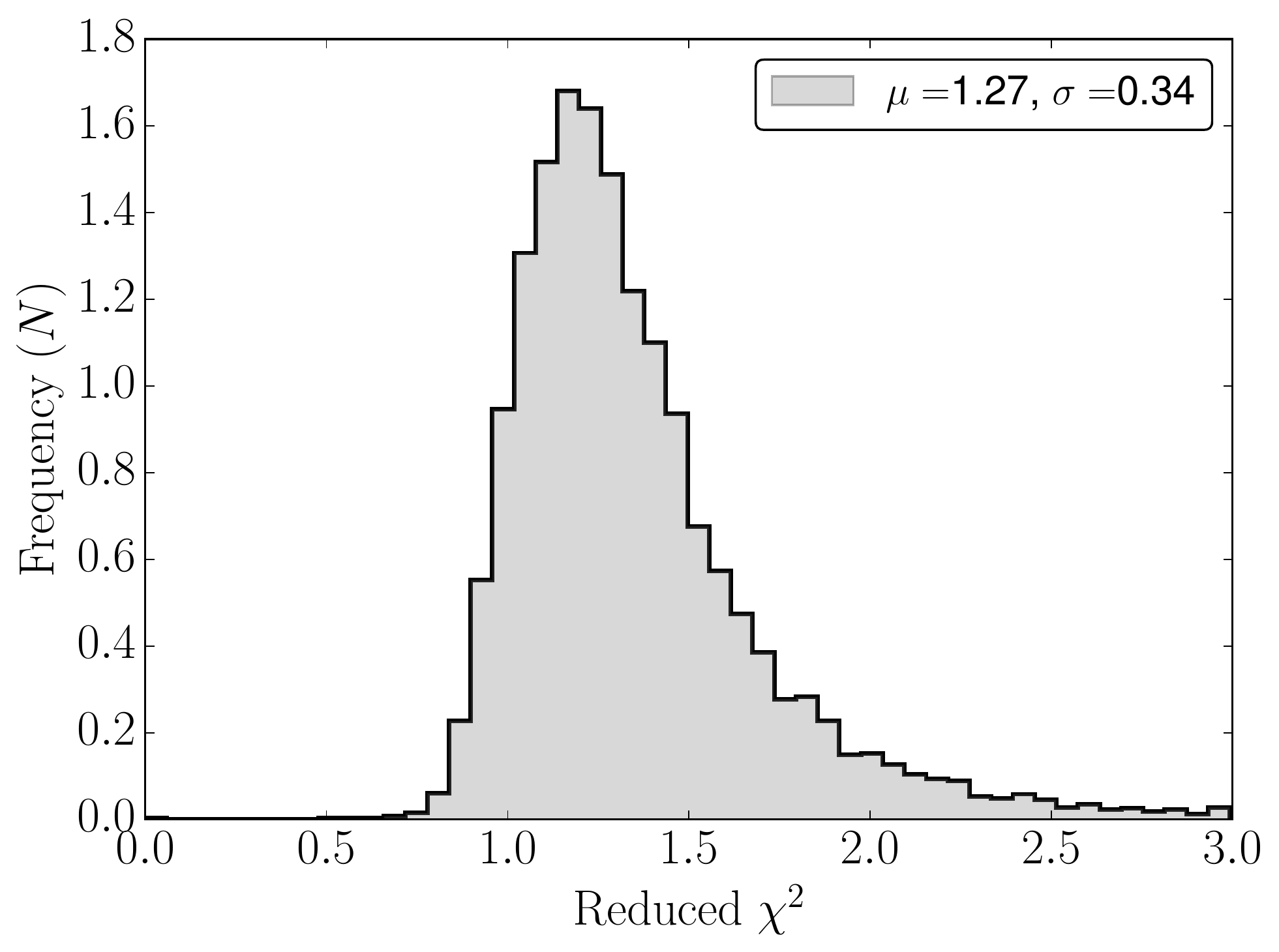}
	\caption{MILES-based models.}
\end{subfigure}
\begin{subfigure}[b]{0.75\linewidth}
	\includegraphics[width=\linewidth]{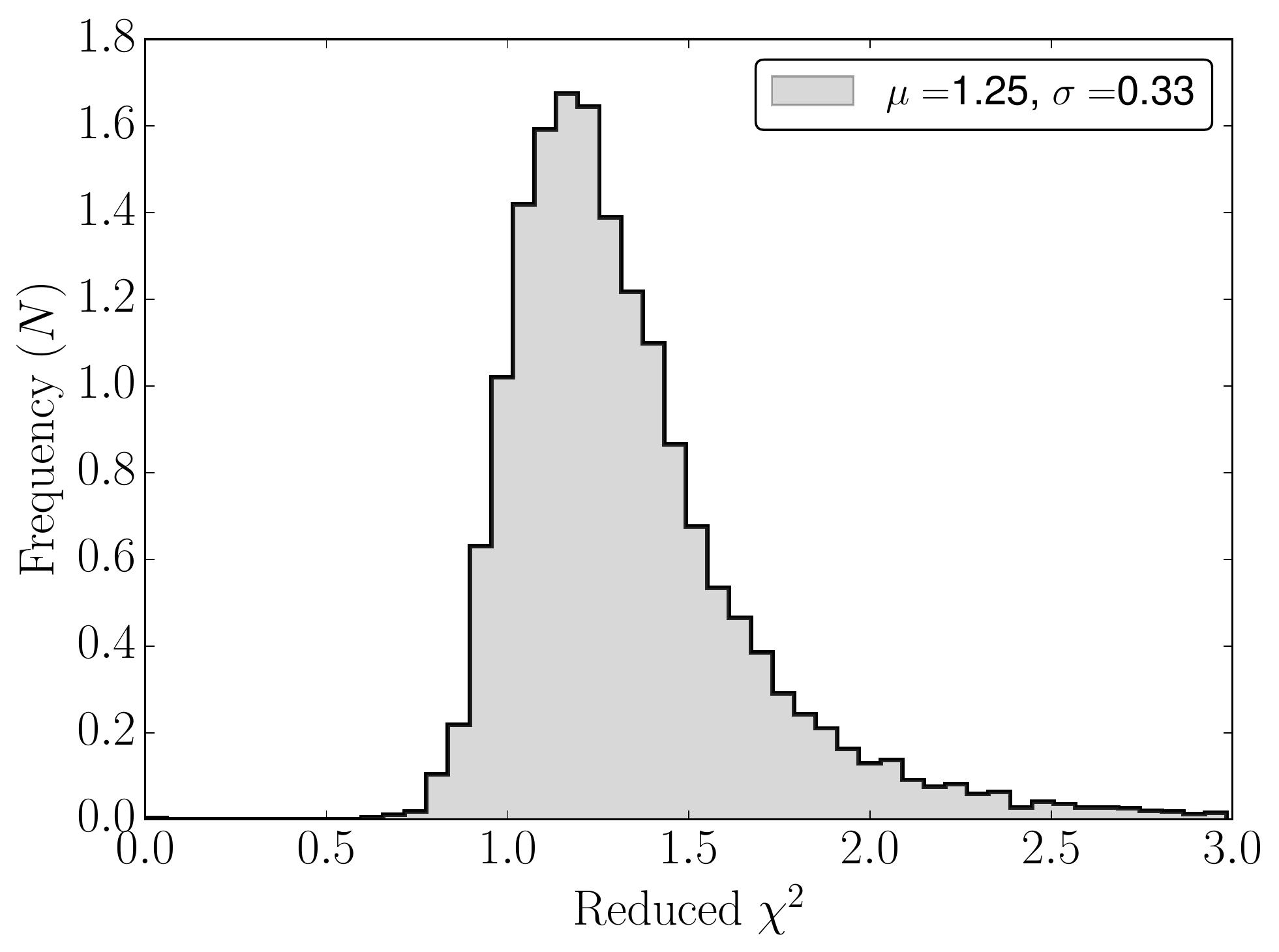}
	\caption{STELIB-based models.}
\end{subfigure}
\begin{subfigure}[b]{0.75\linewidth}
	\includegraphics[width=\linewidth]{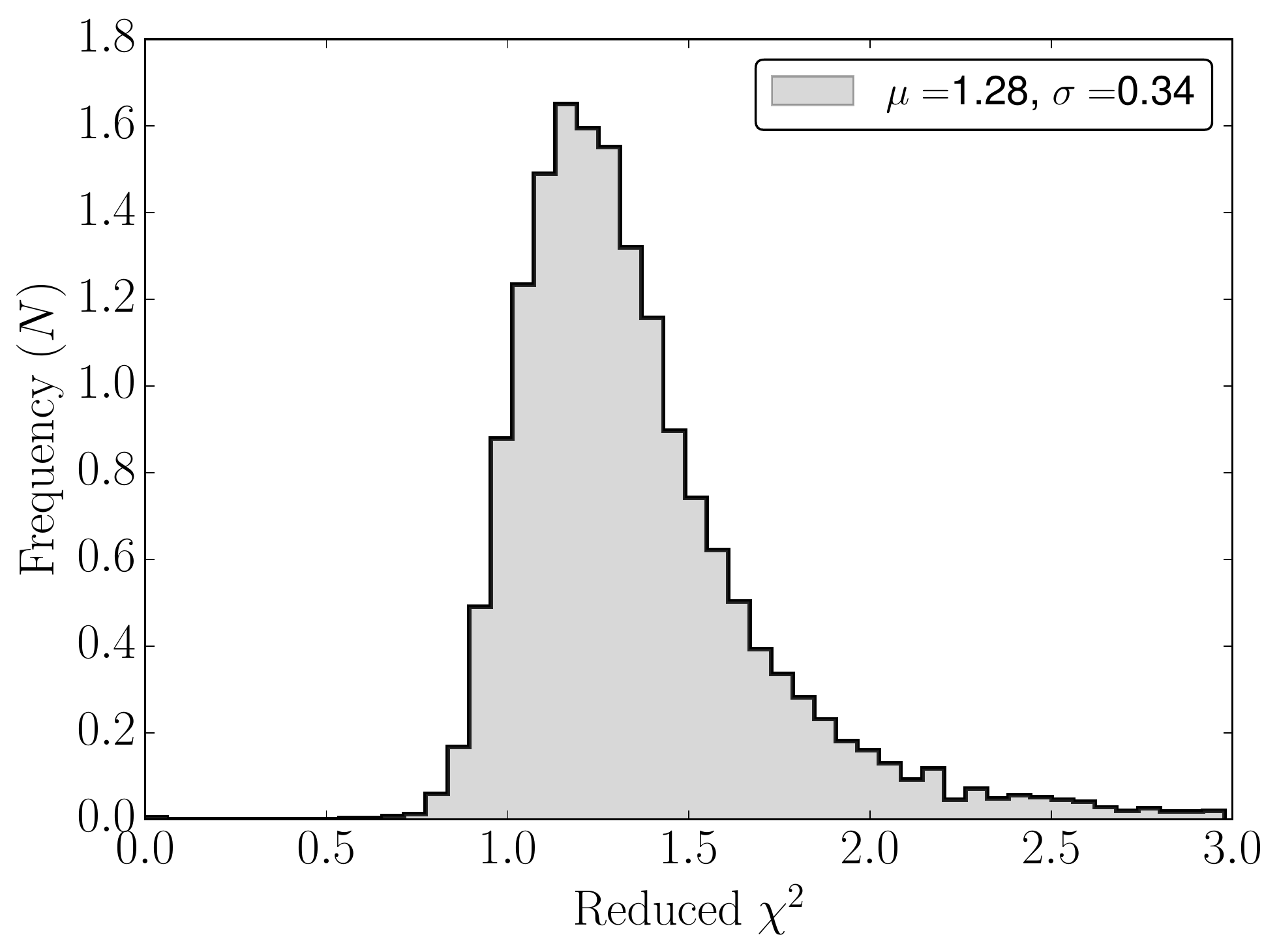}
	\caption{ELODIE-based models.}
\end{subfigure}
\caption[Chi-squared distributions of fitting 3 models to DR7 data.]{Reduced chi-squared distribution of fitting all DR7 galaxies with MILES-, STELIB- and ELODIE-based models, which have corresponding mean averages of 1.13, 1.10, and 1.19 respectively.}
\label{chis_dr7}
\end{figure}
The chi-squared distributions for the three explored models are shown in Figure \ref{chis_dr7}. The distributions are very similar with  median reduced chi-squared values of 1.27, 1.25, and 1.28 for MILES, STELIB and ELODIE-based models respectively. The differences are driven by a small number of galaxies (with no clear pattern for what characterises these galaxies) fitting somewhat more poorly for ELODIE-based models compared to the others, and STELIB-based models having slightly more galaxies with lower reduced chi-squared. Clearly, the $\chi^2$~would not work as discriminator for which model one should use.

\subsection{Comparisons with the Literature}
We compare the results of \FF\ applied to SDSS DR7 galaxies with the results of two popular full spectral fitting codes, the methodologies of which are described in Section \ref{comparecodes}. These codes have publicly accessible and published databases of fitting results from DR7. We note that these databases use different sets of stellar population models, although one of them (VESPA) provides also Maraston-type of models. Hence there is a degeneracy between codes and models which is difficult to separate. Nonetheless, these are still useful comparisons to make.
\\
\\
Firstly, we compare our results to VESPA using their DR7 database \citep{2009ApJS..185....1T}. \footnote{Available at \url{http://www-wfau.roe.ac.uk/vespa/}/} In Figure \ref{vespa_compare_dr7}, we show the results of summing the stellar populations found in all DR7 galaxies for the stellar population models of \citet{2005MNRAS.362..799M} (M05) after VESPA fitting. This distribution should be compared to the ones showed in the right-hand panels of Figure~\ref{dr7sfh}}. The adopted models use the same evolutionary population synthesis code and the same prescriptions for stellar evolution, the only difference being the input stellar library. We note that M11 has greater spectral resolution compared to M05, potentially leading to resolving greater details in galaxies' star formation histories. On the other hand, M05 has a larger extension in wavelength and metallicity. In addition, we note that VESPA is written to give solutions with lower resolution in parameter space than \FF. \\
\begin{figure}
	\includegraphics[width=\linewidth]{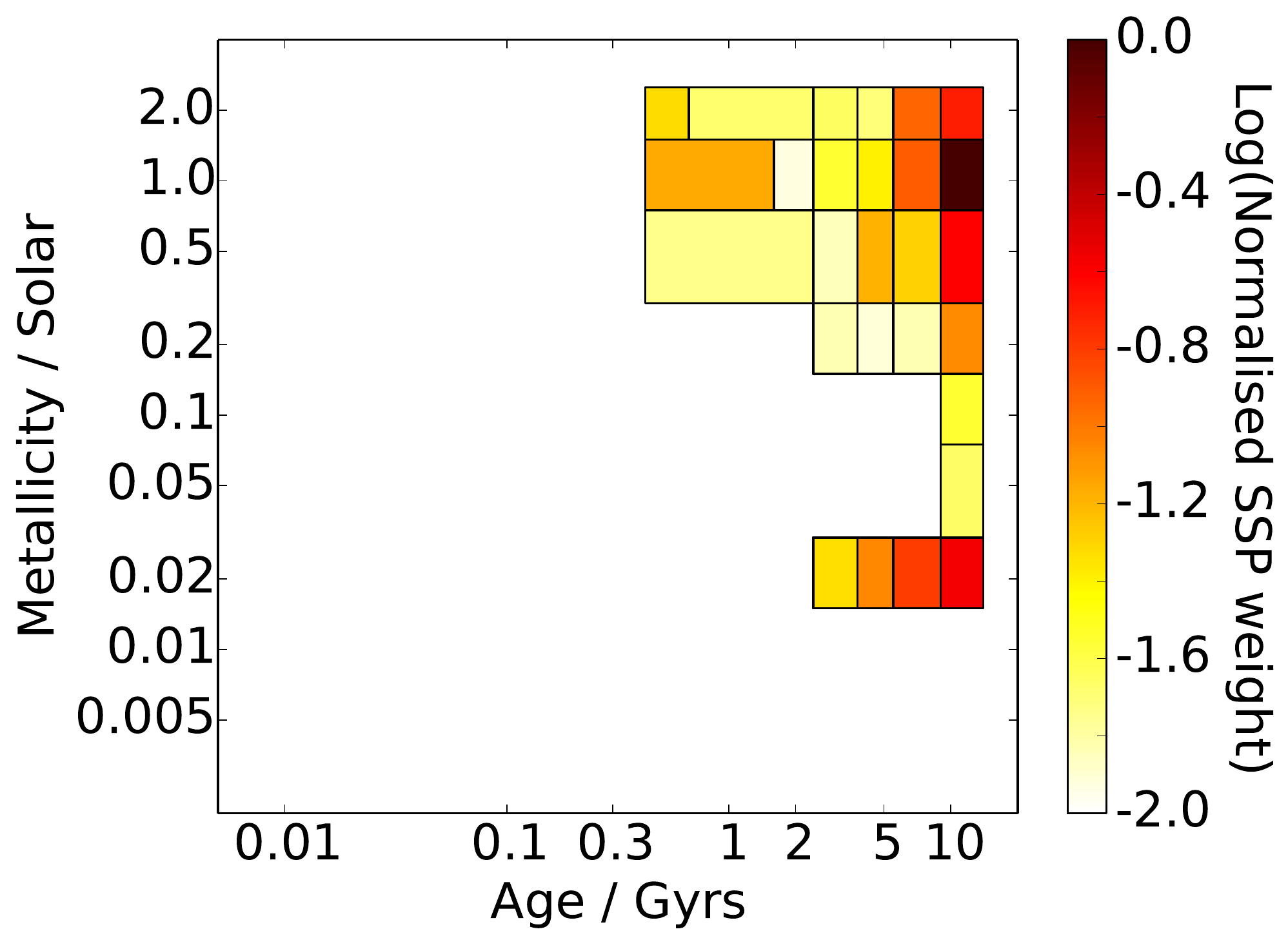}
\caption[Comparison of recovered stellar mass between \FF\ and VESPA.]{Stellar population mass contributions for DR7 galaxies obtained with VESPA and M05 models. The VESPA plot has age-metallicity bins as defined in \cite{2009ApJS..185....1T}, with the colours representing the total weighted mass in that bin, compared to \FF's weights based on the total mass of SSPs in that region of the plot. This plot should be compared to the right-hand panels of Figure~\ref{dr7sfh}.}
\label{vespa_compare_dr7}
\end{figure}
Taking all this into account, the distributions are broadly speaking quite similar, displaying a major peak of galaxies at old ages and high metallicity. The VESPA-M05 distribution is closer to the FF-MILES or ELODIE-based results because of the more similar metallicity coverage. Looking in more detail, the plots show the different priors made in the fitting codes. \FF\ allows for more contributions from individual SSPs compared to VESPA's more strict allowance of introducing more complex star formation histories, hence the age/metallicity distributions are somewhat broader in the \FF\ case. Additionally, \FF\ finds the major, old age contribution at high, super-solar metallicity while the one in VESPA lies at solar metallicity. 

\begin{figure}
\centering
\begin{subfigure}{\linewidth}
	\includegraphics[width=\linewidth]{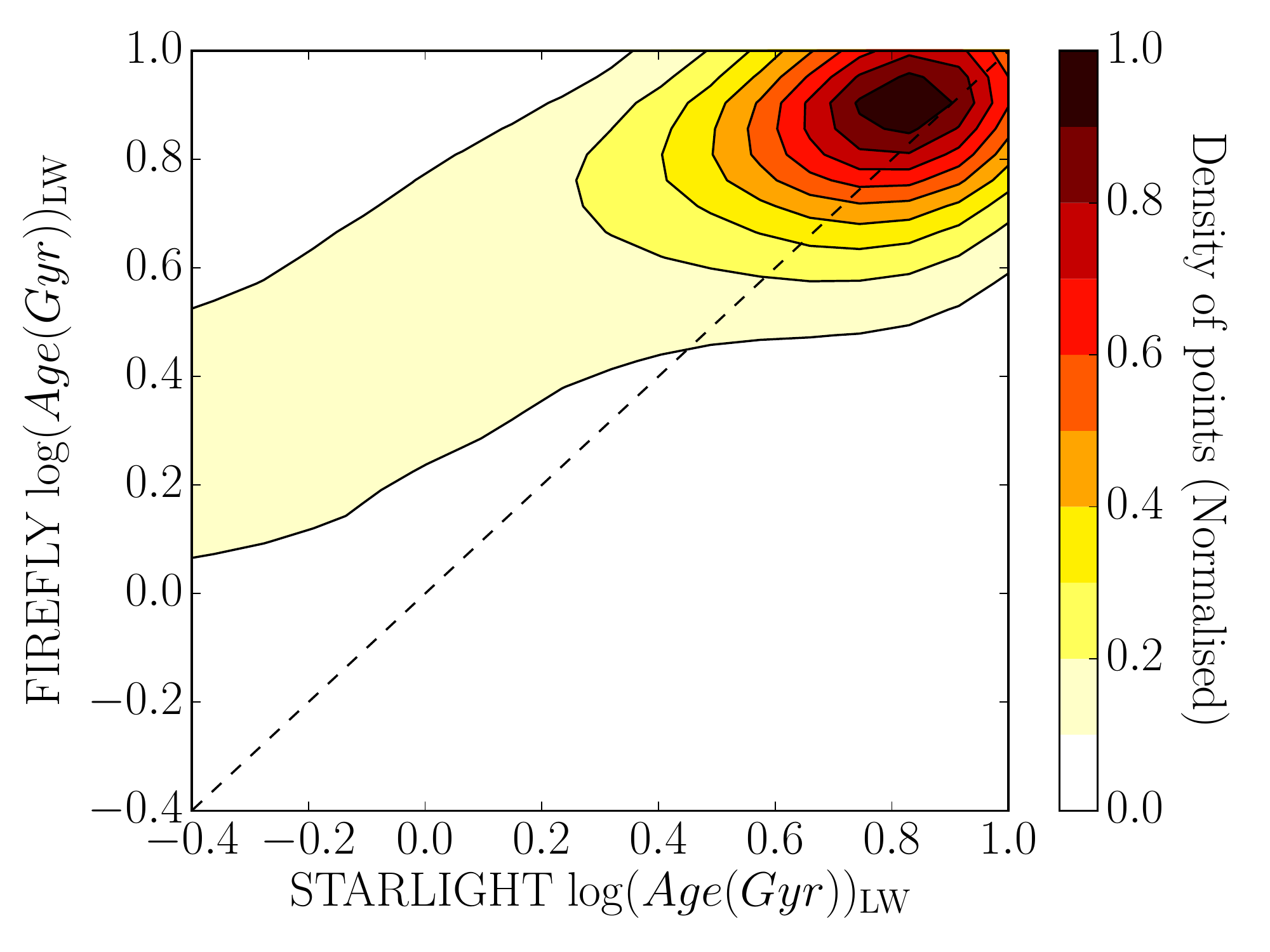}
	\caption{STARLIGHT's light-weighted ages derived from DR7 compared with \FF. Contours represent the density of points.}
\end{subfigure}
\begin{subfigure}{\linewidth}
	\includegraphics[width=\linewidth]{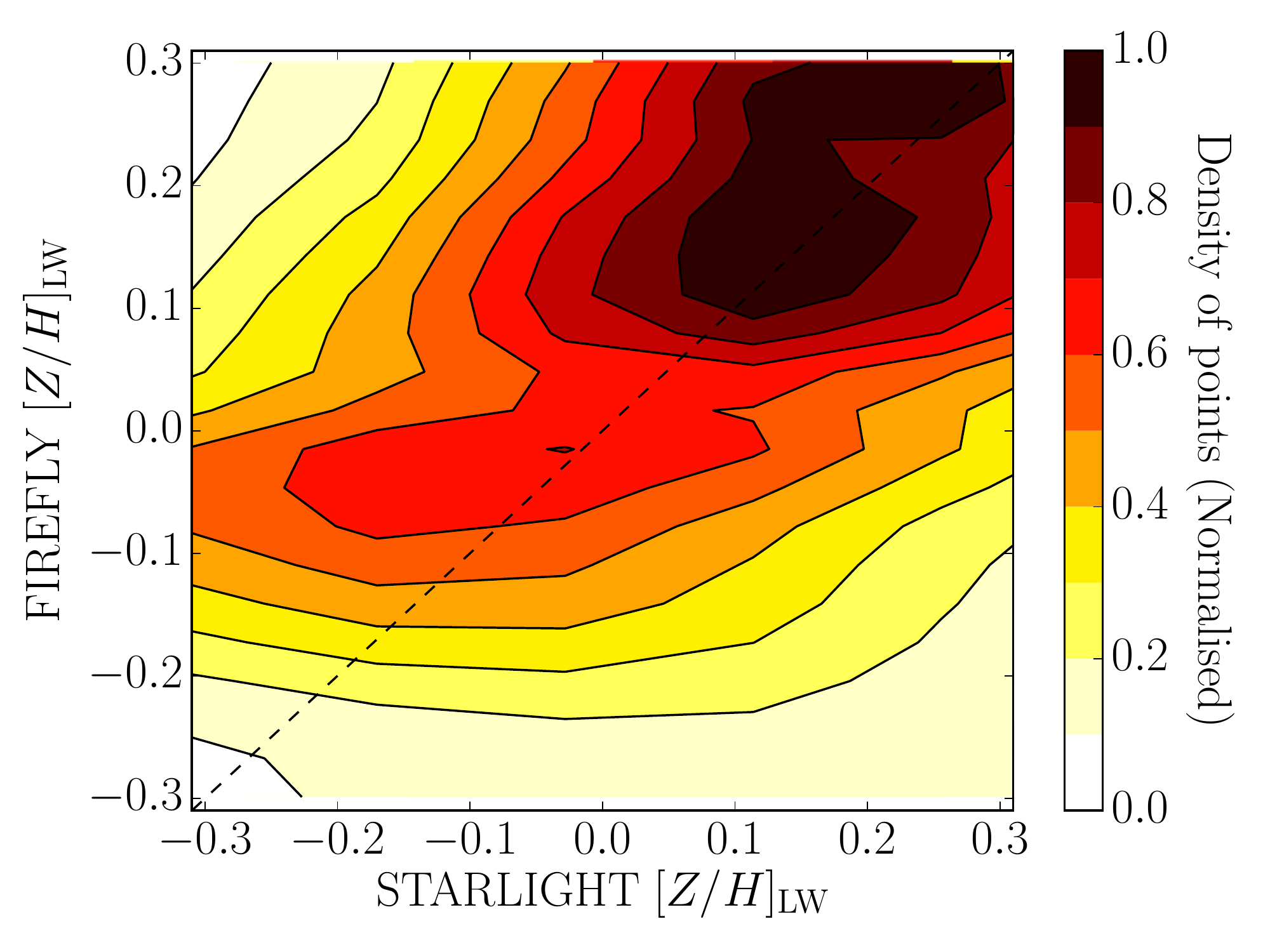}
	\caption{STARLIGHT's light-weighted metallicities derived from DR7 compared with \FF. Contours represent the density of points.}
\end{subfigure}
\caption[Comparisons of recovered age and metallicity between \FF\ and STARLIGHT.]{Comparison of the light-weighted average ages and metallicities obtained with STARLIGHT and \FF. STARLIGHT's results are obtained using BC03 models, and \FF's are obtained with M11 models, but both models are based on the STELIB library.}
\label{starlight_compare_dr7}
\end{figure}

Nonetheless, qualitatively the results agree on a old, high metallicity stellar component being dominant for the DR7 sample, hence the galaxy evolution picture from both codes would probably be similar. The fact that we use the same underlying population model here certainly makes a large part of this consistency.
\\
\\
Secondly, we compare our results with the analysis of the DR7 performed by STARLIGHT \citep{2005MNRAS.358..363C}\footnote{Available at \url{http://casjobs.starlight.ufsc.br/casjobs/}.}, which is based on BC03 STELIB-based models. Given the flexibility of our models, we can compare with M11-STELIB, which should help reducing the differences due to the input models. Unlike VESPA's database, a full breakdown of stellar population components is not available; instead they provide mass and light-weighted average properties, hence we shall compare to those. In Figure \ref{starlight_compare_dr7} we plot the light-weighted properties, which, by spanning a larger range of values, enable a more distinctive comparison. We plot density contours of the $\sim$ 1 million points from directly plotting STARLIGHT results against \FF\ results. \\
We see that - in spite of different procedure and input model - the main density of age and metallicity points are very similar between STARLIGHT and \FF, clustering around 8 Gyr in age and 0.1 to 0.3 dex in [Z/H]. Hence for the total sample, the codes agree very well. Some differences are visible however. Firstly, in the age plot, the extension down to low ages is flatter in \FF\ compared to STARLIGHT, suggesting either that STARLIGHT is more sensitive to small star formation episodes in the star forming sample, or that the BC03 models fit for younger ages compared to our models. Only running STARLIGHT with our models, which is beyond our scope here, could choose between one of these hypotheses. Secondly, also in the metallicity plot, we see a flatter profile from \FF\ out to low metallicity. In \citet{2017MNRAS.466.4731G} we present an extensive comparison of stellar population properties obtained with \FF\ and STARLIGHT, and swap stellar population models between the codes to be able to pin down the interplay of fitting procedure and model.
\\
\\ 
In conclusion, despite differences in the details of the minor stellar population components, we conclude that we have good qualitative agreement with the literature on full spectral fitting of DR7 galaxies even when different stellar population models are used.
\section{Summary and Conclusions}\label{conclusions}
We presented \FF, a full spectral fitting code designed to recover the stellar population properties, such as light-and mass weighted ages, metallicity, reddening and the star formation history, of stellar systems - galaxies and star clusters - from stellar population model fitting to observed spectra. \FF~employs a chi squared minimisation procedure that fits combinations of single-burst stellar population models to spectroscopic data, following an iterative best-fitting process controlled by the Bayesian Information Criterion. No priors are applied, rather all solutions within a statistical cut are retained with their weight. Moreover, no additive or multiplicative polynomia are employed to adjust the spectral shape. This fitting freedom without adjustments is envisaged in order to map out intrinsic degeneracies and explore the effect of changing models and their components. It is a code written from a modeller perspective. 

In addition, we explore an innovative method for including the effects of dust attenuation. In this method reddening is not treated as an additional adjustable free parameter, rather it is determined prior to the fitting by comparing the large modes of data and models. This method was studied in a previous paper (Wilkinson et al. 2015) using IFU data in which dust regions were easily spotted. We plan nonetheless to experiment with different options in future developments of the code.  

We check the performances of our code through comprehensive testing, using a large suite of mock galaxies, representing both simple and complex populations, as well as a sample of well-studied Milky Way globular clusters and galaxies from the Sloan Digital Sky Survey. 

We use Monte Carlo simulations to measure and quantify how offsets and errors in age, metallicity and mass vary as a function of signal-to-noise of the data, which is a crucial information to plan new observations, and input model parameters.
We show that the code is able to recover stellar population properties such as age, metallicity and stellar mass, and even the star formation history, down to a $S/N\sim 5$, for moderately dusty systems. At $S/N\sim 20$ the recovery of the star formation history is remarkably good independently of reddening, unless the star formation is very extended ($\sim 10$ Gyr). Even in the latter case, though, in spite of a lower precision, we find that the code releases multiple generation components, hence it is still able to discriminate between short and extended formation histories.
\\
 For a sample of Milky Way GCs, we find a very good match of their ages and metallicities as independently determined via Colour-Magnitude diagram fit and stellar spectroscopy. We plan to use \FF~to investigate the event of multiple populations in extra-galactic globular clusters. 
 
 Lastly, we have run \FF~on $\sim1$~million galaxies from SDSS DR7 and compared the results with previous analysis based on other fitting codes, such as STARLIGHT \citep{2005MNRAS.358..363C}  and VESPA \citep{2007MNRAS.381.1252T, 2009ApJS..185....1T},. We find generally consistent ages and metallicities for the bulk population of SDSS galaxies. Using SDSS galaxies we have also analysed the effect of input stellar library in the population model. We find that the overall effect is small, although interesting differences in some regions of the age/metallicity diagram emerge, with MILES-based models giving the largest number of old ages and high (super-solar) metallicity, whereas ELODIE-based models are more consistent with a bulk solar metallicity and a metallicity spread. \\
Code, input files and other results are publicly available at www.icg.port.ac.uk/firefly.  
\\
\\

\section{Acknowledgements}
We would like to thank Rita Tojeiro for suggesting to use a high-pass filter function for the treatment of dust. We thank Johan Comparat, Michele Moresco and Zheng Zheng for performing early tests which allowed us to improve the code, and Alice Concas and Pierandrea Guarnieri for testing the code before its public release. We also thank Matthew Bershady, Kevin Bundy and Kyle Westfall for useful discussions on methods and statistics and Violeta Gonzalez-Perez, Jianhui Lian, Sofia Meneses-Goytia and Thomas De Boer for valuable comments. Violeta Gonzalez-Perez is also thanked for leading the preparation of GitHub pages for making the code publicly available. We thank anonymous referees for their careful reading of the manuscript, and for providing comments that improved the presentation. Finally, we are grateful to the MNRAS Editors and Assistant Editors for their support. Numerical computations were done on the Sciama High Performance Compute (HPC) cluster which is supported by the Institute of Cosmology of Gravitation, SEPNet and the University of Portsmouth. DG is supported by an STFC PhD studentship and TP is supported by funding from the University of Portsmouth.
 
 \bibliographystyle{mnras}

\newpage

\appendix

\section{Effect from emission lines}
\label{App:AppendixA}
\begin{figure}
\begin{center}
\includegraphics[width=8.5cm]{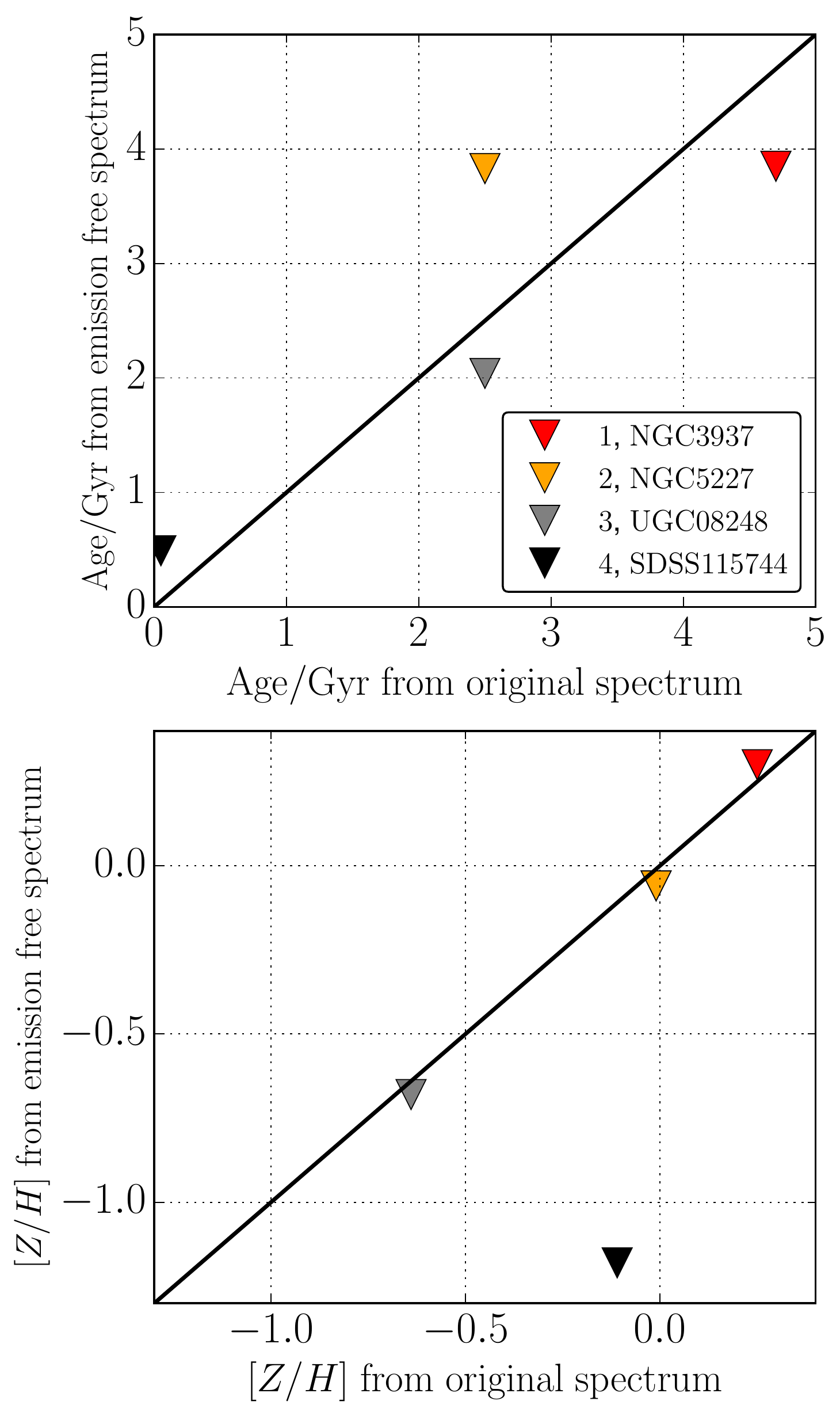}
\caption{Light-weighted ages and metallicities for the four different galaxies, labelled as in the text ($1$ = quiescent elliptical (red),$2$ = spiral galaxy bulge (orange), $3$ = dusty star forming galaxy (grey), $4$ = HII star forming dwarf (black)), as derived from the original spectrum inclusive (x-axis) or not (y-axis) of emission-lines.}
\label{emission_line_analysis}
\end{center}
\end{figure}

To explore the impact of emission-lines on the observed spectrum to be fitted with \FF, we carried out the following exercise. We ran \FF\ on the original observed spectra and on the same after removal of the strongest emission-lines, and compared the resulting ages and metallicities. Emission lines have been removed using a version of the SDSS-IV MANGA Data Analysis Pipeline (Westfall et al. {\it in prep.}), which is based on the code pPXF. This analysis was done on four galaxies from the SDSS covering a range of morphological types and emission-line strengths, namely: a quiescent elliptical (NGC3937, 1), a spiral galaxy bulge (NGC5227, 2), a dusty star forming galaxy (UGC08248, 3) and a HII star forming dwarf (SDSS115744, 4). 
The results are reported in Figure \ref{emission_line_analysis}. Overall, there appears to be no systematic offset in the age and metallicity derived between the original and emission-line free spectrum, which is very interesting. The derived metallicity in particular is remarkably robust independent of emission-lines. It seems that the full spectral fitting by modelling a large number of flux points gives results that are not too dependent on the presence of narrow emission lines randomly distributed along the spectrum. Not surprisingly, the most affected object is the emission-line dominated dwarf (4). These results performed on four objects only should be regarded as indicative, and should be obtained for a larger number of objects in the future.

\end{document}